\documentclass[longauth]{aa}

\usepackage{txfonts}
\usepackage{graphicx}
\usepackage{natbib}
\usepackage{xspace}
\bibpunct{(}{)}{;}{a}{}{,}

\def\mhmpc{\,h^{-1}~{\rm Mpc}}

\def\mhmpcc{\,h^{-3}~{\rm Mpc^3}}
\def\mhgpcc{\,h^{-3}~{\rm Gpc^3}}
\def\mihmpcc{\,h^{3}~{\rm Mpc^{-3}}}

\def\mhmsun{\,h^{-1}~{\rm M_{\odot}}}

\def\mxism{\xi(s,\mu)}
\def\Pdd{P_{\rm{\delta\delta}}}
\def\Pdt{P_{\rm{\delta\theta}}}
\def\Ptt{P_{\rm{\theta\theta}}}
\def\Pgg{P_{\rm{gg}}}
\def\Pgt{P_{\rm{g\theta}}}
\def\d{\mathrm d}
\def\rpe{r_{\perp}}
\def\rpa{r_{\parallel}}

\newcommand{\kms}{$\,{\rm km\, s^{-1}}$\xspace}
\newcommand{\SSR}{${\rm SSR}$\xspace}
\newcommand{\TSR}{${\rm TSR}$\xspace}

\newcommand{\ESR}{${\rm ESR}$\xspace}

\begin{document}

\title{The VIMOS Public Extragalactic Redshift Survey (VIPERS)\thanks{
    Based on observations collected at the European Southern
    Observatory, Cerro Paranal, Chile, using the Very Large Telescope
    under programmes 182.A-0886 and partly 070.A-9007.  Also based on
    observations obtained with MegaPrime/MegaCam, a joint project of
    CFHT and CEA/DAPNIA, at the Canada-France-Hawaii Telescope (CFHT),
    which is operated by the National Research Council (NRC) of
    Canada, the Institut National des Sciences de l’Univers of the
    Centre National de la Recherche Scientifique (CNRS) of France, and
    the University of Hawaii. This work is based in part on data
    products produced at TERAPIX and the Canadian Astronomy Data
    Centre as part of the Canada-France-Hawaii Telescope Legacy
    Survey, a collaborative project of NRC and CNRS. The VIPERS web
    site is http://www.vipers.inaf.it/.}  }

\subtitle{Gravity test from the combination of redshift-space
  distortions and galaxy-galaxy lensing at $0.5<z<1.2$}

\titlerunning{Gravity test from RSD and galaxy-galaxy lensing in VIPERS}

\author{
  S.~de~la~Torre\inst{\ref{lam}}
  \and E.~Jullo\inst{\ref{lam}}
  \and C.~Giocoli\inst{\ref{lam}}
  \and A.~Pezzotta\inst{\ref{brera},\ref{bicocca}}
  \and J.~Bel\inst{\ref{cpt}}
  \and B.~R.~Granett\inst{\ref{brera}}
  \and L.~Guzzo\inst{\ref{brera},\ref{unimi}}
  \and B.~Garilli\inst{\ref{iasf-mi}}
  \and M.~Scodeggio\inst{\ref{iasf-mi}}
  \and M.~Bolzonella\inst{\ref{oabo}}
  \and U.~Abbas\inst{\ref{oa-to}}
  \and C.~Adami\inst{\ref{lam}}
  \and D.~Bottini\inst{\ref{iasf-mi}}
  \and A.~Cappi\inst{\ref{oabo},\ref{nice}}
  \and O.~Cucciati\inst{\ref{unibo},\ref{oabo}}
  \and I.~Davidzon\inst{\ref{lam},\ref{oabo}}
  \and P.~Franzetti\inst{\ref{iasf-mi}}
  \and A.~Fritz\inst{\ref{iasf-mi}}
  \and A.~Iovino\inst{\ref{brera}}
  \and J.~Krywult\inst{\ref{kielce}}
  \and V.~Le Brun\inst{\ref{lam}}
  \and O.~Le F\`evre\inst{\ref{lam}}
  \and D.~Maccagni\inst{\ref{iasf-mi}}
  \and K.~Ma{\l}ek\inst{\ref{warsaw-nucl}}
  \and F.~Marulli\inst{\ref{unibo},\ref{infn-bo},\ref{oabo}}
  \and M.~Polletta\inst{\ref{iasf-mi},\ref{marseille-uni},\ref{toulouse}}
  \and A.~Pollo\inst{\ref{warsaw-nucl},\ref{krakow}}
  \and L.A.M.~Tasca\inst{\ref{lam}}
  \and R.~Tojeiro\inst{\ref{st-andrews}}
  \and D.~Vergani\inst{\ref{iasf-bo}}
  \and A.~Zanichelli\inst{\ref{ira-bo}}
  \and S.~Arnouts\inst{\ref{lam}}
  \and E.~Branchini\inst{\ref{roma3},\ref{infn-roma3},\ref{oa-roma}}
  \and J.~Coupon\inst{\ref{geneva}}
  \and G.~De Lucia\inst{\ref{oats}}
  \and O.~Ilbert\inst{\ref{lam}}
  \and T.~Moutard\inst{\ref{halifax},\ref{lam}}
  \and L.~Moscardini\inst{\ref{unibo},\ref{infn-bo},\ref{oabo}}
  \and J.~A.~Peacock\inst{\ref{roe}}
  \and R.~B.~Metcalf\inst{\ref{unibo}}
  \and F.~Prada\inst{\ref{prada1},\ref{prada2},\ref{prada3}}
  \and G.~Yepes\inst{\ref{duam}}
}

\institute{
  Aix Marseille Univ, CNRS, LAM, Laboratoire d'Astrophysique de Marseille, Marseille, France  \label{lam}
  \and INAF - Osservatorio Astronomico di Brera, Via Brera 28, 20122 Milano --  via E. Bianchi 46, 23807 Merate, Italy \label{brera}
  \and Dipartimento di Fisica, Universit\`a di Milano-Bicocca, P.zza della Scienza 3, I-20126 Milano, Italy \label{bicocca}
  \and Aix Marseille Univ, Univ Toulon, CNRS, CPT, Marseille, France \label{cpt}
  \and Universit\`{a} degli Studi di Milano, via G. Celoria 16, 20133 Milano, Italy \label{unimi}
  \and INAF - Istituto di Astrofisica Spaziale e Fisica Cosmica Milano, via Bassini 15, 20133 Milano, Italy \label{iasf-mi}
  \and INAF - Osservatorio Astronomico di Bologna, via Gobetti 93/3, I-40129, Bologna, Italy \label{oabo}
  \and INAF - Osservatorio Astrofisico di Torino, 10025 Pino Torinese, Italy \label{oa-to}
  \and Laboratoire Lagrange, UMR7293, Universit\'e de Nice Sophia Antipolis, CNRS, Observatoire de la C\^ote d’Azur, 06300 Nice, France \label{nice}
  \and Dipartimento di Fisica e Astronomia - Alma Mater Studiorum Universit\`{a} di Bologna, viale Berti Pichat 6/2, I-40127 Bologna, Italy \label{unibo} 
  \and Institute of Physics, Jan Kochanowski University, ul. Swietokrzyska 15, 25-406 Kielce, Poland \label{kielce}
  \and National Centre for Nuclear Research, ul. Hoza 69, 00-681 Warszawa, Poland \label{warsaw-nucl}
  \and INFN, Sezione di Bologna, viale Berti Pichat 6/2, I-40127 Bologna, Italy \label{infn-bo}
  \and Aix-Marseille Université, Jardin du Pharo, 58 bd Charles Livon, F-13284 Marseille cedex 7, France \label{marseille-uni}
  \and IRAP,  9 av. du colonel Roche, BP 44346, F-31028 Toulouse cedex 4, France \label{toulouse}
  \and Astronomical Observatory of the Jagiellonian University, Orla 171, 30-001 Cracow, Poland \label{krakow}
  \and School of Physics and Astronomy, University of St Andrews, St Andrews KY16 9SS, UK \label{st-andrews}
  \and INAF - Istituto di Astrofisica Spaziale e Fisica Cosmica Bologna, via Gobetti 101, I-40129 Bologna, Italy \label{iasf-bo}
  \and INAF - Istituto di Radioastronomia, via Gobetti 101, I-40129, Bologna, Italy \label{ira-bo}
  \and Dipartimento di Matematica e Fisica, Universit\`{a} degli Studi Roma Tre, via della Vasca Navale 84, 00146 Roma, Italy\label{roma3}
  \and INFN, Sezione di Roma Tre, via della Vasca Navale 84, I-00146 Roma, Italy \label{infn-roma3}
  \and INAF - Osservatorio Astronomico di Roma, via Frascati 33, I-00040 Monte Porzio Catone (RM), Italy \label{oa-roma}
  \and Department of Astronomy, University of Geneva, ch. d’Ecogia 16, 1290 Versoix, Switzerland \label{geneva}
  \and INAF - Osservatorio Astronomico di Trieste, via G. B. Tiepolo 11, 34143 Trieste, Italy \label{oats}
  \and Department of Astronomy \& Physics, Saint Mary's University, 923 Robie Street, Halifax, Nova Scotia, B3H 3C3, Canada \label{halifax}
  \and Institute for Astronomy, University of Edinburgh, Royal Observatory, Blackford Hill, Edinburgh EH9 3HJ, UK \label{roe}
  \and Instituto de F\'isica Te\'orica, (UAM/CSIC), Universidad Aut\'onoma de Madrid, Cantoblanco, E-28049 Madrid, Spain \label{prada1}
  \and Campus of International Excellence UAM+CSIC, Cantoblanco, E-28049 Madrid, Spain \label{prada2}
  \and Instituto de Astrof\'{i}sica de Andaluc\'ia (CSIC), Glorieta de la Astronom\'ia, E-18080 Granada, Spain \label{prada3}
  \and Departamento de F\'isica Te\'orica, M-8, Universidad Aut\'onoma de Madrid, Cantoblanco, E-28049 Madrid, Spain \label{duam}
}

\authorrunning{S. de la Torre et al.}

\offprints{\mbox{S.~de~la~Torre},\\ \email{sylvain.delatorre@lam.fr}}

\abstract{We carry out a joint analysis of redshift-space distortions
  and galaxy-galaxy lensing, with the aim of measuring the growth rate
  of structure; this is a key quantity for understanding the nature of
  gravity on cosmological scales and late-time cosmic acceleration. We
  make use of the final VIPERS redshift survey dataset, which maps a
  portion of the Universe at a redshift of $z\simeq0.8$, and the
  lensing data from the CFHTLenS survey over the same area of the
  sky. We build a consistent theoretical model that combines
  non-linear galaxy biasing and redshift-space distortion models, and
  confront it with observations. The two probes are combined in a
  Bayesian maximum likelihood analysis to determine the growth rate of
  structure at two redshifts $z=0.6$ and $z=0.86$. We obtain
  measurements of $f\sigma_8(0.6)=0.48\pm0.12$ and
  $f\sigma_8(0.86)=0.48\pm0.10$. The additional galaxy-galaxy lensing
  constraint alleviates galaxy bias and $\sigma_8$ degeneracies,
  providing direct measurements of $f$ and $\sigma_8$:
  $\left[f(0.6),\sigma_8(0.6)\right]=[0.93\pm0.22,0.52\pm0.06]$ and
  $\left[f(0.86),\sigma_8(0.86)\right]=[0.99\pm0.19,0.48\pm0.04]$. These
  measurements are statistically consistent with a Universe where the
  gravitational interactions can be described by General Relativity,
  although they are not yet accurate enough to rule out some commonly
  considered alternatives. Finally, as a complementary test we measure
  the gravitational slip parameter, $E_G$, for the first time at
  $z>0.6$. We find values of $\smash{\overline{E}_G}(0.6)=0.16\pm0.09$
  and $\smash{\overline{E}_G}(0.86)=0.09\pm0.07$, when $E_G$ is
  averaged over scales above $3\mhmpc$. We find that our $E_G$
  measurements exhibit slightly lower values than expected for
  standard relativistic gravity in a $\Lambda\rm{CDM}$ background,
  although the results are consistent within $1-2\sigma$.}
 

\maketitle

\section{Introduction}

The origin of the late-time acceleration of the universal expansion is
a major question in cosmology. The source of this acceleration and its
associated energy density are crucial in understanding the properties
of the Universe and its evolution and fate.  In the standard
cosmological model, this cosmic acceleration can be associated with
the presence of a dark energy component, a cosmological fluid with
negative pressure, which opposes the gravitational force on large
scales. However, this apparent acceleration can conversely be
interpreted as a failure of the standard relativistic theory of
gravity. A key goal for cosmology is therefore to investigate the
nature of gravity empirically. To be clear, what can potentially be
falsified is the validity of Einstein's field equations, rather than
General Relativity itself; this sets a broader framework within which
Einstein gravity or modified alternatives can operate.

The large-scale structure of the Universe has proved to be very
powerful for testing the cosmological model through the use of various
observables such as the two-point statistics of the galaxy
distribution and its features \citep[e.g.][and references
  therein]{peacock01,cole05,tegmark04,eisenstein05,guzzo08,percival10,
  beutler11,blake12,anderson14,alam16}. In this context, a unique
probe of gravitational physics is the large-scale component of galaxy
peculiar velocities affecting the observed galaxy distribution in
redshift surveys \citep{guzzo08}, sensitive to the growth rate of
structure $f$ defined as $\d\ln D/\d\ln a$, where $D$ and $a$ are
respectively the linear growth factor and scale factor. In turn, the
growth rate of structure tells us about the strength of gravity acting
on cosmological scales and is a direct prediction of gravity
theories. The distortions induced by peculiar velocities in the
apparent galaxy clustering, the so-called redshift-space distortions
(RSD), are a very important cosmological probe of the nature of
gravity. In the last decade, they have been studied in large galaxy
redshift surveys, showing a broad consistency with $\Lambda \rm{CDM}$
and General Relativity predictions
\citep[e.g.][]{blake12,beutler12,delatorre13a,samushia14,gilmarin16,chuang16}.

Although galaxy redshift surveys are powerful cosmological tools for
understanding the geometry and the dynamics of the Universe, they are
fundamentally limited by the inherent uncertainty related to the bias
of galaxies, the fact that these are not faithful tracers of the
underlying matter distribution. Gravitational lensing represents a
powerful probe that is complementary to galaxy redshift-space
clustering. In the weak regime in particular, the statistical shape
deformations of background galaxies probe the relativistic
gravitational deflection of light by the projected dark matter
fluctuations due to foreground large-scale structure. There are
several techniques associated with weak gravitational lensing; one
that is particularly useful for combining with galaxy clustering is
galaxy-galaxy lensing. This technique consists of studying the weak
deformations of background galaxies around foreground galaxies, whose
associated dark matter component acts as a gravitational lens. This is
particularly useful for probing the galaxy-matter cross-correlation,
which in turn provides insights on the bias of foreground galaxies and
the matter energy density $\Omega_m$, although the projected nature of
the statistic makes it insensitive to redshift-space distortions. The
combination of galaxy-galaxy lensing with redshift-space galaxy
correlations is therefore a very promising way to study gravitational
physics, given that both lensing information on background sources and
spectroscopic information on foreground galaxies are available on the
same field.

Beyond the determination of the growth rate of structure, one can
define consistency tests of gravity that are sensitive to both the
Newtonian and curvature gravitational potentials, $\Psi$ and $\Phi$
respectively \citep[e.g.][]{simpson13}. One is the gravitational slip,
$E_G$, which was originally proposed by \citet{Zhang07} and
implemented by \citet{reyes10} in terms of the ratio between the
galaxy-galaxy lensing signal and the redshift-space distortions
parameter $\beta=f/b$ times the galaxy clustering signal of the
lenses. Here $b$ is the galaxy linear bias. $E_G$ effectively tests
whether the Laplacian of $\Psi+\Phi$, to which gravitational lensing
is sensitive, and that of $\Psi$, to which galaxy peculiar velocities
are sensitive, are consistent with standard gravity predictions. In
the standard cosmological model, $E_G$ asymptotes to $\Omega_m/f$ on
large linear scales. A failure of this test would either imply an
incorrect matter energy density or a departure from standard
gravity. This test has been performed at low redshift in the SDSS
survey by \citet{reyes10} and more recently at redshifts up to
$z=0.57$ by \citet{blake16} and \citet{pullen16}.

The $E_G$ statistic is formally defined as $E_G=\Upsilon_{gm} / (\beta
\Upsilon_{gg})$, where $\Upsilon_{gm}$ and $\Upsilon_{gg}$ are
filtered versions of the real-space projected galaxy-matter and
galaxy-galaxy correlation functions respectively, and $\beta$ is the
RSD parameter. In practice, its implementation involves measuring
$\beta$ and the ratio $\Upsilon_{gm} / \Upsilon_{gg}$ separately, to
finally combine them. But since $\beta$ and $\Upsilon_{gg}$ are
extracted from the same observable, namely the anisotropic two-point
correlation function of lens galaxies, this is suboptimal and does not
account for the covariance between them. In this analysis, we follow a
different approach. We combine the galaxy-galaxy lensing quantity
$\Upsilon_{gm}$ and the redshift-space anisotropic correlation
function monopole and quadrupole moments $\xi_0$ and $\xi_2$ (from
which $\beta$ can be estimated) in a joint likelihood analysis, to
provide constraints on $f$ and gravity at redshifts above $z=0.6$. We
note that we do not include $\Upsilon_{gg}$ because of the redundant
cosmological information shared with $\xi_0$ and $\xi_2$.

The VIMOS Public Extragalactic Redshift Survey (VIPERS) is a large
galaxy redshift survey probing the $z\simeq0.8$ Universe with an
unprecedented density of spectroscopic galaxies of $5 \times
10^{-3}\mihmpcc$ and covering an overall area of $23.5~\rm{deg}^2$ on
the sky. The prime goal of VIPERS is an accurate measurement of the
growth rate of structure at redshift around unity. A first measurement
has been performed using the Public Data Release 1 (PDR-1), setting a
reference measurement of $f\sigma_8$ at $z=0.8$
\citep{delatorre13a}. The survey is now complete and several analyses
including this one are using the final dataset to produce the VIPERS
definitive growth rate of structure measurements, but following a
variety of approaches. The present analysis aims at maximizing the
cosmological information available and takes advantage of the
overlapping lensing information provided by CFHTLenS lensing survey,
to provide a precise gravity test at redshifts $0.5<z<1.2$ by
combining RSD and galaxy-galaxy lensing.

The paper is organized as follows. The data are described in Sect. 2;
Sect. 3 describes our methods for estimating galaxy clustering and
galaxy-galaxy lensing; Sect. 4 describes the theoretical modelling
that is tested in Sect. 5; Sect. 6 presents how the likelihood
analysis is constructed; Sect. 7 describes the cosmological results,
and Sect. 8 summarizes our findings and concludes.

Throughout this analysis and if not stated otherwise, we assume a flat
$\Lambda {\rm CDM}$ ($\Lambda$-Cold Dark Matter) cosmological model
with $(\Omega_m,\Omega_b,n_s)=(0.3,0.045,0.96)$ and a Hubble constant
of $H_0=100~h~\rm{km~s^{-1}~Mpc^{-1}}$.

\section{Data}

\subsection{Combined VIPERS-CFHTLenS dataset}

The VIPERS galaxy target sample was selected from the optical
photometric catalogues of the Canada-France-Hawaii Telescope Legacy
Survey Wide \citep[CFHTLS-Wide,][]{goranova09}.  VIPERS covers $23.5$
deg$^2$ on the sky, divided over two areas within the W1 and W4 CFHTLS
fields. Galaxies are selected to a limit of $i_{\rm AB}<22.5$,
applying a simple and robust $gri$ colour pre-selection to efficiently
remove galaxies at $z<0.5$. Coupled with a highly optimized observing
strategy \citep{scodeggio09}, this allows us to double the galaxy
sampling rate in the redshift range of interest, with respect to a
pure magnitude-limited sample.  At the same time, the area and depth
of the survey result in a relatively large volume, $5 \times
10^{7}\mhmpcc$, analogous to that of the Two Degree Field Galaxy
Redshift Survey (2dFGRS) at $z\simeq0.1$ \citep{colless01,colless03}.
Such a combination of sampling rate and depth is unique amongst
current redshift surveys at $z>0.5$. VIPERS spectra are collected with
the VIMOS multi-object spectrograph \citep{lefevre03} at moderate
resolution ($R=220$) using the LR Red grism, providing a wavelength
coverage of 5500-9500$\rm{\AA}$ and a redshift error corresponding to
a galaxy peculiar velocity error at any redshift of $\sigma_{\rm
  vel}=163$\kms. The full VIPERS area of $23.5$ deg$^2$ is covered
through a mosaic of 288 VIMOS pointings (192 in the W1 area, and 96 in
the W4 area).  A discussion of the survey data reduction and
management infrastructure is presented in \citet{garilli14}. A
complete description of the survey construction, from the definition
of the target sample to the actual spectra and redshift measurements,
is given in the survey description paper \citep{guzzo14}.

The data used here correspond to the publicly released PDR-2 catalogue
\citep{scodeggio16} that includes $86\,775$ galaxy spectra, with the
exception of a small sub-set of redshifts (340 galaxies missing in the
range $0.6 < z< 1.1$), for which the redshift and quality flags were
revised closer to the release date. Concerning the analysis presented
here, this has no effect. A quality flag has been assigned to each
object in the process of determining their redshift from the spectrum,
which quantifies the reliability of the measured redshifts. In this
analysis (as with all statistical analyses presented in the parallel
papers of the final science release), we use only galaxies with flags
2 to 9 inclusive, corresponding to objects with a redshift confidence
level of $96.1\%$ or larger. This has been estimated from repeated
spectroscopic observations in the VIPERS fields
\citep[see][]{scodeggio16}. The catalogue used here, which we will
refer to just as the VIPERS sample in the following, includes $76584$
galaxies with reliable redshift measurements.

In addition to the VIPERS spectroscopic sample, we make use of the
public lensing data from the Canada-France-Hawaii Lensing Survey
\citep{heymans12}, hereafter referred to as CFHTLenS. The CFHTLenS
survey analysis combined weak lensing data processing with {\sc theli}
\citep{erben13}, shear measurement with {\sc lensfit}
\citep{miller13}, and photometric redshift measurement with
PSF-matched photometry \citep{hildebrandt12}. A full systematic error
analysis of the shear measurements in combination with the photometric
redshifts is presented in \citet{heymans12}, with additional error
analyses of the photometric redshift measurements presented in
\citet{benjamin13}.

\subsection{Sample selection}

\begin{figure}
\resizebox{\hsize}{!}{\includegraphics{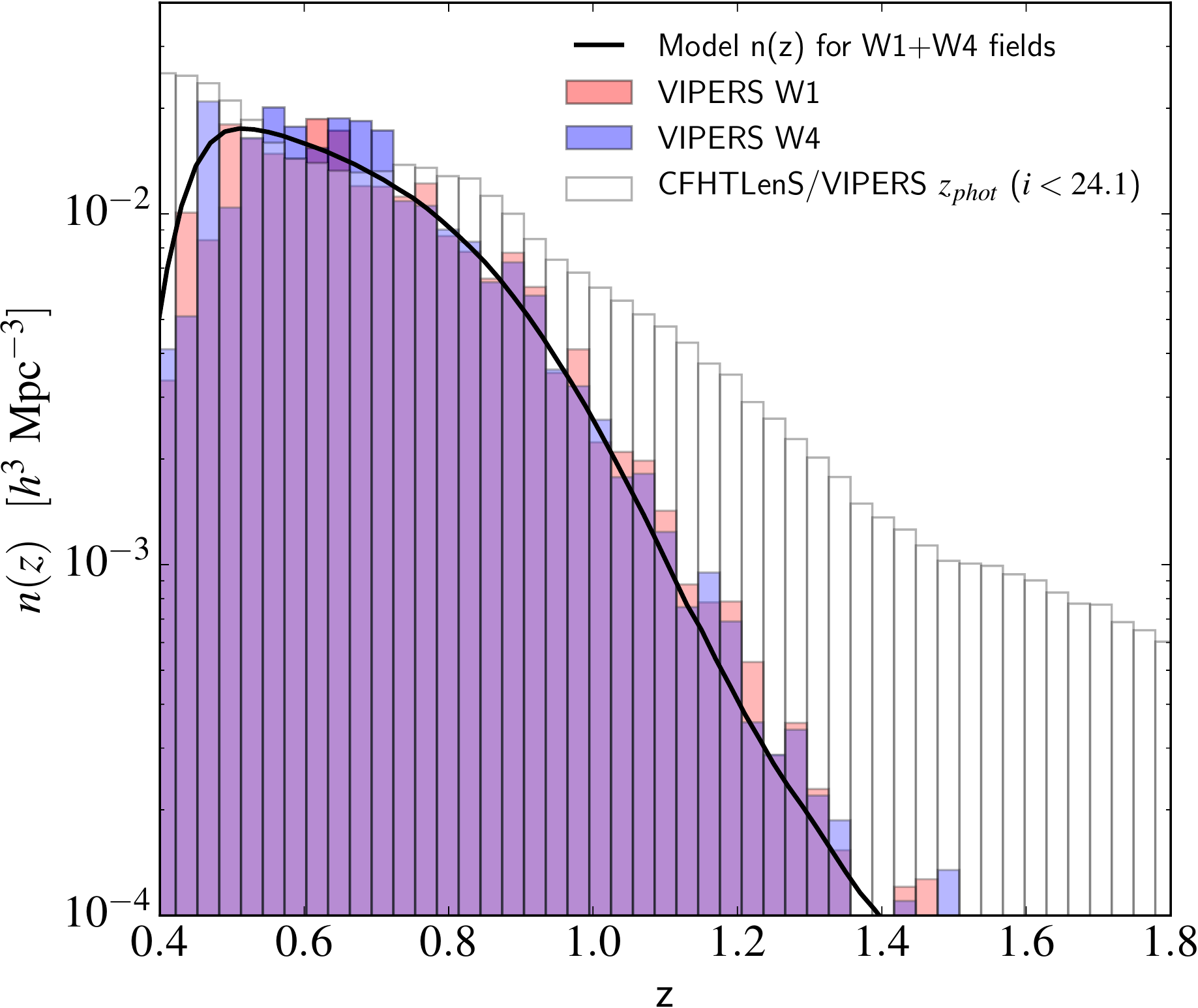}}
\caption{Number densities of VIPERS galaxies in the individual W1 and
  W4 fields and of CFHTLenS/VIPERS photometric redshift galaxies, as a
  function of redshift. The number densities of VIPERS galaxies are
  corrected for the survey incompleteness by weighting each galaxy in
  the counts by its associated inverse completeness weight $w^C$. The
  solid curve corresponds to the model $n(z)$ used in the analysis. It
  was obtained by randomly sampling galaxy redshifts within their
  $V_{\rm max}$ (see text for details).}
\label{fig:nz}
\end{figure}

For this analysis, we define two redshift intervals covering the full
volume of the VIPERS survey: $0.5<z<0.7$ and $0.7<z<1.2$. The number
density of galaxies in the combined W1 and W4 fields is presented in
Fig. \ref{fig:nz}, after correction with survey incompleteness weights
$w^C$ (see Sect. \ref{sec:gcestimation}). It is worth emphasizing that
after application of survey incompleteness corrections, the VIPERS
spectroscopic sample represents a statistically unbiased subset of the
parent $i_{\rm AB}<22.5$ photometric catalogue
\citep{guzzo14,garilli14,scodeggio16}. The redshift distribution is
modelled using the $V_{\rm max}$ method \citep{cole11,delatorre13a}
and shown with the solid curve in the figure. In this method, we
randomly sample 500 times the $V_{\rm max}$ of each galaxy, defined as
the comoving volume between the minimum and maximum redshifts where
the galaxy is observable given its apparent magnitude and the
magnitude limit of VIPERS, $i_{\rm AB}=22.5$. The redshift
distribution thus obtained is regular and can be straightforwardly
interpolated with a smooth function, showed with the solid curve in
Fig. \ref{fig:nz}.

In addition to VIPERS spectroscopic galaxies, photometric galaxies
from the CFHTLenS survey on the overlapping areas with VIPERS survey,
have been used for the galaxy-galaxy lensing. The lens sample
satisfies the VIPERS selection $i_{\rm AB}<22.5$ and uses VIPERS
spectroscopic redshifts when available (i.e. for about $30\%$ of
objects) or CFHTLenS maximum likelihood photometric redshifts
otherwise. The sources have been selected to have $i_{\rm AB}<24.1$
and thus have a higher surface density. Sources inside the mask
delimiting bad photometric areas in the CFHTLenS catalogue have been
discarded. We also make use of the individual source redshift
probability distribution function estimates obtained from {\sc bpz}
\citep{hildebrandt12} as described in
Sect. \ref{sec:wlestimation}. Source galaxies extend above $z_{\rm
  phot}=1.4$ and their number density is represented with the unfilled
histogram in Fig. \ref{fig:nz}.

\section{Galaxy clustering and galaxy-galaxy lensing estimation}

\subsection{Anisotropic galaxy clustering estimation} \label{sec:gcestimation}

We estimate the redshift-space galaxy clustering by measuring the
two-point statistics of the spatial distribution of galaxies in
configuration space. For this we infer the anisotropic two-point
correlation function $\xi(s,\mu)$ using the \citet{landy93} estimator:
\begin{equation}
\xi(s,\mu)=\frac{GG(s,\mu)-2GR(s,\mu)+RR(s,\mu)}{RR(s,\mu)}, \label{eq:xir}
\end{equation}
where $GG(s,\mu)$, $GR(s,\mu)$, and $RR(s,\mu)$ are respectively the
normalized galaxy-galaxy, galaxy-random, and random-random number of
pairs with separation $(s,\mu)$. Since we are interested in
quantifying RSD effects, we have decomposed the three-dimensional
galaxy separation vector $\vec{s}$ into polar coordinates $(s,\mu)$,
where $s$ is the norm of the separation vector and $\mu$ is the cosine
of the angle between the line-of-sight and separation vector
directions. This estimator minimizes the estimation variance and
circumvents discreteness and finite volume effects
\citep{landy93,hamilton93}. A random catalogue needs to be
constructed, whose aim is to accurately estimate the number density of
objects in the sample. It must be an unclustered population of objects
with the same radial and angular selection functions as the data. In
this analysis, we use random samples with 20 times more objects than
in the data to minimize the shot noise contribution in the estimated
correlation functions, and the redshifts of random points are drawn
randomly from the model $n(z)$ presented in Fig. \ref{fig:nz}.

In order to study redshift-space distortions, we further extract the
multipole moments of the anisotropic correlation function
$\mxism$. This approach has the main advantage of reducing the number
of observables, compressing the cosmological information contained in
the correlation function. This eases the estimation of the covariance
matrices associated with the data. We adopt this methodology in this
analysis and use the two first non-null moments $\xi_0(s)$ and
$\xi_2(s)$, where most of the relevant information is contained, and
ignore the contributions of the more noisy subsequent orders. The
multipole moments are related to $\xi(s,\mu)$ as
\begin{equation}
\xi_\ell(s)=\frac{2\ell+1}{2}\int_{-1}^{1}\xi(s,\mu)L_\ell(\mu)\d\mu, \label{eq:xil}
\end{equation}
where $L_\ell$ is the Legendre polynomial of order $\ell$. In practice
the integration of Eq. \ref{eq:xil} is approximated by a Riemann sum
over the binned $\xi(s,\mu)$. We use a logarithmic binning in $s$ with
$\Delta \log(s)=0.1$ and a linear binning in $\mu$ with $\Delta
\mu=0.02$.

VIPERS has a complex angular selection function which has to be taken
into account carefully when estimating the correlation function. This
has been studied in detail for the VIPERS Public Data Release 1
(PDR-1) \citep{guzzo14,garilli14} and particularly for the galaxy
clustering estimation in \citet{delatorre13a} and
\citet{marulli13}. We follow the same methodology to account for it in
this analysis with only small improvements. We summarize it in the
following and refer the reader to the companion paper,
\citet{pezzotta16}, for further details and tests of the method when
applied to the VIPERS final dataset.

The main source of incompleteness in the survey is introduced by the
VIMOS slit positioner, SSPOC, and the VIPERS one-pass observational
strategy. This results in an incomplete and uneven spectroscopic
sampling, described in detail in \citet{guzzo14,garilli14}. In terms
of galaxy clustering, the effect is to introduce an underestimation in
the amplitude of the measured galaxy correlation function, which
becomes scale-dependent on the smallest scales. We demonstrate in
\citet{delatorre13a} that this can be corrected by weighting each
galaxy in the estimation of the correlation function. For this we
define a survey completeness weight, $w^C$, which is defined for each
spectroscopic galaxy as well as an angular pair weight, $w^A$, which
is applied only to GG pair counts. The latter is obtained from the
ratio of one plus the angular correlation functions of targeted and
spectroscopic galaxies, as described in \citet{delatorre13a}.

The improvements compared to the PDR-1 analysis only concern the
estimation of survey completeness weights $w^C$. These in fact
correspond to the inverse effective sampling rate, \ESR, and are defined
for each galaxy as
\begin{equation}
w^C={\rm ESR}^{-1}=({\rm SSR}\times {\rm TSR})^{-1}, \label{eq:weight}
\end{equation}
where \SSR, \TSR are respectively the spectroscopic and target
sampling rates \citep[for details, see][]{guzzo14}. A significant
effort has been invested in improving the estimation of the \SSR and
\TSR. In particular the \SSR, which characterizes our ability of
measuring the redshifts from observed galaxy spectra, has been refined
and now accounts for new galaxy property dependencies, as described in
\citet{scodeggio16}. The \TSR, defined as the fraction of
spectroscopically observed galaxies in the parent target catalogue,
has been recomputed with better angular resolution, on rectangular
apertures of 60 by 100 $\rm{arcsec}^2$ around spectroscopic
galaxies. In order to mitigate the shot noise contribution in the
galaxy counts in such small apertures, we use the Delaunay tesselation
that naturally adapts to local density of points \citep[for details,
  see][]{pezzotta16}. The accuracy of this new set of weights is
tested in the next section and in \citet{pezzotta16}.

By applying these weights we effectively up-weight galaxies in the
pair counts. It is important to note that the spatial distribution of
the random objects is kept consistently uniform across the survey
volume. The final weights assigned to $GG$, $GR$, and $RR$ pairs
combine the survey completeness and angular pair weights as
\begin{align}
GG(s,\mu)&=\sum_{i=1}^{N_G}\sum_{j=i+1}^{N_G}w^C_iw^C_jw^A(\theta_{ij})\Theta_{ij}\left(s,\mu\right) \\
GR(s,\mu)&=\sum_{i=1}^{N_G}\sum_{j=1}^{N_R}w^C_i\Theta_{ij}\left(s,\mu\right) \\
RR(s,\mu)&=\sum_{i=1}^{N_R}\sum_{j=i+1}^{N_R}\Theta_{ij}\left(s,\mu\right),
\end{align}
where $\Theta_{ij}(s,\mu)$ is equal to unity for $\log(s_{ij})$ in
$[\log(s)-\Delta \log(s)/2,\log(s)+\Delta \log(s)/2]$ and $\mu_{ij}$
in $[\mu-\Delta \mu/2,\mu+\Delta \mu/2]$, and null otherwise. We
define the separation associated with each logarithmic bin as the
median pair separation inside the bin. This definition is more
accurate than using the bin centre, particularly at large $s$ when the
bin size is large.

One can also extract real-space clustering information from the
anisotropic redshift-space correlation function. This can be done by
measuring the latter with the estimator of Eq. \ref{eq:xir}, but where
the redshift-space galaxy separation vector is decomposed in two
components, $r_p$ and $\pi$, respectively perpendicular and parallel
to the line-of-sight \citep{fisher94}. This decomposition allows the
isolation of the effect of peculiar velocities as these modify only
the component parallel to the line-of-sight. This way, redshift-space
distortions can then be mitigated by integrating $\xi(r_p,\pi)$ over
$\pi$, thus defining the projected correlation function
\begin{equation}
w_p(r_p)=\int^{\pi_{\rm max}}_{-\pi_{\rm max}} \xi(r_p,\pi)\d\pi.
\end{equation}
We measure $w_p(r_p)$ using an optimal value of $\pi_{\rm
  max}=50\mhmpc$, allowing us to reduce the underestimation of the
amplitude of $w_p(r_p)$ on large scales and at the same time to avoid
including noise from uncorrelated pairs with separations of
$\pi>50\mhmpc$. From the projected correlation function, one can
derive the following quantity
\begin{equation}
\Upsilon_{gg}(r_p,r_0) =  \rho_c \left( \frac{2}{r^2_p} \int_{r_0}^{r_p} r w_p(r)\,dr - w_p(r_p) + \frac{r^2_0}{r^2_p} w_p(r_{0}) \right), \label{eq:upsilongg}
\end{equation}
where $r_0$ is a cut-off radius, $\rho_c=3H^2/(8\pi G)$ is the
critical density, $H(a)=\dot{a}/a$ is the Hubble parameter, and $G$ is
the gravitational constant. This quantity is equivalent to
$\Upsilon_{gm}$, which is measurable from galaxy-galaxy lensing (see
next section), but for galaxy-galaxy correlations instead of
galaxy-matter ones. It enters the definition of the gravitational slip
parameter $E_G$. In order to measure it in practice, since the
logarithmic binning in $r_p$ is rather large in our analysis, we
interpolate $w_p(r_p)$ using cubic spline interpolation before
evaluating the integral in Eq. \ref{eq:upsilongg} numerically. We find
that $\Upsilon_{gg}$ is more accurately measured with this technique
than by modelling $w_p(r_p)$ as a power law to perform the integral,
as is often done \citep[e.g.][]{mandelbaum13}.

\subsection{Galaxy-galaxy lensing estimation} \label{sec:wlestimation}

We use in this analysis the weak lensing technique usually referred to
as galaxy-galaxy lensing, in which one infers the tangential shear of
background sources $\gamma_t$ around foreground objects (lenses)
induced by the projected matter distribution in between. This quantity
is sensitive to the projected cross-correlation between lens galaxies
and the underlying matter distribution. Since the shear signal is weak
and the intrinsic ellipticity of galaxies is unknown, one has to
average the former over a large number of foreground sources. The
quantity that is effectively measured is the differential excess
surface density
\begin{equation}
  \Delta \Sigma_{gm} (r_p)=\Sigma_{\rm crit} \left< \gamma_t(r_p)\right>,
\end{equation}
where
\begin{equation}
  \Sigma_{\rm crit}=\frac{c^2}{4\pi G}\frac{D_{\rm S}}{D_{\rm LS}D_{\rm
      L}}.
\end{equation}
In the above equations, $r_p$ is the comoving transverse distance
between lens and source galaxies, $D_{\rm S}$, $D_{\rm LS}$, $D_{\rm
  L}$ are the angular diameter observer-source, lens-source, and
observer-lens distances, and $c$ is the speed of light in the vacuum.

We use the inverse variance-weighted estimator for the differential
excess surface density \citep[e.g.][]{mandelbaum13}:
\begin{equation}
  \Delta \Sigma_{gm} (r_p)= \frac{\sum^{N_S}_{i=1} \sum^{N_L}_{j=1} w^S_i e_{t,i} \Sigma^{-1}_{{\rm crit},~ij}\Theta_{ij}\left(r_p\right)}{\sum^{N_{\rm S}}_{i=1} \sum^{N_{\rm L}}_{j=1} w^S_i \Sigma^{-2}_{{\rm crit},~ij}\Theta_{ij}\left(r_p\right)}, \label{eq:dsigest}
\end{equation}
where the $i$ and $j$ indices run over source and lens galaxies
respectively, $N_{\rm S}$ and $N_{\rm L}$ are respectively the number
of source and lens galaxies, $e_{t,i}$ is the tangential ellipticity
for each lens-source pair, $w^S$ are statistical weights accounting
for biases in the determination of background source ellipticities,
and $\Theta_{ij}(r_p)$ is equal to unity for $r_{p,~ij}$ in
$[r_p-\Delta r_p/2,r+\Delta r_p/2]$ and null otherwise. The projected
separation $r_p$ is calculated as $r_p=\theta \chi_{\rm L}$, where
$\theta$ and $\chi_{\rm L}$ are respectively the angular distance
between the lens and the source, and the radial comoving distance of
the lens. This estimator includes an inverse-variance weight for each
lens-source pair $\smash{\Sigma^{-2}_{\rm crit}}$, which downweights
the pairs at close redshifts that contribute little to the weak
lensing signal \citep[][]{mandelbaum13}.

This estimator is unbiased if the redshifts of the sources are
perfectly known, but here we have only photometric redshift estimates:
the maximum likelihood photometric redshift and the normalized
redshift probability distribution function for each source
$p_s(z)$. Using the maximum likelihood photometric redshift of sources
in Eq. \ref{eq:dsigest} and restricting the sum to pairs with $z_{\rm
  S}>z_{\rm L}$ can possibly lead to a dilution of the signal induced
by the non-negligible probability that $z_{\rm S}<z_{\rm L}$. This
effect can be mitigated by replacing $\Sigma^{-1}_{\rm crit}$ in
Eq. \ref{eq:dsigest} by its average over the source redshift
probability distribution function $p_s$
\begin{equation}
 \left< \Sigma^{-1}_{\rm crit} \right> = \int_{z_{\rm L}}^\infty dz_{\rm S} p_s(z_{\rm S})\Sigma^{-1}_{\rm crit}(z_{\rm L},z_{\rm S}),
\end{equation}
which leads to the following estimator \citep[e.g.][]{miyatake15,blake16}:
\begin{equation}
  \Delta \Sigma_{gm} (r_p)= \frac{\sum^{N_S}_{i=1} \sum^{N_L}_{j=1} w^S_i e_{t,i} \left<\Sigma^{-1}_{{\rm crit},~ij}\right>\Theta_{ij}\left(r_p\right)}{\sum^{N_S}_{i=1} \sum^{N_L}_{j=1} w^S_i \left<\Sigma^{-1}_{{\rm crit},~ij}\right>^2\Theta_{ij}\left(r_p\right)} \label{eq:dsigest2}.
\end{equation}

\begin{figure}
\resizebox{\hsize}{!}{\includegraphics{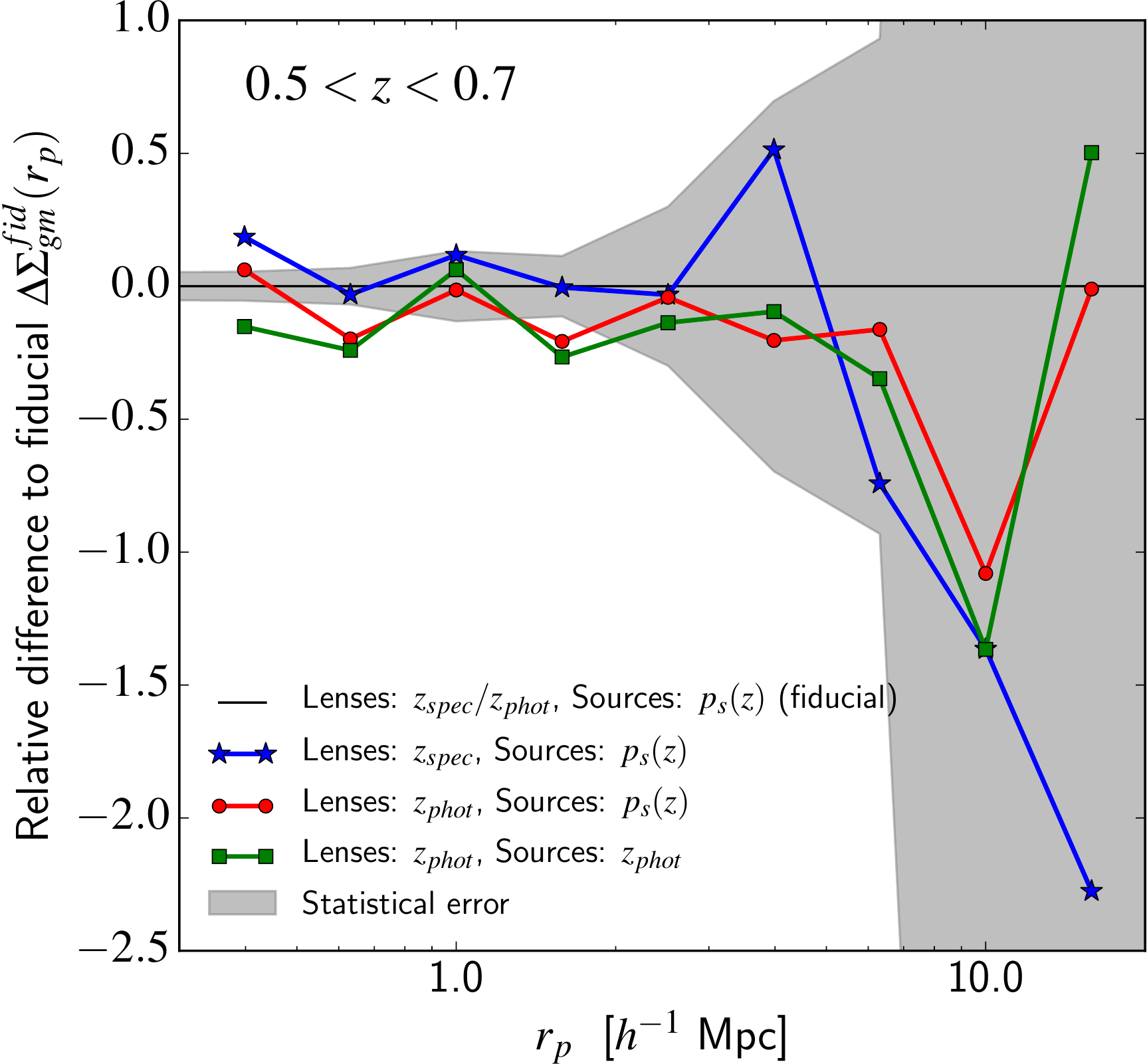}}
\caption{Relative difference between various estimates of $\Delta
  \Sigma_{gm}$, based on different assumptions for source and lens
  redshifts, and the fiducial estimate in the data at $0.5<z<0.7$. The
  quantity shown in the figure is $\Delta \Sigma_{gm}/\Delta
  \Sigma^{fid}_{gm}-1$ as a function of the projected separation
  $r_p$. The fiducial estimate $\Delta \Sigma^{fid}_{gm}$ is that
  obtained by using Eq. \ref{eq:dsigest2}, which includes the
  individual redshift probability distribution function $p_s(z)$ of
  the sources, and for the lenses, the VIPERS spectroscopic redshift
  ($z_{spec}$) when available or the CFHTLenS maximum likelihood
  photometric redshift ($z_{phot}$) otherwise (see text). It
  corresponds to the adopted estimate for the analysis. The grey
  shaded area represents the relative statistical error expected in
  the survey.}
\label{fig:dsigsyst}
\end{figure}

 In principle, those estimators hold in the limit where the lens
 redshift distribution is narrow and lens redshifts accurate
 \citep{nakajima12}. To better understand the importance of the
 effects introduced by an imperfect knowledge of the source and lens
 redshifts in the data, we perform a comparison of different estimates
 using Eq. \ref{eq:dsigest} or Eq. \ref{eq:dsigest2}, and various
 assumptions on the source and lens redshifts. This is presented in
 terms of the relative difference with respect to a fiducial estimate
 in Fig. \ref{fig:dsigsyst}.  The fiducial estimate is that obtained
 by using Eq. \ref{eq:dsigest2}, which includes the individual
 redshift probability distribution function $p_s(z)$ of the sources,
 and for the lenses, the VIPERS spectroscopic redshift when available
 or the CFHTLenS maximum likelihood photometric redshift otherwise.
  
We find that the estimate based on Eq. \ref{eq:dsigest}, which only
uses maximum likelihood photometric redshifts for both lenses and
sources, underestimate the signal on all probed scales by about $15\%$
with respect to the fiducial case. Here, we impose $z_S > 0.1 + z_L$,
including the additive term of $0.1$ to account for typical
photometric redshift errors \citep[e.g.][]{coupon15}. Further
including the source redshift probability distribution function
through the estimator of Eq. \ref{eq:dsigest2} allows a slight
improvement, reaching an underestimation of about $10\%$ with respect
to the fiducial case. The two previous estimates are still affected by
the uncertainty on the lens redshifts, which effectively tends to
dilute the overall signal. If we now use as lenses only VIPERS
spectroscopic galaxies, which represents about $30\%$ of all galaxies
with $i_{AB}<22.5$, we find a remarkably good agreement with the
fiducial estimate. In principle, this estimate may be considered as
the reference unbiased estimate, however on the largest scales probed
by the data, i.e. at $r_p = 10-20\mhmpc$, the signal drops
significantly. This can be imputed to the lack of source-lens pairs
induced by the reduced number of lenses, directly affecting our
ability to probe the largest scales signal. However, we find that this
effect can be mitigated by adding photometric lenses from the CFHTLenS
catalogue, taking the maximum likelihood photometric redshifts: this
corresponds to the fiducial estimate.  We note that the expected
statistical uncertainty, which is shown in Fig. \ref{fig:dsigsyst}
with the grey shaded area, is not negligible particularly above $r_p =
10\mhmpc$, and higher than any residual systematic effect. This test
makes us confident that our fiducial estimate of $\Delta
\Sigma_{gm}(r_p)$ is robust, given the expected level of statistical
error in the data. Similar results are found at $0.7<z<1.2$, leading
to the same conclusions.

A non-negligible source of systematics in weak lensing measurements is
related to the measurement of background galaxy shapes. This can lead
to systematic biases in the lensing measurements. The CFHTLenS
collaboration has studied these extensively in \citet{miller13} and
\citet{heymans12}, and we follow their method to correct our
measurements. We used the additive and multiplicative shear
calibration corrections $c$ and $m$, as well as the optimal weights
$w^S$ provided by {\sc lensfit}, which are available in the CFHTLenS
catalogue. In particular, to correct for the multiplicative bias we
applied the correction factor $\left(1+K(r_p)\right)^{-1}$ to $\Delta
\Sigma_{gm}(r_p)$ as described in \citet{miller13} and
\citet{velander14}. We found this correction to boost the
galaxy-galaxy lensing signal by about $5\%$ independently of the
scale.

For the purpose of constraining the cosmological model, it can be
difficult to use $\Delta \Sigma_{gm}$ as its modelling is
non-linear. One of the difficulties is to model the non-linear scales
and the intrinsic mixing of small-scale non-linear and large-scale
linear information \citep{baldauf10}. This is achievable but at the
expense of introducing additional nuisance parameters in the model
\citep[e.g.][]{cacciato13,more15}. An alternative approach, which we
use in this analysis, consists of using a derived statistic that
allows the mitigation of non-linearities: the annular differential
surface density $\Upsilon_{gm}$, which is defined as \citep{baldauf10}
\begin{equation}
\Upsilon_{gm}(r_p, r_0) = \Delta \Sigma_{gm}(r_p) -
\frac{r^2_0}{r^2_p} \Delta \Sigma_{gm}(r_{0}). \label{eq:upsilon}
\end{equation}
This statistic removes the small-scale non-linear contribution of
$\Delta \Sigma_{gm}$ below a cut-off radius $r_0$. We use this
quantity in our analysis and study the impact of the choice of $r_0$
in Sect. \ref{sec:mocks}.

\section{Theoretical modelling} \label{sec:modelling}

\subsection{Galaxy biasing}

Galaxies are not faithful tracers of the underlying matter
distribution and this has to be taken into account in cosmological
analyses, since cosmological models primarily predict matter
observables. The modelling of galaxy biasing is simplified when
focusing on large scales, where bias can be considered as linear and
simply be represented as a constant multiplicative factor in front of
the matter power spectrum. This is a common assumption in RSD
analyses. In our case, however, the relatively small survey volume
means that much of our information lies below fully linear scales; for
this reason, and because of the intrinsic non-linearities in the
excess surface density $\Delta \Sigma_{gm}$, additional care must be
taken to model galaxy biasing. We use a non-linear prescription for
galaxy bias based on the cosmological perturbation theory that allows
describing it more accurately down to translinear scales. We adopt the
non-linear non-local bias model of \citet{mcdonald09} that relates the
galaxy overdensity $\delta_{g}$ and matter overdensity $\delta$ as:
\begin{eqnarray} \label{eq:bias}
\delta_{g}({\bf x}) &=& b_1 \delta({\bf x}) + \frac{1}{2}b_2
      [\delta^2({\bf x})-\sigma^2]+\frac{1}{2}b_{s^2}[s^2({\bf
          x})-\langle s^2 \rangle] \nonumber \\
      && + O(s^3({\bf x})),
\end{eqnarray}
where $b_1$ and $b_2$ are the linear and second-order non-linear bias
terms, $b_{s^2}$ the non-local bias term, $s$ is the tidal tensor term
from which non-locality originates. The $\sigma^2$ and $\langle
s^2\rangle$ terms ensure the condition $\langle \delta_g\rangle=0$.

\subsection{Annular differential excess surface density}

The galaxy-galaxy lensing quantity that we observe is the differential
excess surface density. It is defined as
\begin{equation}
  \Delta \Sigma_{gm} (r_p) = \overline{\Sigma}_{gm}(r_p) - \Sigma_{gm}(r_p),
\end{equation}
where
\begin{equation}
  \overline{\Sigma}_{gm}(r_p) = \frac{2}{r^2_p} \int^{r_p}_0 \Sigma_{gm} (r)\,r\,dr
\end{equation}
and $\Sigma_{gm}(r_p)$ is the projected surface density defined as
\begin{equation}
  \Sigma_{gm}(r_p) = \Omega_m\rho_c\int^{\infty}_{-\infty} \left(1+\xi_{gm}(\sqrt{r^2_p+\chi^2}\right)d\chi.
\end{equation}
In the above equation, $\Omega_m$ is matter energy density and $\chi$ is the radial
comoving coordinate. $\Upsilon_{gm}$ can be predicted from $\Delta
\Sigma_{gm}$ by using Eq. \ref{eq:upsilon} or directly from the
galaxy-matter cross-correlation function as \citep{baldauf10}
\begin{equation}
\Upsilon_{gm}(r_p) = \int_0^\infty \xi_{gm}(x)W_\Upsilon(x,r_p,r_0)dx,
\end{equation}
where $W_\Upsilon(x,r_p,r_0)$ is the window function \citep{baldauf10}:
\begin{eqnarray}
  W_\Upsilon(x,r_p,r_0)&=&\frac{4x}{r^2_p}\left(\sqrt{x^2-r^2_0}\Theta(x-r_0) - \sqrt{x^2-r_p^2}\Theta(x-r_p) \right) \nonumber \\ 
  && -\frac{2x}{r^2_p}\left( \frac{r^2_p\Theta(x-r_p)}{\sqrt{x^2-r^2_p}} - \frac{r_0^2\Theta(x-r_0)}{\sqrt{x^2-r^2_0}}\right),
\end{eqnarray}
where $\Theta(x)$ is the Heaviside step function.

From these equations one can see explicitly that $\Upsilon_{gm}$ is
related to the galaxy-matter cross-correlation function $\xi_{gm}$ or
cross-power spectrum $P_{gm}$. If we assume the biasing model of Eq.
\ref{eq:bias}, $P_{gm}$ can be written as \citep{mcdonald09}
\begin{eqnarray}
  P_{gm}(k) &=& b_1 P_{\delta\delta}(k) + b_2 P_{b2,\delta}(k) + b_{s^2} P_{bs2,\delta}(k) \nonumber \\
  && + b_{3nl}\sigma^2_3(k)P_{\rm lin}(k),
\end{eqnarray}
where $\Pdd$ is the non-linear matter density-density power spectrum,
$b_{3nl}$ is a third-order non-local bias term, $P_{\rm lin}$ is the
linear matter power spectrum, and $P_{b2,\delta}$, $P_{bs2,\delta}$
are 1-loop integrals given in Appendix \ref{appendix}. In the local
Lagrangian picture where one assumes no initial non-local bias, one
can predict that the non-local bias terms at later time are related to
$b_1$ such that \citep{chan12,saito14}
\begin{eqnarray}
b_{s^2} &=& - \frac{4}{7} (b_1-1) \\
b_{3nl} &=& \frac{32}{315} (b_1-1).
\end{eqnarray}
We adopt these relations and our model has finally two galaxy biasing
parameters: $b_1$ and $b_2$, $b_1$ being the standard linear bias
parameter.

\subsection{Redshift-space distortions}

The most general formalism describing the redshift-space anisotropies
in the power spectrum derives from writing the matter density
conservation in real and redshift space \citep{kaiser87}. In
particular, in the plane-parallel approximation that is assumed in
this analysis, the anisotropic power spectrum of matter has the
general compact form \citep{scoccimarro99}
\begin{eqnarray}
P^s(k,\nu)&=&\int \frac{\d^3\vec{r}}{(2\pi)^3} e^{-i\vec{k} \cdot \vec{r}}\left<e^{-ikf\nu \Delta u_\parallel} \times \right. \nonumber \\ 
&& \left. [\delta(\vec{x})+f \partial_{_\parallel} u_{_\parallel}(\vec{x})][\delta(\vec{x}^\prime)+f \partial_{_\parallel} u_{_\parallel}(\vec{x}^\prime)]\right> \label{eq:rspk}
\end{eqnarray}
where $\nu=k_\parallel/k$,
$u_\parallel(\vec{r})=-v_\parallel(\vec{r})/(f aH(a))$,
$v_\parallel(\vec{r})$ is the line-of-sight component of the peculiar
velocity, $\delta$ is the matter density field, $\Delta
u_\parallel=u_\parallel(\vec{x})-u_\parallel(\vec{x}^\prime)$ and
$\vec{r}=\vec{x}-\vec{x}^\prime$. It is worth noting that in Fourier
space, for an irrotational velocity field,
$\smash{\partial_{_\parallel} u_{_\parallel}}$ is related to the
divergence of the velocity field $\theta$ via
$\smash{\partial_{_\parallel}
  u_{_\parallel}(\vec{k})=\nu^2\theta(\vec{k})}$. Although exact,
Eq. \ref{eq:rspk} is impractical and we use the approximation proposed
by \citet{taruya10}. In the case of perfect matter tracers, the latter
model takes the form
\begin{eqnarray}
  P^s(k,\nu) &=& D(k\nu\sigma_v)\left[\Pdd(k)+2\nu^2 f \Pdt(k) + \nu^4 f^2 \Ptt(k) \right. \nonumber \\ 
    && \left. + C_A(k,\nu,f) + C_B(k,\nu,f) \right],
\end{eqnarray}
where $D(k\nu\sigma_v)$ is a damping function, $\Pdd$, $\Pdt$, $\Ptt$
are respectively the non-linear matter density-density,
density-velocity divergence, and velocity divergence-velocity
divergence power spectra, and $\sigma_v$ is an effective pairwise
velocity dispersion that we can fit for and then treat as a nuisance
parameter. The expressions for $C_A(k,\nu,f)$ and $C_B(k,\nu,f)$ are
given in \citet{taruya10,delatorre12}. This phenomenological model can
be seen in configuration space as a convolution of a pairwise velocity
distribution, the damping function $D(k\mu\sigma_v)$ that we assume to
be Lorentzian in Fourier space, i.e.
\begin{equation}
  D(k\nu\sigma_v)=(1+k^2\nu^2\sigma^2_v)^{-1},
\end{equation}
and a term involving the density and velocity divergence correlation
functions and their spherical Bessel transforms.

This model can be generalized to the case of biased tracers by
including a biasing model. By introducing that of Eq. \ref{eq:bias},
one obtains for the redshift-space galaxy power spectrum
\citep{beutler14,gilmarin14}
\begin{eqnarray}
  P_g^s(k,\nu) &=& D(k\nu\sigma_v)\left[\Pgg(k)+2\nu^2 f \Pgt(k) + \nu^4 f^2 \Ptt(k) \right. \nonumber \\ 
    && \left. + C_A(k,\nu,f,b_1) + C_B(k,\nu,f,b_1) \right] \label{eq:pkrsd}
\end{eqnarray}
where
\begin{eqnarray}
  P_{gg}(k)&=&b_1^2 P_{\delta\delta}(k)+2b_2b_1P_{b2,\delta}(k)+2b_{s^2}b_1P_{bs2,\delta}(k) \nonumber \\
  && +b_2^2P_{b22}(k) +2b_2b_{s^2}P_{b2s2}(k)+b_{s^2}^2P_{bs22}(k) \nonumber \\
  && +2b_1b_{3\rm nl}\sigma_3^2(k)P_{\rm lin}(k) + N, \\
  P_{g\theta}(k)&=&b_1P_{\delta\theta}(k)+b_2P_{b2,\theta}(k)+b_{s^2}P_{bs2,\theta}(k) \nonumber \\
  && +b_{3\rm nl}\sigma_3^2(k)P_{\rm lin}(k).
\end{eqnarray}
In the above equations $\Pdt$ is the non-linear matter
density-velocity divergence power spectrum, $P_{\rm lin}$ is the
matter linear power spectrum, and $P_{b2,\delta}$, $P_{bs2,\delta}$,
$P_{b2,\theta}$, $P_{bs2,\theta}$, $P_{b22}$, $P_{b2s2}$, $P_{bs22}$,
$\sigma_3^2$ are 1-loop integrals given in Appendix \ref{appendix}.

The final model for $\xi^s_\ell(s)$ is obtained from its Fourier
counterpart $P^s_\ell(k)$ defined as
\begin{equation} \label{expmomK}
P^s_\ell(k)=\frac{2\ell+1}{2} \int_{-1}^1 P_g^s(k,\nu) L_\ell(\nu)\,\d\nu,
\end{equation}
where
\begin{equation} \label{expmom}
\xi^s_\ell(s)=i^\ell \int \frac{k^2}{2\pi^2} P^s_\ell(k)j_\ell(ks)\,\d k.
\end{equation} 
In the above equation, $j_\ell$ denotes the spherical Bessel
functions.

The ingredients of the model are the non-linear power spectra of
density and velocity divergence at the effective redshift of the
sample. These power spectra can be predicted from perturbation theory
or simulations for different cosmological models. The non-linear
matter power spectrum can also be obtained to a great accuracy from
semi-analytical prescriptions such as {\sc HALOFIT} \citep{smith03},
for various cosmologies. In particular, {\sc HALOFIT} allows the
prediction of $\Pdd$ from the linear matter power spectrum and the
knowledge of the scale of non-linearity at the redshift of interest,
$k_{\rm nl}(z)$. We note that at fixed linear matter power spectrum
shape, variations of $\sigma_8(z)$ can be straightforwardly mapped
into variations of $k_{\rm nl}(z)$ \citep[see][]{smith03}. In this
analysis, the linear matter power spectrum is predicted using the {\sc
  CLASS} Boltzmann code \citep{lesgourgues11}, and we use the latest
calibration of {\sc HALOFIT} by \citet{takahashi12} to obtain
$\Pdd$. To predict $\Ptt$ and $\Pdt$, we use the nearly universal
fitting functions of \citet{bel17} that depend on the linear power
spectrum and $\sigma_8(z)$ as
\begin{eqnarray}
  \Ptt(z)&=&P_{\rm lin}(z) e^{-k m_1\sigma^{m_2}_8(z)}
  \\ \Pdt(z)&=&\left(\Pdd(z) P_{\rm lin}(z) e^{-k
      n_1\sigma^{n_2}_8(z)}\right)^{1/2},
\end{eqnarray}
where $P_{\rm lin}$ is the linear power spectrum and
$(m_1,m_2,n_1,n_2)$ are free parameters calibrated on simulations. We
adopt here the values
$(m_1,m_2,n_1,n_2)=(1.906,2.163,2.972,2.034)$. These predictions for
$\Ptt$ and $\Pdt$ are accurate at the few percent level up to
$k\simeq0.7$ \citep{bel17}. Therefore, the overall degree of
non-linearity in $\Pdd$, $\Pdt$ and $\Ptt$ is solely controlled by
$\sigma_8(z)$, which is left free when fitting the model to
observations.

In the model, the linear bias and growth rate parameters, $b_1$ and
$f$, are degenerate with the normalization of the matter power
spectrum parameter $\sigma_8$. Generally with RSD, only the
combination of $b_1\sigma_8$ and $f\sigma_8$ can be constrained if no
assumption is made on the actual value of $\sigma_8$. However in the
\citet{taruya10} model, $b_1^2f\sigma_8^4$, $b_1f^2\sigma_8^4$, and
$f^3\sigma_8^4$ terms appear in the correction term $C_A$
\citep[see][]{taruya10,delatorre12}. Accordingly, in the general case,
$(f,b_1,b_2,\sigma_8,\sigma_v)$ are treated as separate parameters in
the fit and we provide marginalized constraints on the derived
$f\sigma_8$.

\subsection{Redshift errors}

Redshift errors can potentially affect the anisotropic RSD signal. In
the anisotropic correlation function they have a similar effect as
galaxy random motions in virialized objects: they introduce a smearing
of the correlation function along the line of sight at small
transverse separations. If the probability distribution function of
redshift errors is known, their effect can be forward modelled by
adding another multiplicative damping function in the redshift-space
galaxy power spectrum of Eq. 19. In that case, the damping function
should be the Fourier transform of the error probability distribution
function. We follow this approach and the final model is obtained by
multiplying Eq. 19 by a Gaussian with standard deviation set to the
estimated pairwise redshift dispersion of VIPERS galaxies such that
the final RSD model $\smash{\widehat{P_g^s}}$ reads
\begin{equation} \label{eq:rsdpkzerr}
  \widehat{P_g^s}(k,\nu)=G(k\nu\sigma_z)P_g^s(k,\nu),
\end{equation}
where $P_g^s(k,\nu)$ is taken from Eq. \ref{eq:pkrsd}, $G$ is the
Fourier transform of the Gaussian kernel
\begin{equation}
  G(k\nu\sigma_z)=\exp\left(-\frac{k^2\nu^2\sigma^2_z}{2}\right),
\end{equation}
and $\sigma_z$ is the pairwise standard deviation associated with the
redshift error probability distribution function.

\begin{figure}
\resizebox{\hsize}{!}{\includegraphics{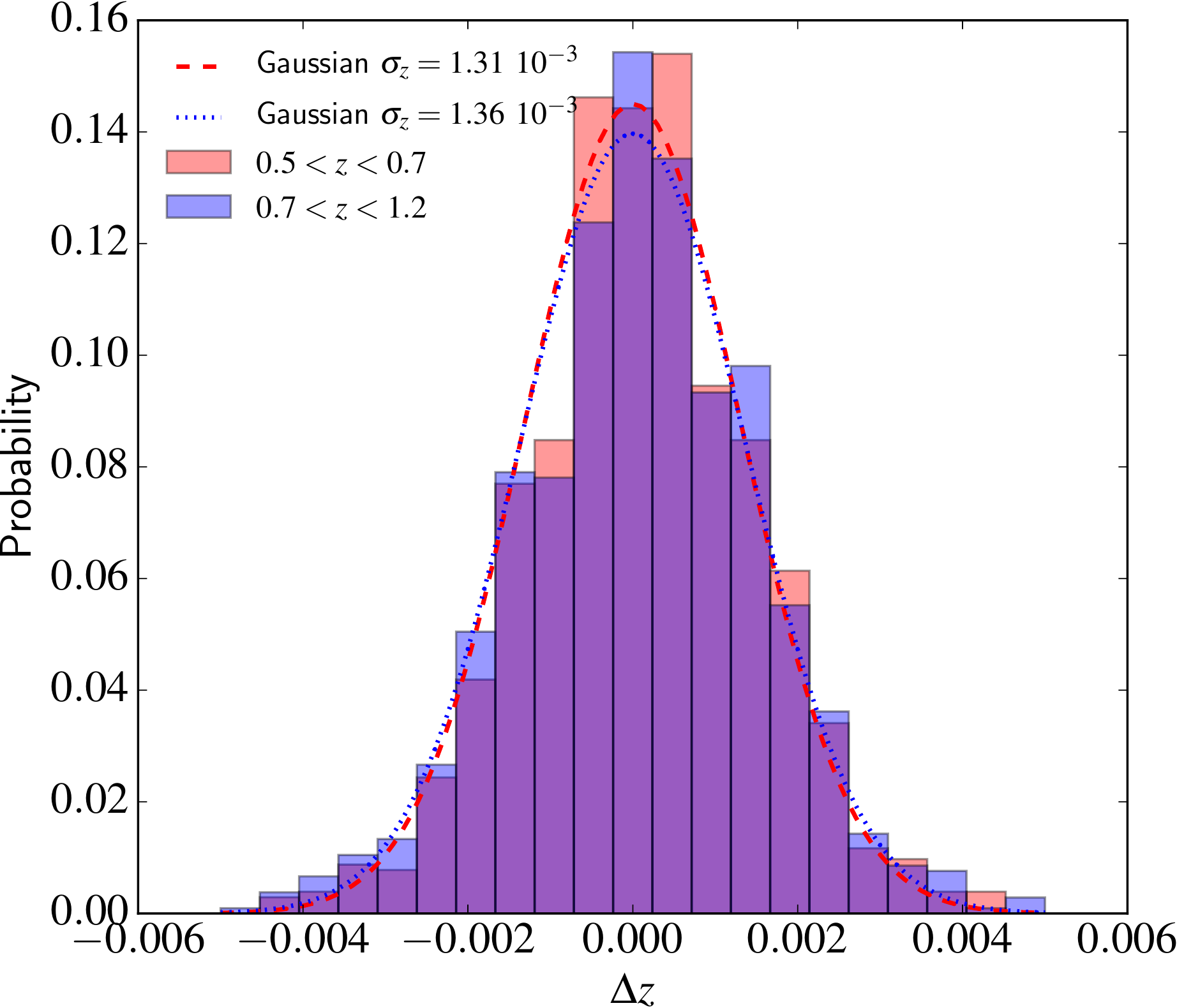}}
\caption{Probability distribution function of redshift errors at
  $0.5<z<0.7$ and $0.7<z<1.2$ in the VIPERS data. This is obtained
  from the redshift differences of reobserved galaxies, for which
  there are two independent redshift measurements. The dotted and
  dashed curves are best-fitting Gaussians for the redshift intervals
  $0.5<z<0.7$ and $0.7<z<1.2$ respectively.}
\label{fig:pdferr}
\end{figure}

The Gaussian form is motivated by the data themselves as shown in
Fig. \ref{fig:pdferr}. In this figure are shown the distributions of
redshift differences at $0.5<z<0.7$ and $0.7<z<1.2$ in VIPERS
reobservations ($1061$ at $0.5<z<0.7$ and $1086$ at $0.7<z<1.2$), for
which we have two independent redshift measurements for the same
galaxies \citep[see][]{scodeggio16}. These distributions can be rather
well modelled by Gaussians, and by doing so, we obtain values of
$\sigma_z=1.31 \times 10^{-3}$ and $\sigma_z=1.36 \times 10^{-3}$ for
the pairwise redshift standard deviations at $0.5<z<0.7$ and
$0.7<z<1.2$ respectively. These are further converted in comoving
length assuming the fiducial cosmology to enter the model in
Eq. \ref{eq:rsdpkzerr}.

\subsection{Alcock-Paczynski effect}

Additional distortions can arise in galaxy clustering because of the
need to assume a fiducial cosmology to convert redshift and angular
positions into comoving distances, and the fact that this fiducial
cosmology is not necessarily the true one. This is the
\citet{alcock79} effect (AP). More specifically, since the
line-of-sight separations require the knowledge of the Hubble
parameter, $H(z)$, and transverse separations that of the angular
diameter distance, $D_A(z)$, any difference in $H(z)$ and $D_A(z)$
between the fiducial and true cosmologies, translates into an
anisotropic clustering, independently of RSD. Although AP and RSD
anisotropies are degenerate to some extent in the observables
\citep{ballinger96,matsubara96}, they have a fundamentally different
origin: AP is sensitive to the geometry whereas RSD are sensitive to
the growth of cosmological perturbations.

We follow \citet{xu13} and model AP distortions using the $\alpha$ and
$\epsilon$ parameters, which characterize respectively the isotropic
and anisotropic distortion components associated with AP. These are
given by
\begin{eqnarray}
  \alpha &=& \left( \frac{D^2_A}{D^{\prime 2}_A} \frac{H^{\prime}}{H} \right)^{1/3} \\
  \epsilon &=& \left( \frac{D^{\prime}_A}{D_A} \frac{H^{\prime}}{H} \right)^{1/3} - 1,
\end{eqnarray}
where quantities calculated in the fiducial cosmology are denoted with
primes. Those parameters modify the scales at which the correlation
function is measured such that
\begin{eqnarray}
  \rpa^{\prime} &=& \alpha(1+\epsilon)^2 \rpa \\
  \rpe^{\prime} &=& \alpha(1+\epsilon)^{-1} \rpe.
\end{eqnarray}
Therefore, for the model correlation function monopole and quadrupole
in a tested cosmology, the corresponding quantities in the fiducial
cosmology are obtained as \citep{xu13}
\begin{eqnarray}
\xi^{\prime}_0(s^{\prime}) &=& \xi _0(\alpha s) + \frac{2}{5}\epsilon
\left[ 3\xi_2(\alpha s) + \frac{d\xi_2(\alpha s)}{d\ln(s)} \right] 
\label{eqn:mono} \\
\xi^{\prime}_2(s^{\prime}) &=& 2\epsilon \frac{d\xi_0(\alpha s)}{d\ln(s)}
+\bigg( 1 + \frac{6}{7}\epsilon \bigg)\xi_2(\alpha s)
+\frac{4}{7}\epsilon \frac{d\xi_2(\alpha s)}{d\ln(s)} \nonumber \\
&& + \frac{4}{7}\epsilon \bigg[ 5\xi_4(\alpha s) + 
\frac{d\xi_4(\alpha s)}{d\ln(s)} \bigg].
\label{eqn:quad}
\end{eqnarray}
In the case of the galaxy-galaxy lensing statistic that we are
considering, since it is a function of the transverse separation
$r_p$, the corresponding $\Upsilon_{gm}$ in the fiducial cosmology is
simply given by
\begin{equation}
  \Upsilon^\prime_{gm}(r^\prime_p)=\Upsilon_{gm}\left(\alpha(1+\epsilon)^{-1}r_p\right).
\end{equation}
    
\subsection{Cosmological insights from galaxy clustering and galaxy-galaxy lensing}

Gravitational physics on cosmological scales can be tested from
measurements of the growth rate of structure, which is well measured
from RSD in the galaxy clustering pattern. We have seen that in
practice, the correlation function multipole moments depend not only
on the growth rate of structure $f$, but also on the shape and
amplitude $\sigma_8$ of the matter power spectrum, the galaxy bias
parameters $b_1$ and $b_2$, and the pairwise velocity dispersion
$\sigma_v$. To derive the growth rate of structure, one then needs to
marginalise over those nuisances. This is of course a source of
uncertainty in the determination of the growth rate of
structure. Moreover, since there is a degeneracy between the amplitude
of the matter power spectrum $\sigma_8$, the growth rate of structure
$f$, and the linear bias parameter $b_1$, RSD alone are sensitive to
the $f\sigma_8$ and $b_1\sigma_8$ parameter combinations.

On the other hand, galaxy-galaxy lensing probes the real-space
galaxy-matter correlations that are described by the shape and
amplitude $\sigma_8$ of the matter power spectrum, the galaxy bias
parameters $b_1$ and $b_2$, and the matter density parameter
$\Omega_m$. Projected galaxy-galaxy correlations are also sensitive to
$\sigma_8$, $b_1$, and $b_2$. But by looking in detail at those
dependencies, we can see that in the linear regime $\Upsilon_{gm}
\propto \Omega_m b_1 \sigma^2_8$, while $\Upsilon_{gg} \propto b^2_1
\sigma^2_8$, such that by combining the two we can break the
degeneracy between $b_1$ and $\sigma_8$. We note that $\mxism$, from
which $\xi_0$ and $\xi_2$ are derived, has the same parameter
dependences as $\Upsilon_{gg}$, except for the additional $f$
dependence. Therefore, additional galaxy-galaxy lensing information
brings an independent handle on the bias parameters $b_1$ and $b_2$,
and the power spectrum amplitude $\sigma_8$, reducing the
uncertainties on the growth rate of structure induced by the lack of
knowledge on the bias of galaxies, as well as a supplementary
sensitivity to $\Omega_m$.

\section{Tests on simulated data} \label{sec:test}

\subsection{Simulated data} \label{sec:mocks}

To test the robustness of redshift-space galaxy clustering,
galaxy-galaxy lensing, and associated error estimates, we make use of
a large number of mock galaxy samples, which are designed to be a
realistic match to the VIPERS final dataset. We used the mock lensing
lightcones presented in \citet{giocoli16}. These have been built upon
the Big MultiDark dark matter N-body simulation \citep{klypin16},
which assumes a flat $\Lambda {\rm CDM}$ cosmology with
$(\Omega_m,~\Omega_\Lambda,~\Omega_b,~h,~n,~\sigma_8) = (0.307, 0.693,
0.0482, 0.678, 0.960, 0.823)$ and covers a volume of $15.625\mhgpcc$.
These lightcones contain the shear information associated with
simulated background galaxies distributed uniformly on the sky but
following the redshift distribution of CFHTLenS galaxies.  More
specifically, the lightcones have been built to match the effective
number density and redshift distribution of the CFHTLenS lensing
catalogue. We added Gaussian random errors with standard deviation
$\smash{\sigma_e=(\sigma_{e_1}^2+\sigma_{e_2}^2)^{1/2}=0.38}$ to the
ellipticities to mimic those in the CFHTLenS data. The size of the
simulation allowed us to create $54$ independent lightcones for W1 and
W4, spanning the redshift range $0<z<2.3$ \citep[for details,
  see][]{giocoli16}.

We populate these lightcones with foreground galaxies using the halo
occupation distribution (HOD) technique and apply the detailed VIPERS
selection function and observational strategy. The haloes were
identified in the simulation using a friends-of-friends algorithm with
a relative linking length of $b=0.17$ times the inter-particle
separation. The mass limit to which the halo catalogues are complete
is $10^{11.95}\mhmsun$. Because this limiting mass is too large to
host the faintest galaxies observed with VIPERS, we use the method of
\citet{delatorre13b} to reconstruct haloes below the resolution
limit. This method is based on stochastically resampling the halo
number density field using constraints from the conditional halo mass
function. For this, one needs to assume the shapes of the halo bias
factor and halo mass function at masses below the resolution limit and
use the analytical formulae obtained by
\citet{tinker08,tinker10}. With this method we are able to populate
the simulation with low-mass haloes with a sufficient accuracy to have
unbiased galaxy two-point statistics in the simulated catalogues
\citep[for details, see][]{delatorre13a}. The minimum reconstructed
halo mass we consider for the purpose of creating VIPERS mocks is
$10^{10}\mhmsun$.

In this process, we populate each halo with galaxies according to its
mass, the mean number of galaxies in a halo of a given mass being
given by the HOD. It is common usage to differentiate between central
and satellite galaxies in haloes. While the former are put at rest at
halo centres, the latter are randomly distributed within each halo
according to a NFW radial profile \citep{navarro96,navarro97}. The
halo occupation function and its dependence on redshift and
luminosity/stellar mass must be precisely chosen in order to obtain
mock catalogues with realistic galaxy clustering properties. We
calibrated the halo occupation function directly on the VIPERS data,
as presented in \citet{delatorre13a}. We add velocities to the
galaxies and measure their redshift-space positions. While the central
galaxies are assigned the velocity of their host halo, satellite
galaxies have an additional random component for which each Cartesian
velocity component is drawn from a Gaussian distribution with a
standard deviation that depends on the mass of the host halo. Details
about the galaxy mock catalogue construction technique are given in
Appendix A of \citet{delatorre13a}.

The final step in obtaining fully realistic VIPERS mocks is to add the
detailed survey selection function. We start by applying the magnitude
cut $i_{\rm AB}<22.5$ and the effect of the colour selection on the
radial distribution of the mocks. This is achieved by depleting
the mocks at $z<0.6$ so as to reproduce the VIPERS colour sampling
rate \citep[see][for detail]{guzzo14}. The mock catalogues that we
obtain are then similar to the parent photometric sample in the
data. We next apply the slit-positioning algorithm with the same
setting as for the data. This allows us to reproduce the VIPERS
footprint on the sky, the small-scale angular incompleteness and the
variation of \TSR across the fields. Finally, a random redshift error
is added to the redshifts as in the data. We are thus able to produce
realistic mock galaxy catalogues that contain the detailed survey
completeness function and observational biases of VIPERS, which we
refer to as the `observed' mock catalogues in the following.

We note that another set of VIPERS mock catalogues spanning the
redshift range of $0.4<z<1.2$ have been constructed. This set, which
comprises $306$ and $549$ lightcones of W1 and W4 fields respectively,
has not been explicitly used in this analysis, but in accompanying
VIPERS PDR-2 analyses
\citep[e.g.][]{hawken16,pezzotta16,wilson16,rota16}.

\subsection{Systematics on the correlation function monopole and quadrupole}

\begin{figure}
\resizebox{\hsize}{!}{\includegraphics{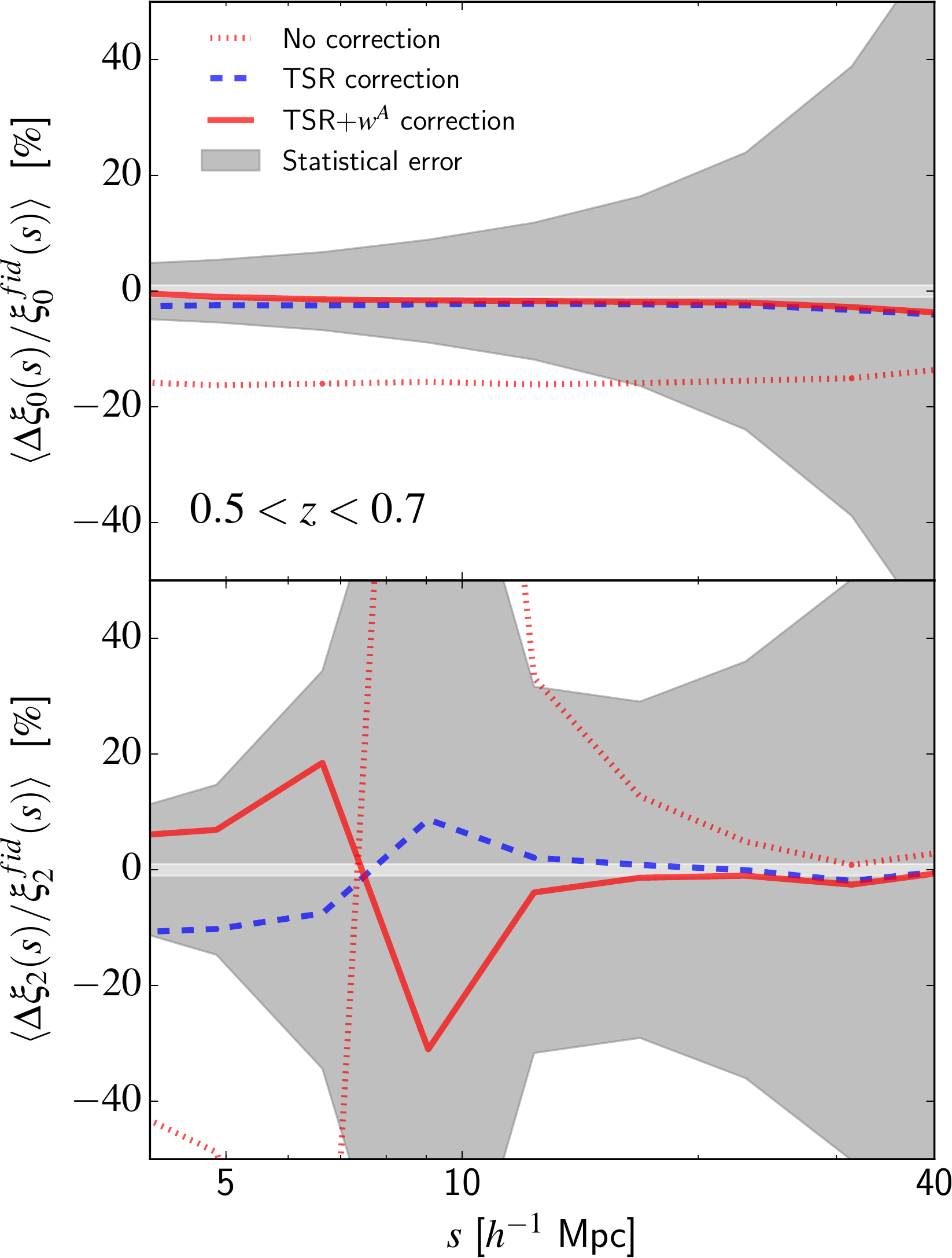}}
\caption{ Relative systematic errors on the correlation function
  monopole (top panel) and quadrupole (bottom panel) at $0.5<z<0.7$
  and effects of target sampling rate (\TSR) and angular pair
  weighting ($w^A$) corrections. The grey shaded areas represent the
  relative statistical error expected in the survey, while light grey
  band mark $\pm 1\%$ relative uncertainties for reference.}
\label{fig:xilsystlow}
\end{figure}

\begin{figure}
\resizebox{\hsize}{!}{\includegraphics{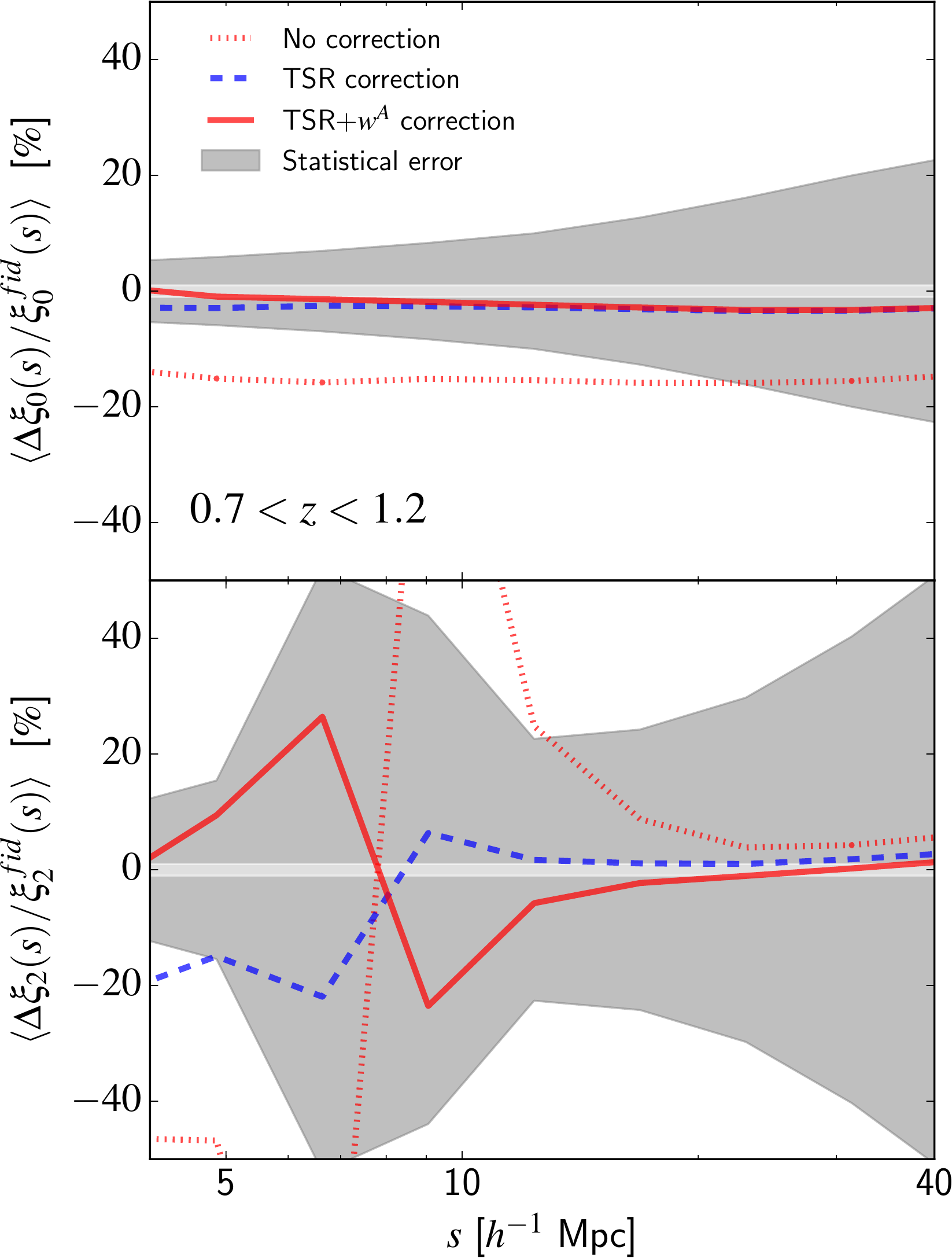}}
\caption{Same as in Fig. \ref{fig:xilsystlow} but for the redshift
  interval $0.7<z<1.2$.}
\label{fig:xilsysthigh}
\end{figure}

The mock samples are crucial for testing the redshift-space clustering
estimation in VIPERS, which is not trivial given the complex selection
function of the survey. We first study the impact of the survey
selection function on the measurement of the monopole and quadrupole
correlation functions. We measured these quantities in the observed
mocks, applying the different weights defined in
Sect. \ref{sec:gcestimation}, and compare them to the reference
measurements obtained from the parent mocks, including VIPERS typical
spectroscopic redshift errors. The relative differences in $\xi_0$ and
$\xi_2$ as a function of separation and averaged over the mocks are
shown in Figs. \ref{fig:xilsystlow} and \ref{fig:xilsysthigh},
respectively for the two samples at $0.5<z<0.7$ and $0.7<z<1.2$. First
of all, it is clear from these figures that without any correction the
spectroscopic strategy introduces biases in the estimation of the
galaxy clustering. But when applying the survey completeness weights
$w^C$, one can recover within a few percent the correct amplitude of
the correlation functions on scales above $5\mhmpc$. By further
applying the angular weights $w^A$, we obtain an almost unbiased
estimate of the monopole and quadrupole down to a few $\mhmpc$. The
statistical relative error induced by sample variance and estimated
from the dispersion among the mock samples, is shown with the shaded
area in these figures. It is important to note that it is much larger
than any residual systematics over the range of scales
considered. Finally, it is worth mentioning that in the quadrupole,
the apparent higher level of systematics at around $s=10 \mhmpc$ is an
artefact due to the zero crossing of the functions at slightly
different separations.

\subsection{Systematics on the growth rate of structure}

\begin{figure}
\resizebox{\hsize}{!}{\includegraphics{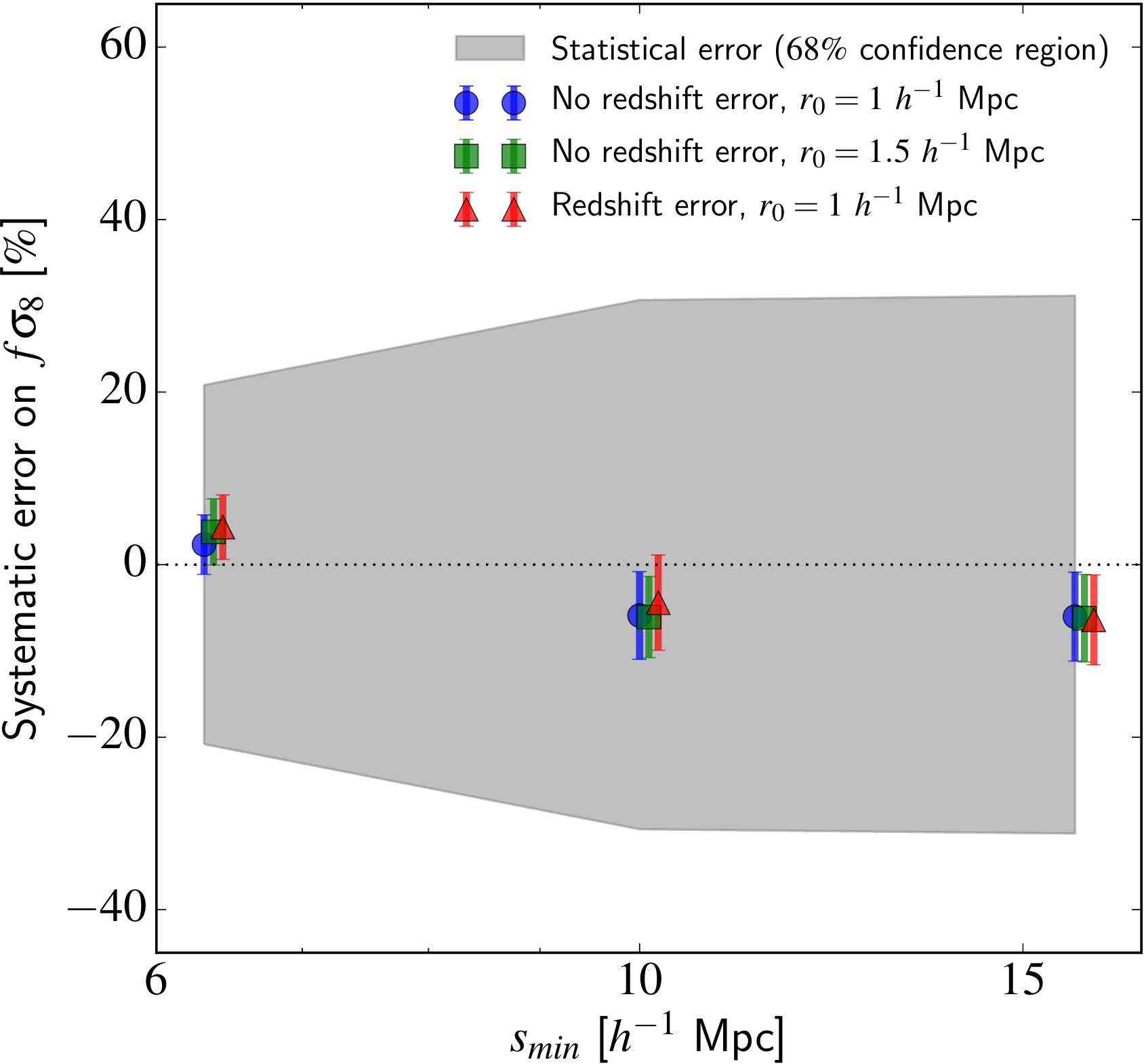}}
\caption{Relative systematic error on $f\sigma_8$ at $0.5<z<0.7$ as a
  function of $s_{\rm min}$, for different values of $r_0$
  ($r_0=1\mhmpc$ and $r_0=1.5\mhmpc$) and when including or not
  redshift error. The error bars represent the relative statistical
  error associated to analysing the mean mock predictions. The shaded
  area shows the $1\sigma$ confidence region associated with the
  relative statistical error expected in VIPERS. The squares and
  triangles are artificially shifted along $s_{\rm min}$ axis to
  improve the clarity of the figure.}
\label{fig:fsig8systlow}
\end{figure}

We further study our ability to determine $f\sigma_8$ when combining
RSD and galaxy-galaxy lensing measurements in a maximum likelihood
analysis.  For this purpose we perform several analyses of the mean
RSD and galaxy-galaxy lensing measurements in the observed mocks, for
different minimum separations $s_{\rm min}$ in the correlation
functions and different cut-off scale $r_0$ in the annular
differential excess surface density. These analyses are performed on
the mean quantities to reduce the impact of statistical errors and
concentrate on systematics. The precision matrix is estimated from the
mocks as explained in Sect. \ref{sec:analysis}, except that each
element is further divided by the number of mocks to characterize the
error on the mean. As an illustration, we present in this section only
the case of the sample at $0.5<z<0.7$. The sample at $0.7<z<1.2$
provides very similar systematic levels.

\begin{figure}
\resizebox{\hsize}{!}{\includegraphics{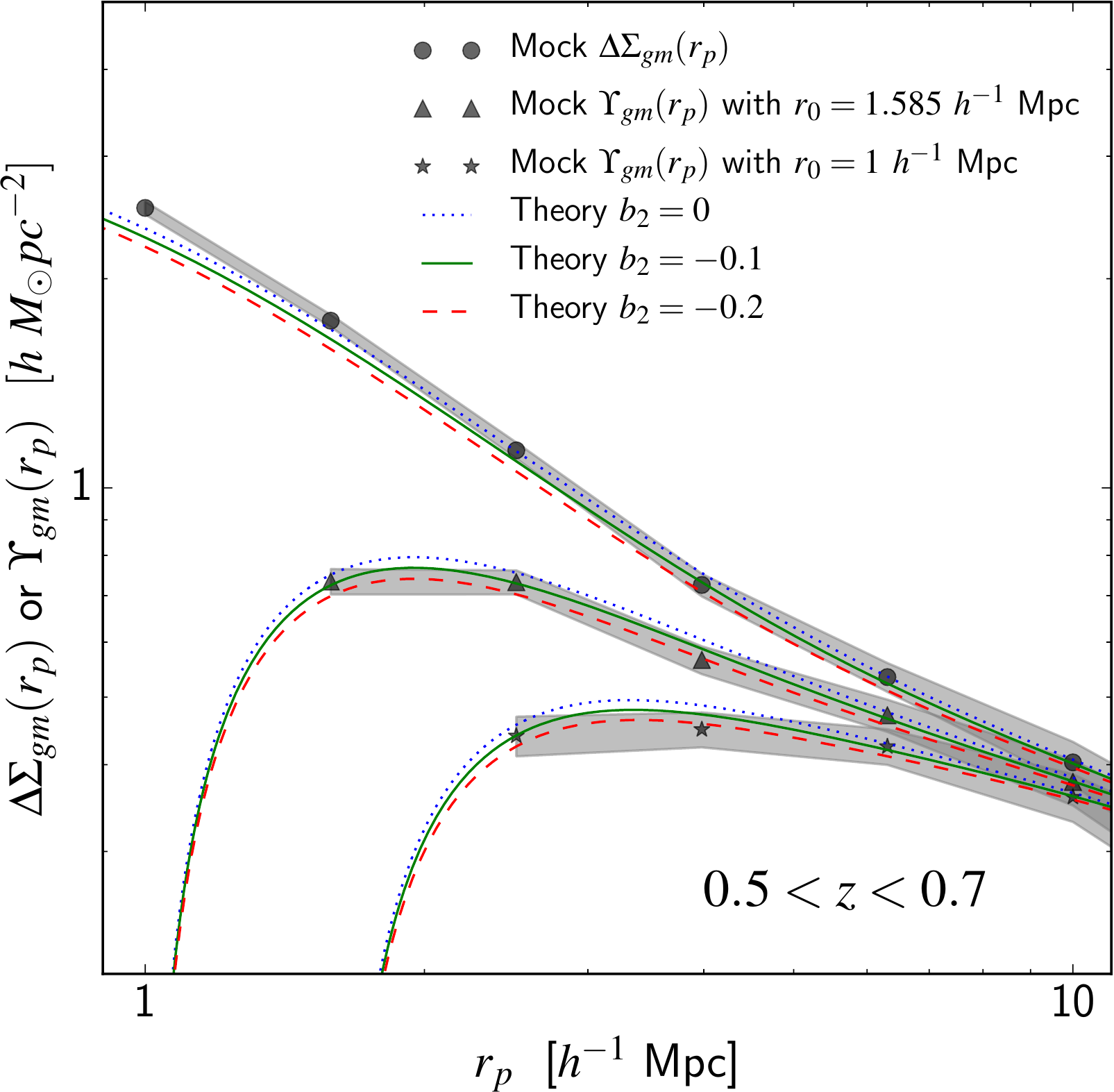}}
\caption{Comparison of differential excess surface density
  $\Delta_{gm}$ and annular differential excess surface density
  $\Upsilon_{gm}$ predictions in the mocks (points and shaded regions)
  and by our theoretical model (curves) at $0.5<z<0.7$. The mock
  predictions correspond to the mean signal among the mock
  realizations (points) and its associated $1\sigma$ error (shaded
  region). The curves show the theoretical predictions for the
  fiducial parameters of the mocks, varying only the $b_2$ parameter
  as labeled.}
\label{fig:dsigmodel}
\end{figure}

Fig. \ref{fig:fsig8systlow} presents the systematic errors on
$f\sigma_8$, i.e. the relative difference of recovered values with
respect to the fiducial value of the mocks, as a function of $s_{\rm
  min}$ and for $r_0=(1 h^{-1}{\rm Mpc}, 1.5 h^{-1}{\rm Mpc})$. We
consider rather small minimum scales and cut-off radii to explore the
extent to which our modelling is robust in the translinear regime. We
can see in this figure that our model allows the recovery of the
fiducial value of $f\sigma_8$ down to $s_{\rm min}=6.3\mhmpc$, with
systematic errors below $5\%$, independently of the choice of
$r_0$. In principle, values of $r_0$ smaller than the typical radius
of haloes hosting these galaxies may not be optimal, since the
non-linear contribution to correlations may dominate on those scales,
which are more difficult to describe \citep{baldauf10}. However, this
also depends on the galaxy type and the redshift. For VIPERS galaxies
and the considered biasing model, we find that $r_0=1\mhmpc$ can be
well described by our model \citep[see also][]{blake16}. This can be
seen in Fig. \ref{fig:dsigmodel} where is shown the comparison between
the mean mock $\Delta\Sigma_{gm}$ and $\Upsilon_{gm}$ obtained with
$r_0=(1h^{-1}{\rm Mpc},1.5h^{-1}{\rm Mpc})$ and the predictions of our
model, when $b_2$ is allowed to vary and $b_1$ is fixed to its
fiducial value. We can see that although the model fails to reproduce
$\Delta\Sigma_{gm}$ on scales below about $3\mhmpc$, it provides a
good description of $\Upsilon_{gm}$ for $b_2=-0.1$.

We finally test the impact of redshift errors on the recovery of
$f\sigma_8$ in Fig.  \ref{fig:fsig8systlow}. This figure shows the
relative systematic error on $f\sigma_8$ as a function $s_{\rm min}$
in the case where $r_0$ is fixed to $1\mhmpc$ and typical VIPERS
redshift errors are added randomly to mock galaxy redshifts. By
comparing it to the case without redshift errors, we can see that for
the RSD model where a Gaussian damping term is added to account for
redshift errors, the recovery of $f\sigma_8$ is achieved without
additional bias, with only a small relative bias of about $3\%$ at
$s_{\rm min}=6.3\mhmpc$ and $-5\%$ above. We note that this is the
ideal case where the redshift error probability distribution function
is perfectly known.

Overall, these tests demonstrate that our model with $s_{\rm
  min}=6.3\mhmpc$ and $r_0=1\mhmpc$ is robust enough to provide a
precise measurement of $f\sigma_8$ with VIPERS data, with residual
systematics of the order of a few per cent only, but only representing
about one fifth of expected statistical error on the measurement as
shown with the grey shaded region in
Fig. \ref{fig:fsig8systlow}. Based on these tests, we adopt $(s_{\rm
  min},r_0)=(6.3~h^{-1}{\rm Mpc}, 1~h^{-1}{\rm Mpc})$ values for the
following analysis.

\section{Likelihood analysis and precision matrix} \label{sec:analysis}

In order to derive cosmological parameters from the combination of RSD
and galaxy-galaxy lensing measurements, we perform a maximum
likelihood analysis in which we define the likelihood function
$\mathcal{L}$ such that
\begin{equation}
  -2\ln{\mathcal{L}}=\sum_{i=1}^{N_p}\sum_{j=1}^{N_p}\Delta_i \hat{C}^{-1}_{ij} \Delta_j,
\end{equation}
where $N_p$ is the number of data points in the fit, $\bf \Delta$ is
the data-model difference vector, and $\smash{{\bf \hat{C}^{-1}}}$ is
the inverse data covariance matrix. $\bf \Delta$ is defined such that
each element is $\Delta_i=d_i-m_i$, where $\bf d$ and $\bf m$ are
respectively the data and model prediction vectors. In our case, $\bf
d$ is the concatenation of $\xi_0$, $\xi_2$, and $\Upsilon_{gm}$, for
the set of considered separations. The parameter space of the model is
explored using a Monte Carlo Markov Chain (MCMC) method implementing
the Metropolis-Hastings algorithm \citep{metropolis53}.

A robust estimation of the inverse data covariance matrix, or
precision matrix, is crucial in order to achieve realistic posterior
likelihood functions of the parameters. The different bins in $\xi_0$,
$\xi_2$, and $\Upsilon_{gm}$ are correlated to some degree and this
must be allowed for in the likelihood analysis. We measure these three
quantities in the $54$ mocks and estimate the covariance matrix $\bf
C$. The generic elements of the matrix can be evaluated as
\begin{equation}
C_{ij}=\frac{1}{N_{m}-1}\sum_{k=1}^{N_m}\left(d^k_i-\bar{d_i}\right)\left(d^k_j-\bar{d_j}\right)\, ,
\end{equation}
where $N_m$ is the number of mock realizations and the indices $i,j$
run over the data vector $\bf d$ elements. An unbiased estimate of the
inverse covariance matrix, ${\bf \hat{C}^{-1}}$, is obtained as
\citep{hartlap07}
\begin{equation} \label{eq:covmat}
{\bf \hat{C}^{-1}}=\frac{N_{m}-N_{p}-2}{N_{m}-1}{\bf C}^{-1},
\end{equation}
for $N_m>N_p-2$. The resulting inverse covariance matrix obtained from
mock realizations can be noisy, depending on how large $N_m$ is with
respect to $N_p$. In our case, $N_m=54$ and $N_p=16$, which suggests
the presence of a non-negligible noise in the inverse covariance
matrix. In order to reduce the level of noise, we adopt the tapering
technique of \citet{kaufman08}. This technique has been introduced in
the context of cosmological analysis by \citet{paz15}. This technique
relies on the assumption that large-scale covariances vanish, and
consists of tapering the covariance matrix around the diagonal using a
specific positive and compact taper function. Contrary to other
estimators such as shrinkage \citep[e.g.][]{pope08}, the two-tapers
estimator has the advantage of being unbiased. The inverse tapered
covariance matrix is obtained as
\begin{equation}
  {\bf \hat{C_T}^{-1}}=\left( \frac{N_{m}-N_{p}-2}{N_{m}-1}\right) ({\bf C}\circ {\bf T})^{-1}\circ {\bf T},
\end{equation}
where `$\circ$' denotes the element-wise matrix product and ${\bf T}$
is the tapering matrix. We follow \citet{paz15} and use the tapering
matrix defined as
\begin{equation}
  T_{ij} = K(|{\bf x}_i-{\bf x}_j|),
\end{equation}
where $x_i$ is the $i^{th}$ measurement position in the data vector,
and $K$ is the taper function that we take to be a Wendland function:
\begin{align}
K(x)=\left\{ \nonumber
\begin{array}{lcl}
\left(1-\frac{x}{T_p}\right)^4 \left(4\frac{x}{T_p}+1\right)\hspace{0.5cm}{\rm if}~x<T_p
\\
\\
0 \hspace{3cm} \rm{otherwise}.
\end{array}
\right.
\end{align}
This taper function has one free parameter, the tapering scale $T_p$,
which essentially represents the typical scale difference above which
covariances are nullified.

\begin{figure}
\resizebox{\hsize}{!}{\includegraphics{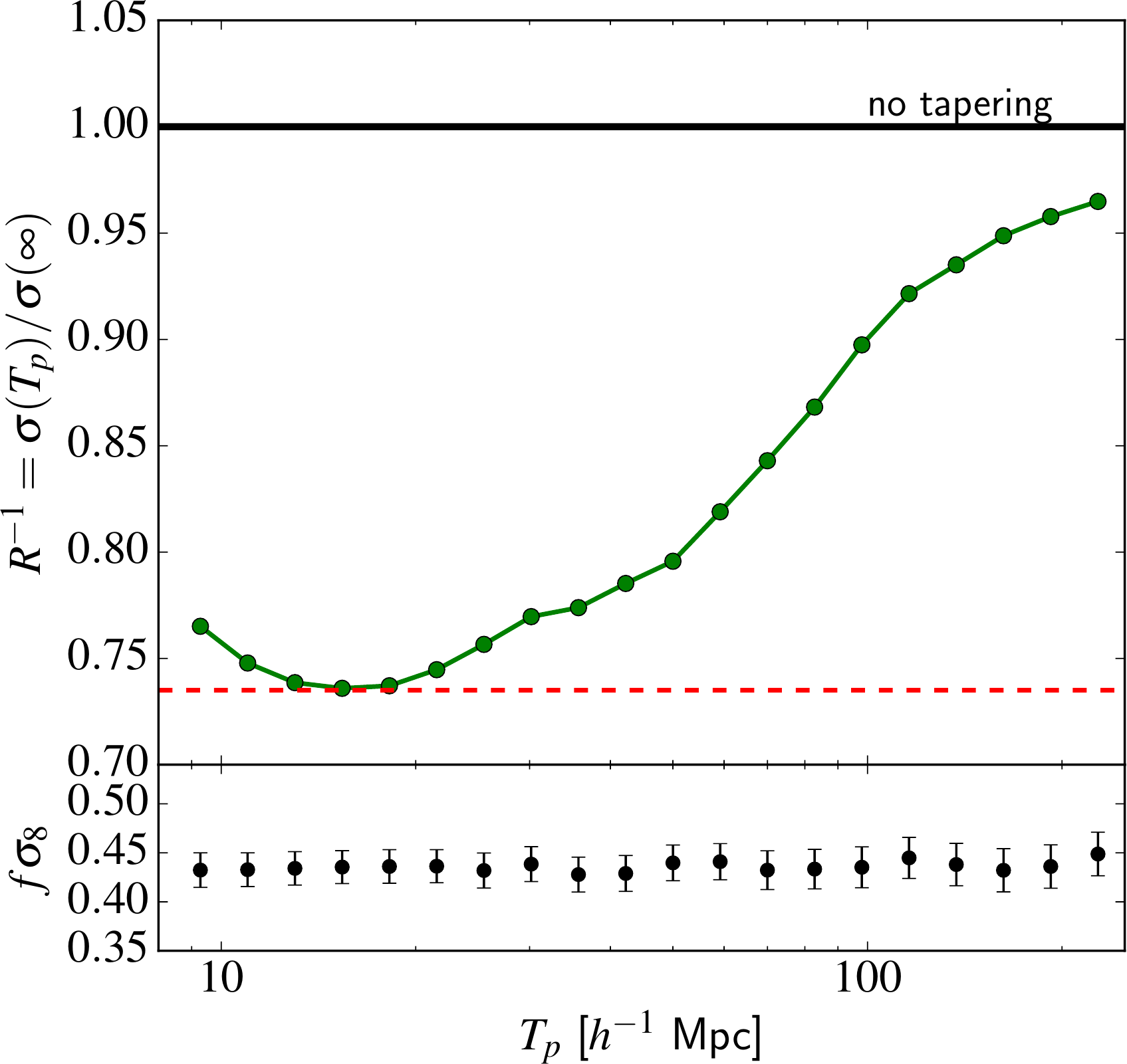}}
\caption{Top panel: recovered errors on $f\sigma_8$ normalized to that
  obtained without tapering (equivalent to applying tapering with
  $T_p=\infty$), as a function of the tapering scale $T_p$ used in the
  estimation of the precision matrix. This is obtained from the mocks
  at $0.5<z<0.7$. Bottom panel: recovered maximum likelihood values
  for $f\sigma_8$ and associated $1\sigma$ error as a function of the
  tapering scale $T_p$.}
\label{fig:optts}
\end{figure}

\begin{figure*}[ht]
\centering
\includegraphics[width=9cm]{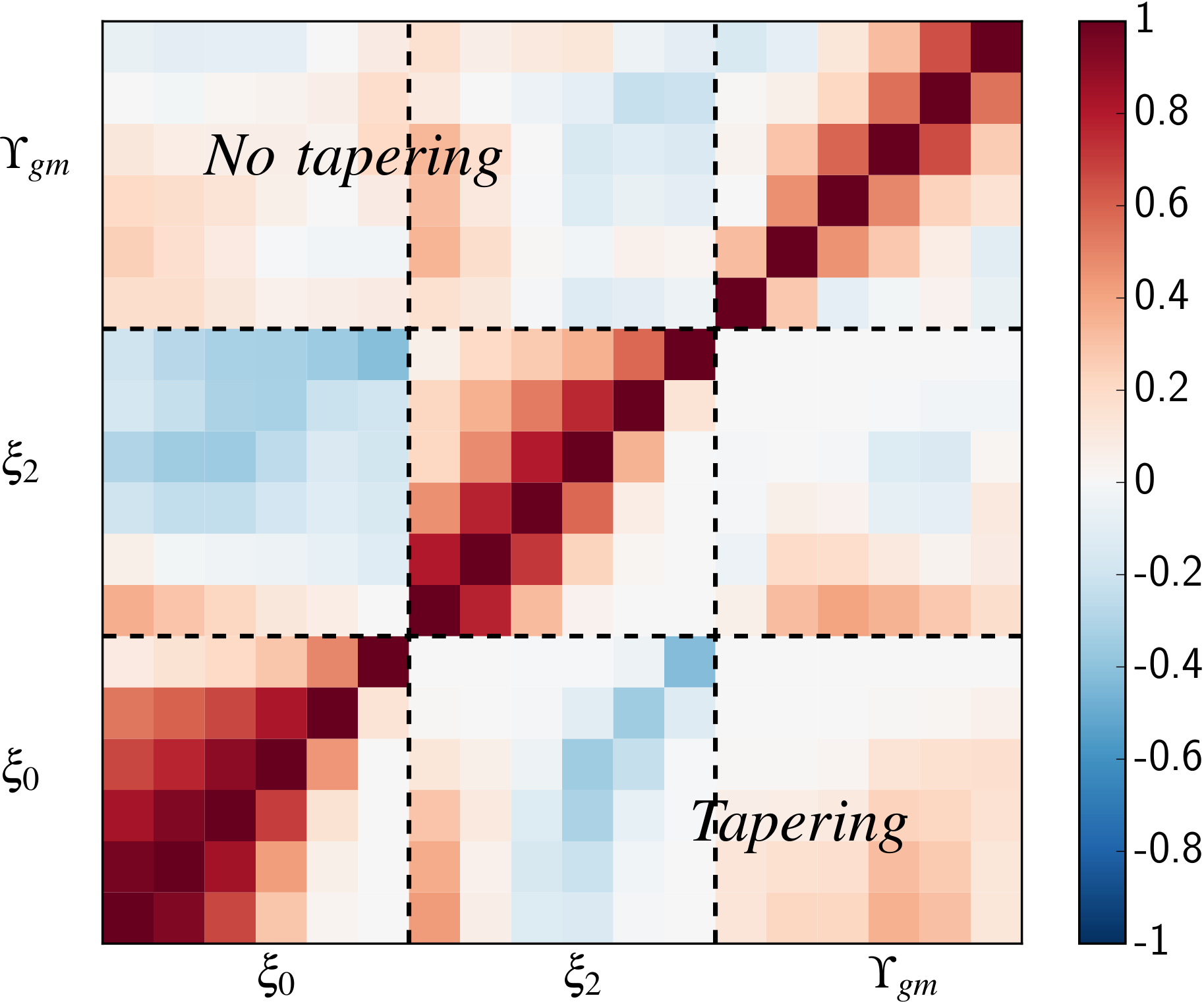}
\includegraphics[width=9cm]{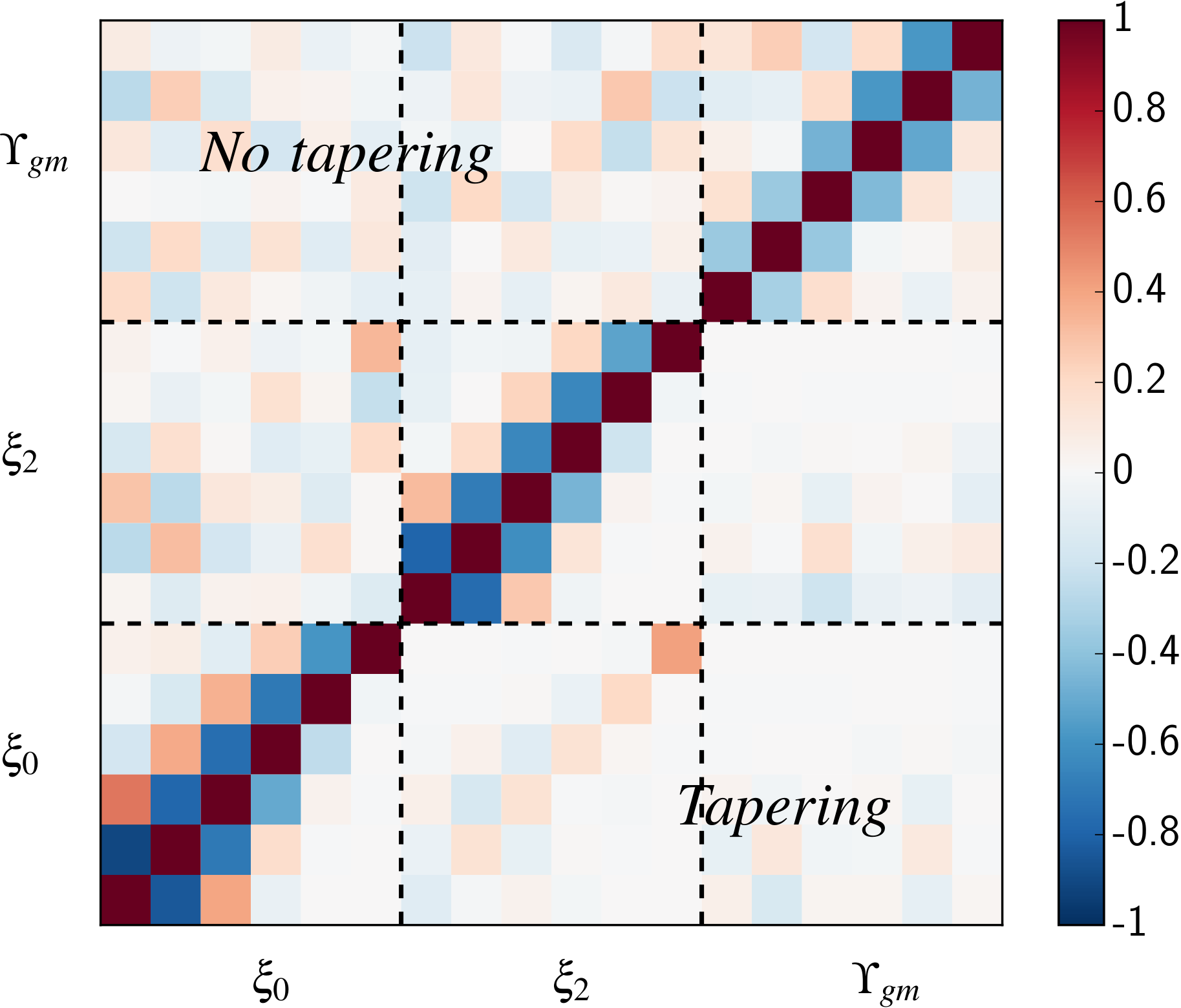}
\caption{Correlation matrix (left panel) and normalized precision
  matrix (right panel) for galaxy clustering and galaxy-galaxy lensing
  data in the redshift interval $0.5<z<0.7$. These are defined as
  $C_{ij}/\sqrt{C_{ii}C_{jj}}$ and
  $C^{-1}_{ij}/\sqrt{C^{-1}_{ii}C^{-1}_{jj}}$ respectively, where
  $C_{ij}$ and $C^{-1}_{ij}$ refer to covariance and precision matrix
  elements respectively. In both panels, the upper triangular matrix
  represents the case without tapering, while the lower panel the case
  with tapering. The precision matrix is normalized such that diagonal
  elements are unity.}
\label{fig:cormat}
\end{figure*}

In our case, the covariance matrix is associated with three different
quantities as well as two different separation types, $s$ and
$r_p$. One would then potentially need to use a combination of several
taper functions, since one does not expect the large-scale covariance
to vanish at the same scales for all quantities. Although it may be
sub-optimal to use a single taper function, we still expect to
increase the signal-to-noise, and since the estimator is unbiased, one
cannot introduce additional bias or error. We therefore decided to use
a single taper function for simplicity.

In the general case, it is not straightforward to define a priori the
optimal tapering scale. \citet{paz15} introduced a simple empirical
method, which consists of performing several maximum likelihood
analyses of the data varying only the tapering scale, and taking as
the optimal $T_p$ the one that minimizes the error on the parameter of
interest. We perform the same exercise on the mean mock
predictions. The marginalized $1\sigma$ error on $f\sigma_8$ as a
function of $T_p$ is presented in the top panel of
Fig. \ref{fig:optts}. We can see that the $T_p$ value that minimizes
the error is around $15\mhmpc$, and we adopt this value in our
analysis. We also verified that the maximum likelihood values for
$f\sigma_8$ remain unchanged for any value of $T_p$ as shown in the
bottom panel of Fig. \ref{fig:optts}.

To illustrate the method, we present in Fig.  \ref{fig:cormat} the
correlation matrix and normalized precision matrix, for the combined
RSD and galaxy-galaxy lensing data in the redshift interval
$0.5<z<0.7$, when applying or not the tapering technique (lower and
upper triangles respectively). Those matrices are defined as
$C_{ij}/\sqrt{C_{ii}C_{jj}}$ and
$C^{-1}_{ij}/\sqrt{C^{-1}_{ii}C^{-1}_{jj}}$ respectively, where
$C_{ij}$ and $C^{-1}_{ij}$ refer to covariance and precision matrix
elements respectively. We can see the reduction of noise, which is
particularly clear in the normalized precision matrix for most
off-diagonal terms.

The tapering technique allows a significant reduction of the noise
level in the precision matrix, but cannot completely remove it. The
remaining noise can propagate through the likelihood analysis into
derived parameter uncertainties. In order to obtain realistic
confidence limits on parameters one needs to account for the
additional uncertainties coming from the precision matrix estimation
\citep{taylor14}. \citet{percival14} showed that this additional error
can be described as a rescaling of the target parameter covariance, in
the case when the precision matrix is estimated with the standard
estimator of Eq. \ref{eq:covmat}. But the appropriate degree of
rescaling is unclear when the tapering estimator is used. The
improvement on the error that we find with the tapering estimator
(i.e. $26.5\%$) is similar to or larger than what we would expect with
the standard estimator using $300$ mocks or more as predicted by
\citep{dodelson13,percival14}. This gives us confidence that only a
small correction, if any, would be necessary.

\section{Cosmological results} \label{sec:results}

The comprehensive tests of the methodology described in previous
sections make us confident that we can perform a robust combined
analysis of RSD and galaxy-galaxy lensing with VIPERS and CFHTLenS
dataset, and infer cosmology from it. We present in this section the
data measurements, growth rate of structure constraints, and derived
gravitational slip parameters at $0.5<z<0.7$ and $0.7<z<1.2$.

\subsection{Galaxy clustering and galaxy-galaxy lensing measurements}

\begin{figure}
\resizebox{\hsize}{!}{\includegraphics{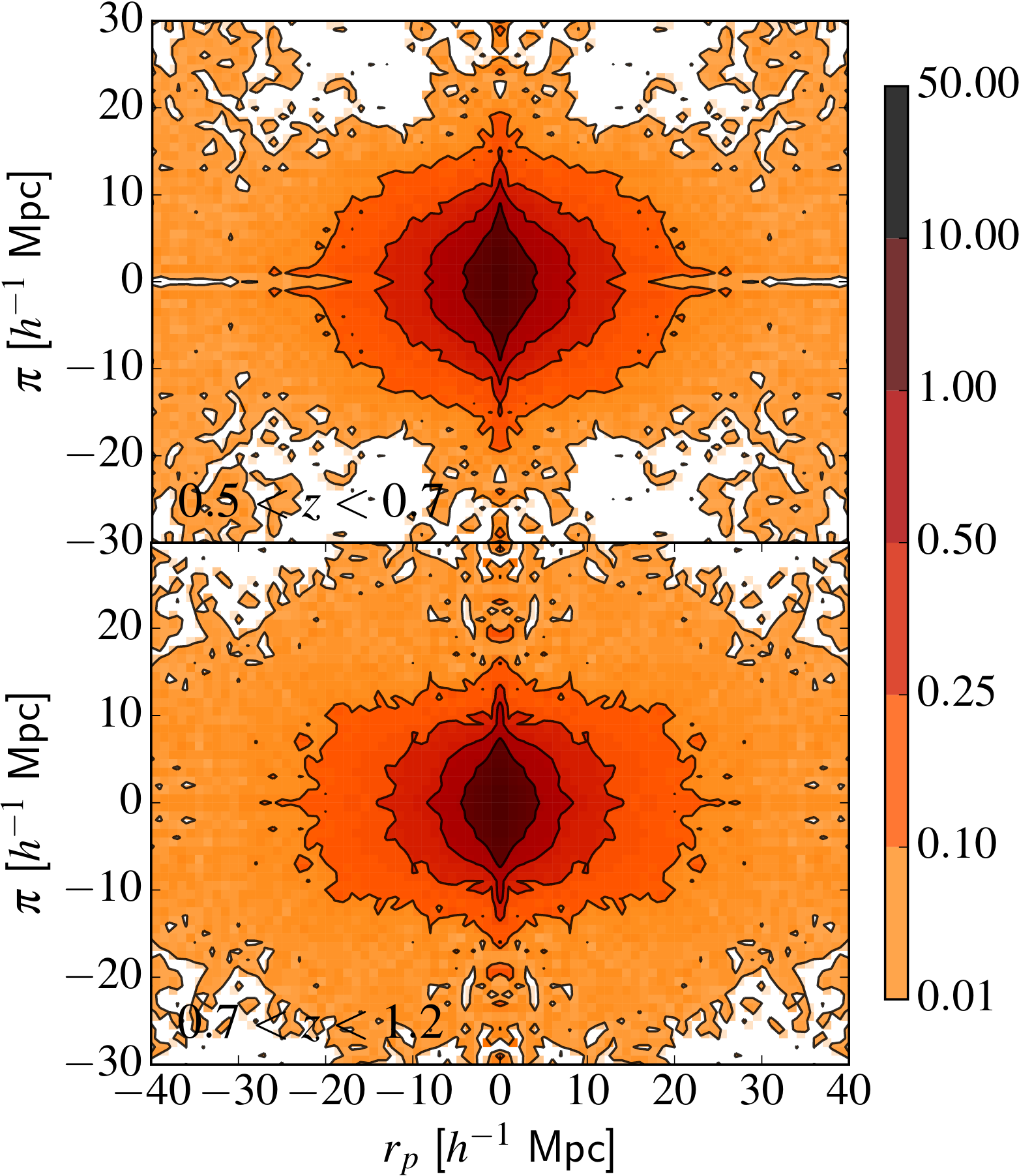}}
\caption{Anisotropic correlation functions of VIPERS galaxies at
  $0.5<z<0.7$ (top panel) and $0.7<z<1.2$ (bottom panel) as a function
  parallel and transverse to the line-of-sight separations.}
\label{fig:xirppi}
\end{figure}

\begin{figure}
\resizebox{\hsize}{!}{\includegraphics{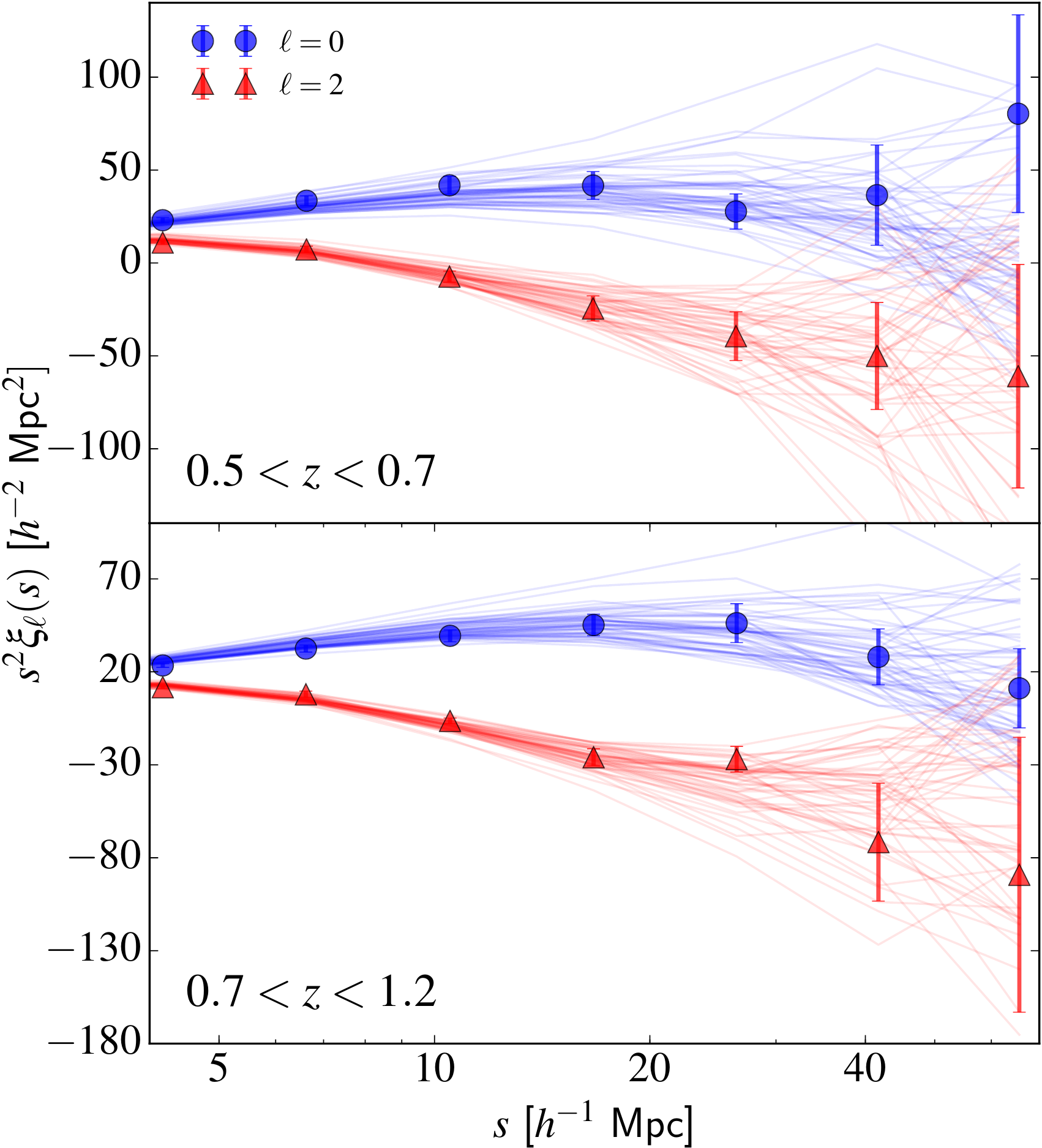}}
\caption{Monopole (circles) and quadrupole (triangles) correlation
  functions of VIPERS galaxies at $0.5<z<0.7$ (top panel) and
  $0.7<z<1.2$ (bottom panel). Solid curves correspond to individual
  mock measurements.}
\label{fig:multi}
\end{figure}

The correlation function measurements are performed on the full VIPERS
galaxy sample in the redshift intervals $0.5<z<0.7$ and
$0.7<z<1.2$. We select all VIPERS galaxies above the limiting
magnitude of the survey, and measure the monopole and quadrupole
correlation functions in both W1 and W4 fields. The combined W1+W4
measurements are obtained by summing up the pairs in the two fields,
contributing to the anisotropic two-point correlation functions
$\mxism$, before deriving $\xi_0$ and $\xi_2$ from
Eq. \ref{eq:xil}. The full anisotropic two-point correlation functions
are presented in Fig. \ref{fig:xirppi}, and the monopole and
quadrupole moments in Fig. \ref{fig:multi}. In the latter figure, the
individual mock measurements are superimposed, giving a visual
appreciation of the error associated with these measurements in
VIPERS. We can see that the combined W1+W4 monopole and quadrupole
correlation function measurements enable us to probe accurately the
redshift-space galaxy clustering signal on scales below about
$s=50\mhmpc$.

\begin{figure}
\resizebox{\hsize}{!}{\includegraphics{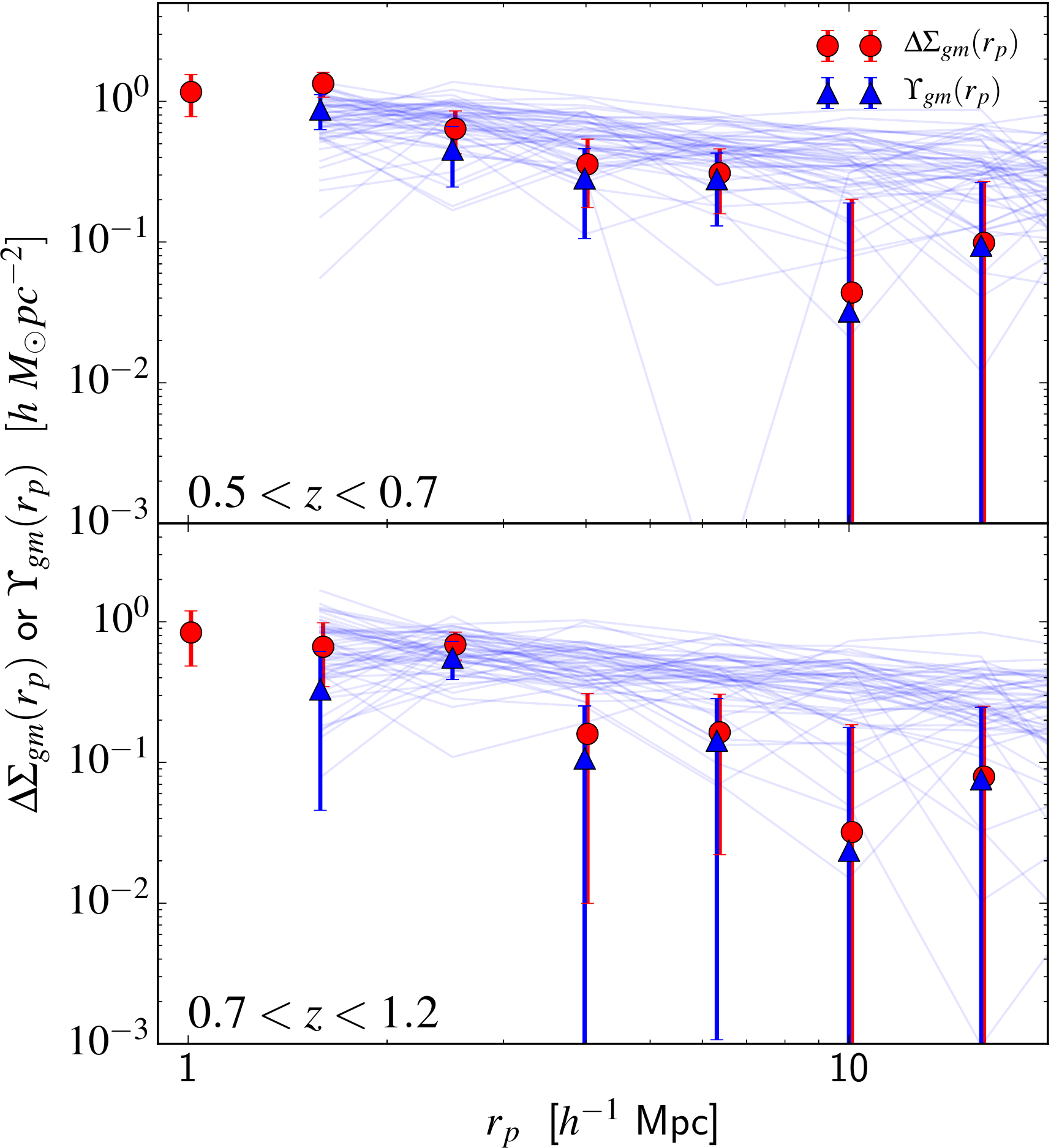}}
\caption{Differential excess surface density (circles) and annular
  differential excess surface density (triangles) at $0.5<z<0.7$ (top
  panel) and $0.7<z<1.2$ (bottom panel). Solid curves correspond to
  individual mock measurements.}
\label{fig:ups}
\end{figure}

The differential excess surface density measurements are obtained by
combining W1 and W4 individual field measurements in a similar
fashion. The lens galaxies are taken from the VIPERS catalogue or the
CFHTLenS catalogue if no spectroscopic redshift is available. They are
selected to have $i_{\rm AB}<22.5$ and a redshift in the intervals
$0.5<z<0.7$ and $0.7<z<1.2$. The source galaxies are taken from the
CFHTLenS catalogue and are selected to have $i_{\rm AB}<24.1$. The
differential excess surface density and annular differential excess
surface density measurements for $r_0=1\mhmpc$ are presented in
Fig. \ref{fig:ups}. As in Fig. \ref{fig:multi}, the individual mock
measurements are superimposed. We can see that with the combined W1+W4
annular differential excess surface density measurements we can reach
scales up to about $r_p=20\mhmpc$.

Our $\Upsilon_{gm}$ measurements are more uncertain than the $\xi_0$
and $\xi_2$ ones. This is essentially related to the way the former
are estimated. Weak lensing is fundamentally limited by the unknown
intrinsic ellipticity of the sources, which dominates the error
budget. This can be mitigated by means of a larger number of
sources. Given the surface density of sources in our sample and its
rather modest angular coverage of $23.5~\rm{deg}^2$, we obtain
relative errors on $\Upsilon_{gm}$ of about $25\%$, estimated from the
mock samples. In contrast, the typical relative error that we obtain
on $\xi_0$ is of $5\%$. Therefore, in our combined analysis of the RSD
and galaxy-galaxy lensing we expect $\Upsilon_{gm}$ to have a much
lower weight in the likelihood. We finally remark that the observed
$\Upsilon_{gm}$ tend to exhibit lower amplitudes than expected in the
mock samples, in particular in the highest redshift interval. We
discuss the cosmological implications of this in
Sect. \ref{sec:gravslip}.

\subsection{Growth of structure constraints}

We perform a combined maximum likelihood analysis of the monopole,
quadrupole, and annular differential excess surface density to derive
constraints on the growth rate of structure at $0.5<z<0.7$ and
$0.7<z<1.2$. The effective redshifts associated with these intervals
are $z=0.6$ and $z=0.86$. They correspond to the average redshift of
pairs contributing the most to monopole and quadrupole correlation
functions in these redshift intervals \citep{samushia14}. The
theoretical model that we use is described in
Sect. \ref{sec:modelling}; it depends on 11 parameters, ${\bf
  p}=(f,b_1,b_2,\sigma_v,\sigma_8,\epsilon,\alpha,\Omega_m,\Omega_mh^2,\Omega_bh^2,n_s)$. The
last three describe the shape of the matter power spectrum and these
are determined most accurately by CMB data. Since our galaxy
clustering and weak lensing measurements cannot provide such tight
constraints on these parameters, we fix them to the best-fitting
Planck 2015 TT+lowP+lensing parameters \citep{planck15}. Consistently,
$\Omega_m$ is kept fixed to the Planck value in
$\Upsilon_{gm}$. Possible departures from those parameter values are
only allowed through variations of the AP distortion parameters
$\epsilon$ and $\alpha$. In the following, we first consider
measurements of $f\sigma_8$, as a derived parameter, and later study
the possibility of deriving independent measurements of $f$ and
$\sigma_8$. All those measurements are obtained by marginalizing over
the nuisance parameters: ${\bf
  p_n}=(b_1,b_2,\sigma_v,\epsilon,\alpha)$. The adopted uniform priors
on the likelihood parameters are summarized in Table \ref{tab} and the
full posterior likelihood contours for the cases presented in the next
section are given in Appendix \ref{appendix2}.

\begin{table}
  \caption{Adopted priors on the likelihood parameters.}
  \centering
  \begin{tabular}{cc}
    \hline\hline
    Parameters & Uniform prior \\
    \hline
    $b_1$ & $[0.5, 2]$ \\
    $b_2$ & $[-1, 1]$ \\
    $\sigma_v$ & $[0, 8]$ \\
    $f$ & $[0.2, 1.4]$ \\
    $\sigma_8$ & $[0, 1.2]$ \\
    $\epsilon$ & $[-0.1, 0.1]$ \\
    $\alpha$ & $[0.9, 1.1]$ \\
    \hline
  \label{tab}
\end{tabular}
\end{table}

\subsubsection{$f\sigma_8$ measurements} \label{sec:fsig8meas}

In our standard configuration, the linear matter power spectrum shape
is fixed to the best-fitting $\Lambda\rm{CDM}$ model from Planck 2015
TT+lowP+lensing data \citep{planck15}. AP distortion parameters are
set to $(\epsilon,\alpha)=(0,1)$ and are not allowed to vary. In this
configuration we obtain $f\sigma_8$ values of
\begin{eqnarray}
  f\sigma_8(z=0.6)&=&0.48\pm0.12 \\
  f\sigma_8(z=0.86)&=&0.48\pm0.10,
\end{eqnarray}
after marginalizing over other parameters. Associated reduced
chi-squared values are $\chi^2_\nu=1.52$ and $\chi^2_\nu=1.62$
respectively.  These measurements use both RSD and galaxy-galaxy
lensing information. It is instructive to see the impact of adding the
galaxy-galaxy lensing on the measurement of $f\sigma_8$. Thus if we
use the standard RSD approach without including galaxy-galaxy lensing
information, we obtain
\begin{eqnarray}
  f\sigma_8(z=0.6)&=&0.48\pm0.11 \\
  f\sigma_8(z=0.86)&=&0.46\pm0.09,
\end{eqnarray}
with a reduced chi-squared value of $\chi^2_\nu=1.12$ for both
redshifts. In that case, we fixed $b_2=b_{s^2}=b_{3nl}=0$ in the RSD
model, as bias non-linearities are negligible for VIPERS galaxies bias
given the minimum scale used in the fit \citep{pezzotta16}. Moreover,
the shape of non-linear power spectra in the model is fixed by setting
$\sigma_8$ to its fiducial value at the effective redshift of the
sample, as is commonly done \citep[e.g.][]{delatorre13a}. The
recovered values and associated errors are very similar to the
previous case. We do not find an improvement on $f\sigma_8$ accuracy
when galaxy-galaxy lensing is included, in fact errors are marginally
larger. This can be explained by the lower number of degrees of
freedom in the RSD-only case and the significant uncertainty
associated with our galaxy-galaxy lensing measurements compared to the
galaxy clustering ones in the VIPERS fields. In fact the real gain is
on contraining $f$ and $\sigma_8$ separately as discussed in
Sect. \ref{sec:breakdegen}.

\subsubsection{Inclusion of Alcock-Paczynski distortions}

As a robustness test, we relax the assumption on the shape of the
linear matter power spectrum. We allow the AP distortion parameters
$(\epsilon,\alpha)$ to vary, considering flat priors on
$\epsilon,\alpha$ parameters, extending by $\pm 0.1$ around
$(\epsilon,\alpha)=(0,1)$. After marginalizing over those parameters
as well, we obtain the following $f\sigma_8$ measurements:
\begin{eqnarray}
  f\sigma_8(z=0.6)&=&0.51\pm0.13, \\
  f\sigma_8(z=0.86)&=&0.52\pm0.11,
\end{eqnarray}
with reduced chi-squared values of $\chi^2_\nu=1.58$ and
$\chi^2_\nu=1.3$ respectively. As expected from the additional
degrees of freedom introduced in the likelihood, the marginalized
$68\%$ errors on $f\sigma_8$ are increased, although the constraints
remain completely compatible with previous measurements when
$\epsilon$ and $\alpha$ were fixed. This test thus removes any
potential concern that our measurements of $f\sigma_8$ might lack
robustness though being dependent on the assumption of a
$\Lambda\rm{CDM}$ expansion history.

\subsubsection{Comparison with other measurements}

\begin{figure*}[ht]
\centering
\includegraphics[width=13cm]{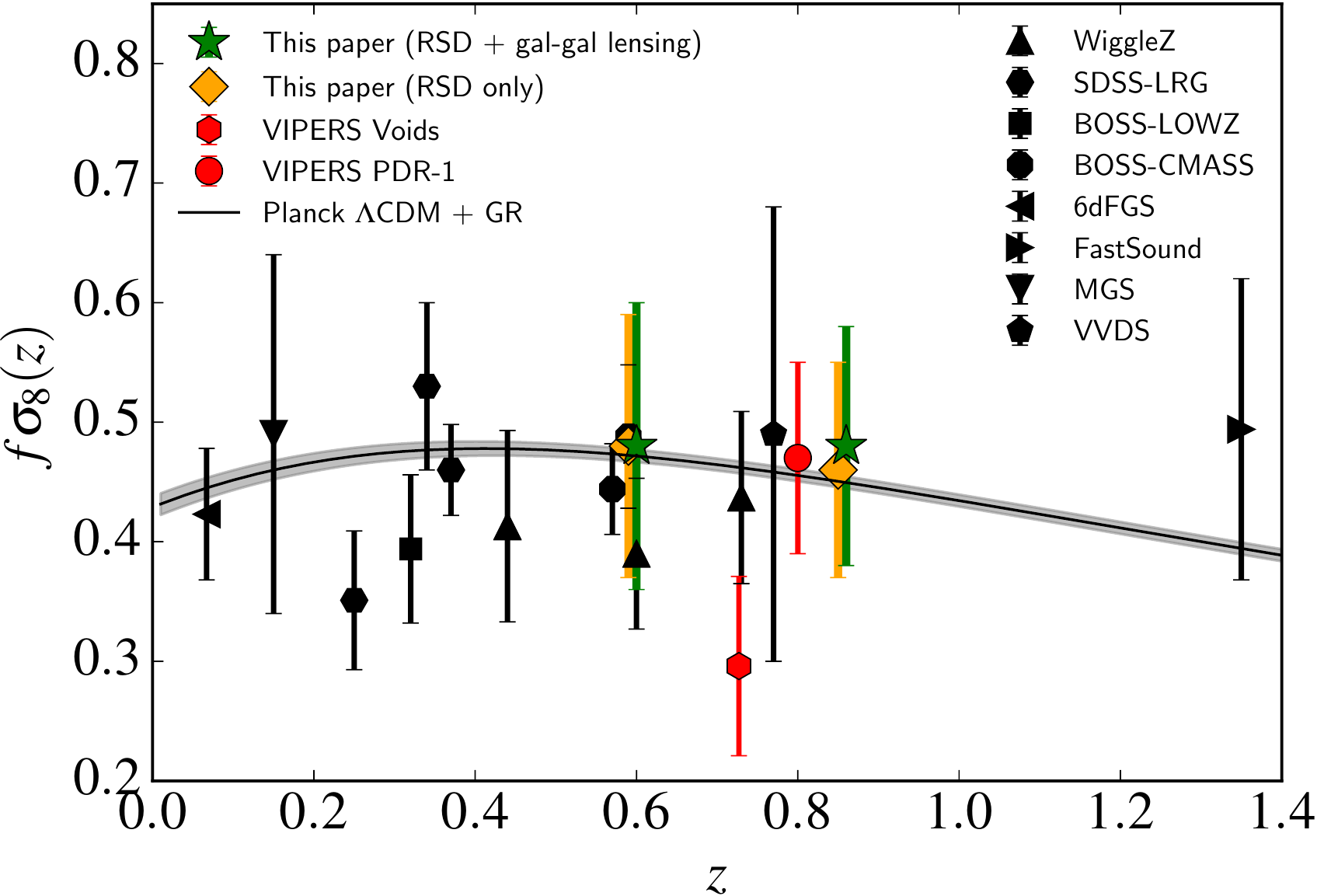}
\caption{$f\sigma_8$ as a function of redshift, showing VIPERS results
  contrasted with a compilation of recent measurements. The previous
  results from VVDS \citep{guzzo08}, SDSS LRG
  \citep{cabre09,samushia12}, WiggleZ \citep{blake12}, 6dFGS
  \citep{beutler12}, VIPERS PDR-1 \citep{delatorre13a}, MGS
  \citep{howlett15}, FastSound \citep{okumura16}, BOSS-LOWZ
  \citep{gilmarin16}, BOSS-CMASS \citep{gilmarin16,chuang16}, and
  VIPERS PDR-2 voids \citep{hawken16} are shown with the different
  symbols (see labels). The solid curve and associated shaded area
  correspond to the expectations and $68\%$ uncertainty for General
  Relativity in a $\Lambda \rm{CDM}$ background model set to
  TT+lowP+lensing Planck 2015 predictions \citep{planck15}.}
\label{fig:fsigma8}
\end{figure*}

In Fig. \ref{fig:fsigma8} we compare our $f\sigma_8$ measurements with
previous measurements from the literature, as well as predictions of
the standard relativistic model for gravity. Our measurements are
consistent with previous measurements at lower or similar redshifts
from VVDS \citep{guzzo08}, SDSS LRG \citep{cabre09,samushia12},
WiggleZ \citep{blake12}, 6dFGS \citep{beutler12}, VIPERS PDR-1
\citep{delatorre13a}, MGS \citep{howlett15}, FastSound
\citep{okumura16}, BOSS-LOWZ \citep{gilmarin16}, and BOSS-CMASS
\citep{gilmarin16,chuang16}. In particular, our measurement at $z=0.6$
is compatible within $1\sigma$ with the WiggleZ $z=0.6$
\citep{blake12} and BOSS-CMASS $z=0.57$ \citep{gilmarin16,chuang16}
measurements. Our results are also very close to the standard
cosmological model predictions: they are consistent within $1\sigma$
with General Relativity predictions in a $\Lambda \rm{CDM}$ model with
cosmological parameters set to Planck CMB results
\citep[][]{planck15}.

These results are part of a combined effort of the VIPERS
collaboration to estimate the growth rate of structure from the same
data but using different complementary techniques. Specifically, in
\citet{pezzotta16} we provide a thorough investigation of the
performances of different RSD models in configuration space, using a
general consistent modelling of non-linear RSD; in \citet{wilson16} we
use the clipping technique in Fourier space to minimise the impact of
non-linearities; finally in \citet{hawken16} we use cosmic voids as
RSD tracers.  In particular in \citet{hawken16}, we make use of the
void catalogue built from the VIPERS PDR-2 data and resulting from the
earlier work by \citet{micheletti14}, to estimate the void-galaxy
cross-correlation function in redshift space. By modelling its
anisotropy we obtain an estimate of $f\sigma_8$ at $z = 0.73$ and
derive a value of $f\sigma_8(z=0.73) = 0.296^{+0.075}_{-0.078}$, which
is lower than those obtained here. However, this technique is still in
its infancy, with potential systematic errors not yet fully
understood.  This and the other VIPERS measurements are all fully
compatible within statistical errors. More discussion is presented in
the specific papers.

\subsubsection{$f, b_1, \sigma_8$ degeneracy breaking} \label{sec:breakdegen}

\begin{figure}
\resizebox{\hsize}{!}{\includegraphics{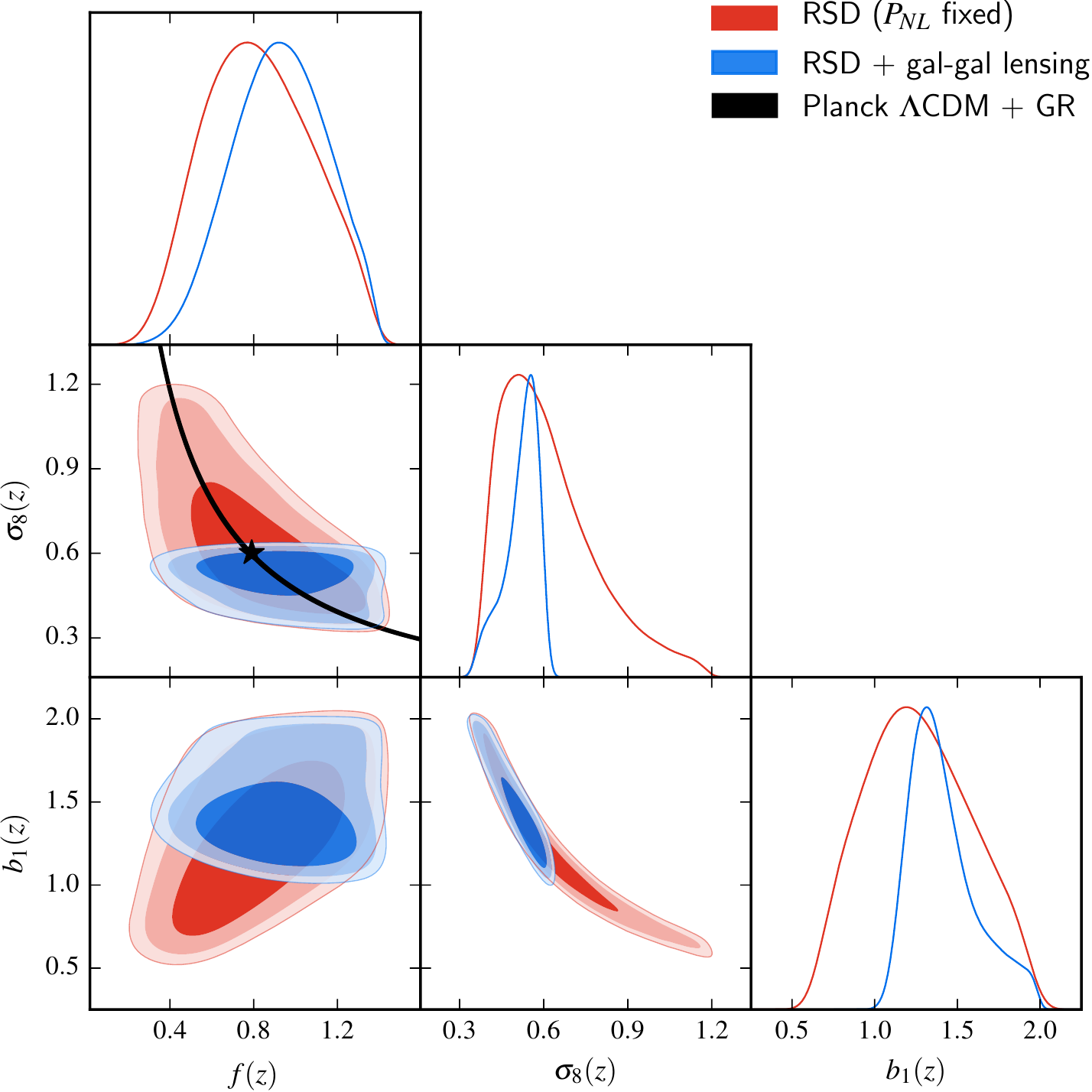}}
\caption{Two-dimensional marginalized posterior likelihood contours
  for $f$ and $\sigma_8$ at $0.5<z<0.7$, showing the impact of the
  additional galaxy-galaxy lensing constraint on the $f-\sigma_8$
  degeneracy. The black curve shows the region of constant $f\sigma_8$
  associated with \citet{planck15} ${\rm \Lambda CDM+GR}$ best-fit,
  while the combined $(f,\sigma_8)$ constraint is marked with the
  star.}
\label{fig:fvssig8_low}
\end{figure}

\begin{figure}
\resizebox{\hsize}{!}{\includegraphics{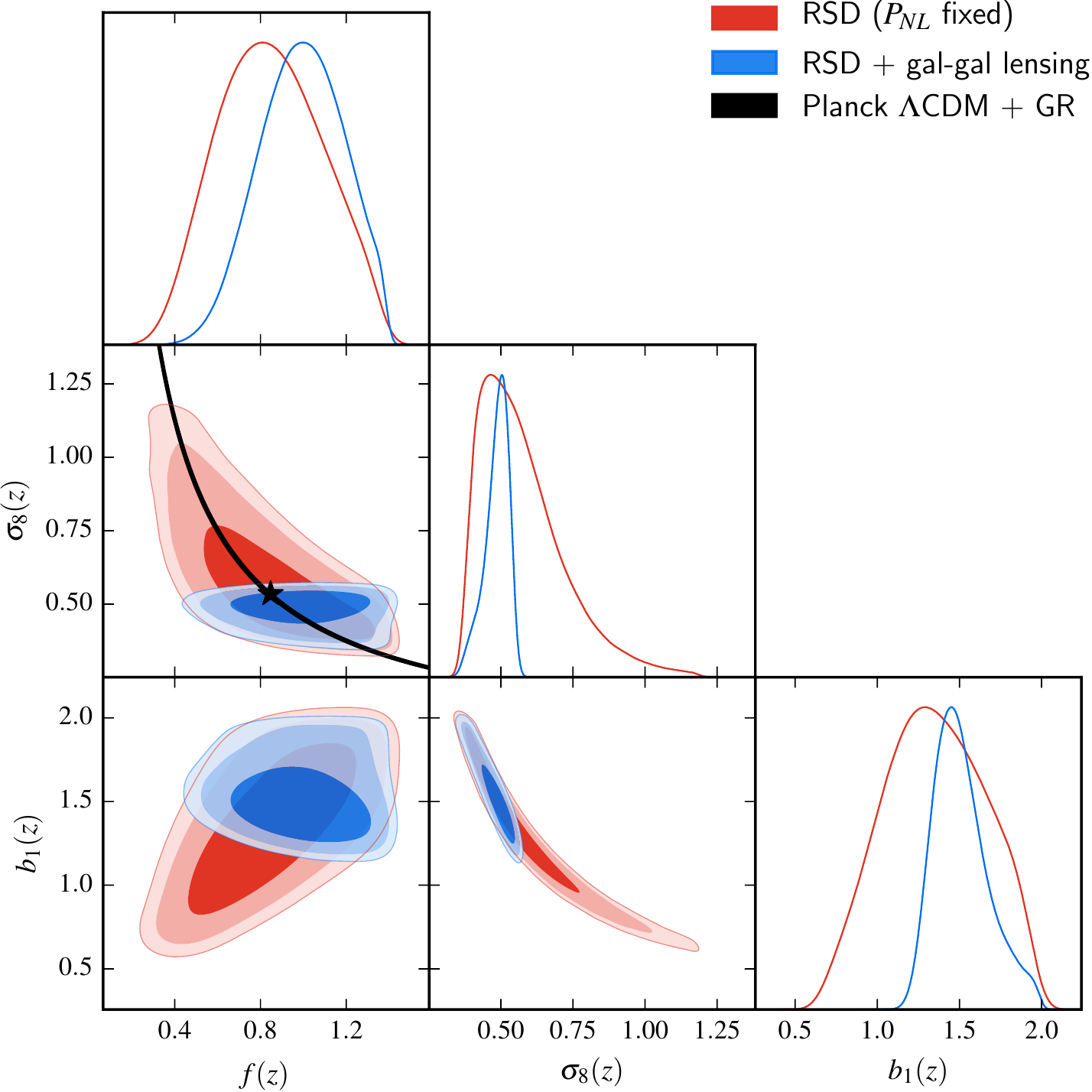}}
\caption{Same as Fig. \ref{fig:fvssig8_low} but for the redshift
  interval $0.7<z<1.2$.}
\label{fig:fvssig8_high}
\end{figure}

As discussed in Sect. \ref{sec:modelling}, the use of RSD in the
galaxy clustering pattern allows a measurement of the parameter
combinations $f\sigma_8$, $b_1\sigma_8$, or $\beta=f/b_1$. But with
the additional constraint of galaxy-galaxy lensing, which exhibits
different parameter dependencies, we expect to be able to break the
$f-b_1-\sigma_8$ degeneracy inherent to galaxy-galaxy correlations. We
investigate this by studying the posterior likelihood contours at
$68\%,95\%,99\%$ for the various pairs of $f$, $b$, $\sigma_8$
parameters in our data. This is done for the likelihood analyses
presented in the previous sections, i.e. when including or not
galaxy-galaxy lensing. The posterior likelihood contours are presented
in Figs. \ref{fig:fvssig8_low} and \ref{fig:fvssig8_high} for the two
considered redshift intervals.

These figures show strong degeneracies in the $f-b_1$, $f-\sigma_8$,
and $b_1-\sigma_8$ planes when considering only RSD. In particular, we
can see in the $f-\sigma_8$ plane the distribution of the likelihood
contours along the regions with constant $f\sigma_8$, marked with
solid and dashed curves in the figures. Now with the inclusion of
galaxy-galaxy lensing, we can see a shrinking of the contours, in
particular along the $\sigma_8$ direction, and to a lesser extent
along the $b_1$ one. Galaxy-galaxy lensing thus effectively provides a
strong handle on the $\sigma_8$ parameter. This allows the
$f-\sigma_8$ degeneracy to be broken and therefore leads to the
possibility of a direct measurement of the growth rate of structure,
$f$. The $f-b_1$ degeneracy is also partially broken, even if the
effect is milder.

\begin{figure}
\resizebox{\hsize}{!}{\includegraphics{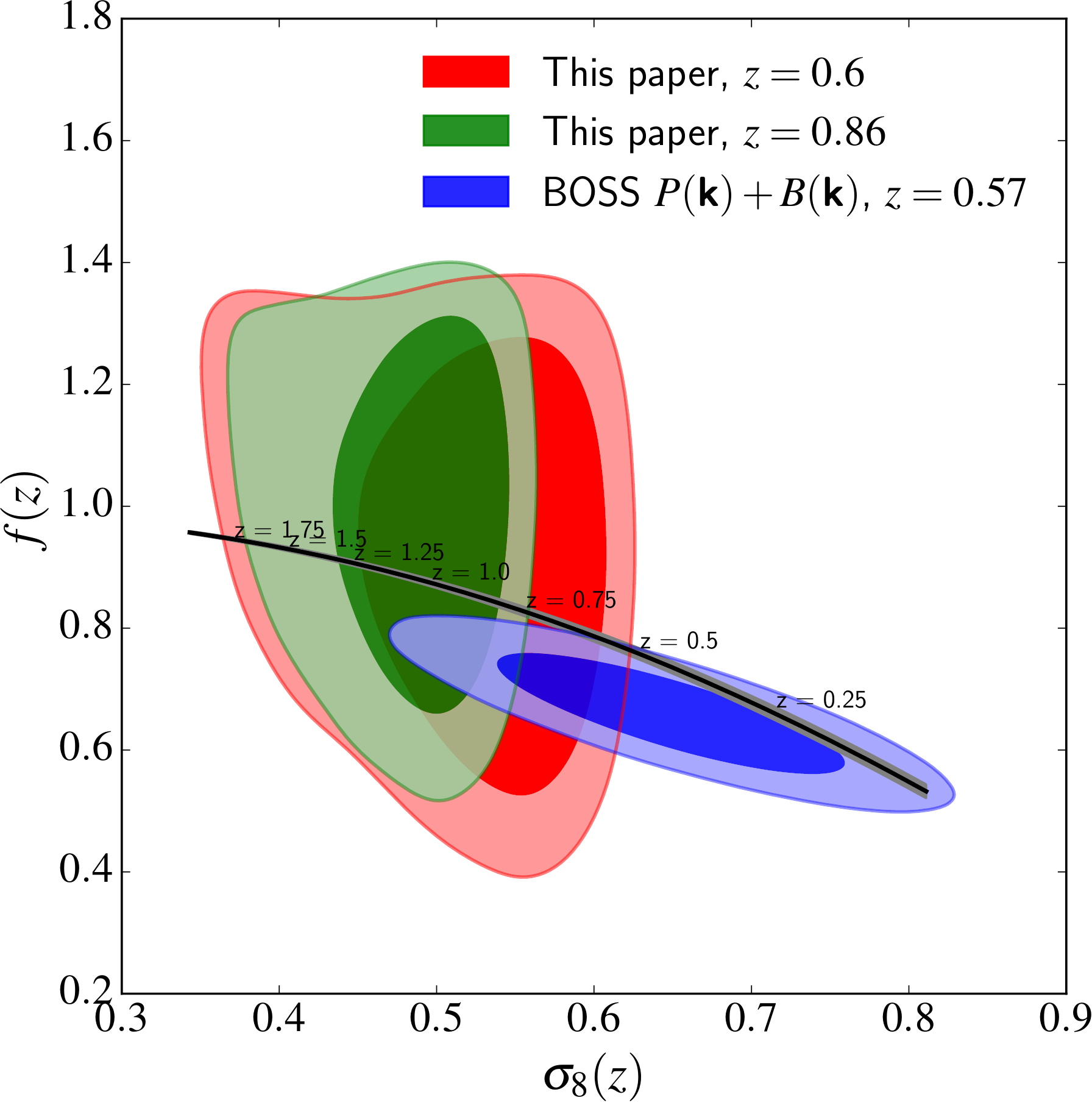}}
\caption{Joint $(f,\sigma_8$) constraints at different redshifts. The
  combined RSD and galaxy-galaxy lensing posterior likelihood contours
  at $1\sigma$ and $2\sigma$ and those from \citet{gilmarin17},
  obtained by combining redshift-space power spectrum and bispectrum
  information in the BOSS survey, are presented. The solid curve and
  associated grey shaded area correspond to the expectations and
  $68\%$ uncertainty for General Relativity in a $\Lambda \rm{CDM}$
  background model set to TT+lowP+lensing Planck 2015 predictions
  \citep{planck15}, as a function of redshift from $z=2$ to $z=0$.}
\label{fig:fsigma8lit}
\end{figure}

We find that the $f-\sigma_8$ degeneracy breaking is more efficient in
the high-redshift interval, with measurements of
$(f,\sigma_8)=(0.93\pm0.22,0.52\pm0.06)$ and
$(f,\sigma_8)=(0.99\pm0.19,0.48\pm0.04)$ at $z=0.6$ and $z=0.86$
respectively. These direct measurements of the growth rate of
structure and $\sigma_8$ are in agreement within $1\sigma$ with Planck
${\rm \Lambda CDM+GR}$ predictions, which are
$(f,\sigma_8)=(0.79,0.60)$ and $(f,\sigma_8)=(0.85,0.53)$ respectively
at $z=0.6$ and $z=0.86$. Planck ${\rm \Lambda CDM+GR}$ predictions are
represented with the stars in Figs. \ref{fig:fvssig8_low} and
\ref{fig:fvssig8_high}. In Fig. \ref{fig:fsigma8lit}, we compare our
$(f,\sigma_8)$ constraints with those from \citet{gilmarin17},
obtained by combining redshift-space galaxy power spectrum and
bispectrum information in the BOSS survey at $z=0.57$. In
\citet{gilmarin17}, they use the galaxy bispectrum instead of
galaxy-galaxy lensing to bring additional constraints on galaxy
bias. Although those measurements are quite uncertain, this parameter
space and how it can be used as a cosmological model diagnostic, will
be very interesting to explore for next-generation cosmological
surveys, such as Euclid, which will allow a dramatical improvement on
such measurement accuracy.
  
Independent measurements of $\sigma_8$ at different redshifts also
carry information about the growth rate of structure. Since $\sigma_8$
grows with time proportionally to the growth factor, the growth rate
can be written as $\d\ln \sigma_8/\d\ln a$. In the case of two
$\sigma_8$ measurements at $a_1$ and $a_2$, as in our analysis, this
equation can be approximated through finite difference by
\begin{equation}
f\simeq\frac{\ln\left(\sigma_8(a_1)/\sigma_8(a_2)\right)}{\ln\left(a_1/a_2\right)}.
\end{equation}
By applying this to our $\sigma_8$ measurements we obtain an
additional, independent measurement of $f=0.57\pm0.96$ at the mean
redshift of $z=0.73$. It is clear that this type of measurement is not
compelling in our dataset, but can potentially be useful as an
additional constraint to be combined with direct measurements in
next-generation cosmological surveys.

Finally, we notice in Figs. \ref{fig:fvssig8_low} and
\ref{fig:fvssig8_high} that the addition of galaxy-galaxy lensing
constraints significantly modifies the posterior probability
distribution function of the linear bias parameter, $b_1$, becoming
more compact and skewed towards larger values.  This means that adding
galaxy-galaxy lensing information reduces the uncertainties on $b_1$,
and pushes its maximum likelihood value towards values that are in
excellent agreement with previous linear bias estimates that are not
solely based on two-point statistics
\citep{diporto14,cappi15,granett15}.

\subsection{Gravitational slip} \label{sec:gravslip}

In addition to the growth rate of structure, we can measure the
gravitational slip parameter $E_G$. This is done by taking the ratio
of the measured $\Upsilon_{gm}$ and $\Upsilon_{gg}$, and multiplying
it by $\beta^{-1}$. The RSD distortion parameter $\beta$ is estimated
from the combined maximum likelihood analysis of the monopole and
quadrupole correlation functions (the same as for the RSD-only case
presented in Sect. \ref{sec:fsig8meas}). After marginalizing over
nuisance parameters we obtain
\begin{eqnarray}
  \beta(z=0.6)&=&0.66\pm0.17 \\
  \beta(z=0.86)&=&0.63\pm0.14.
\end{eqnarray}
The $68\%$ error on the $E_G$ measurements is obtained by adding in
quadrature the fractional error on $\Upsilon_{gm}/\Upsilon_{gg}$
estimated from mock samples and the fractional error on $\beta^{-1}$.

\begin{figure}
\resizebox{\hsize}{!}{\includegraphics{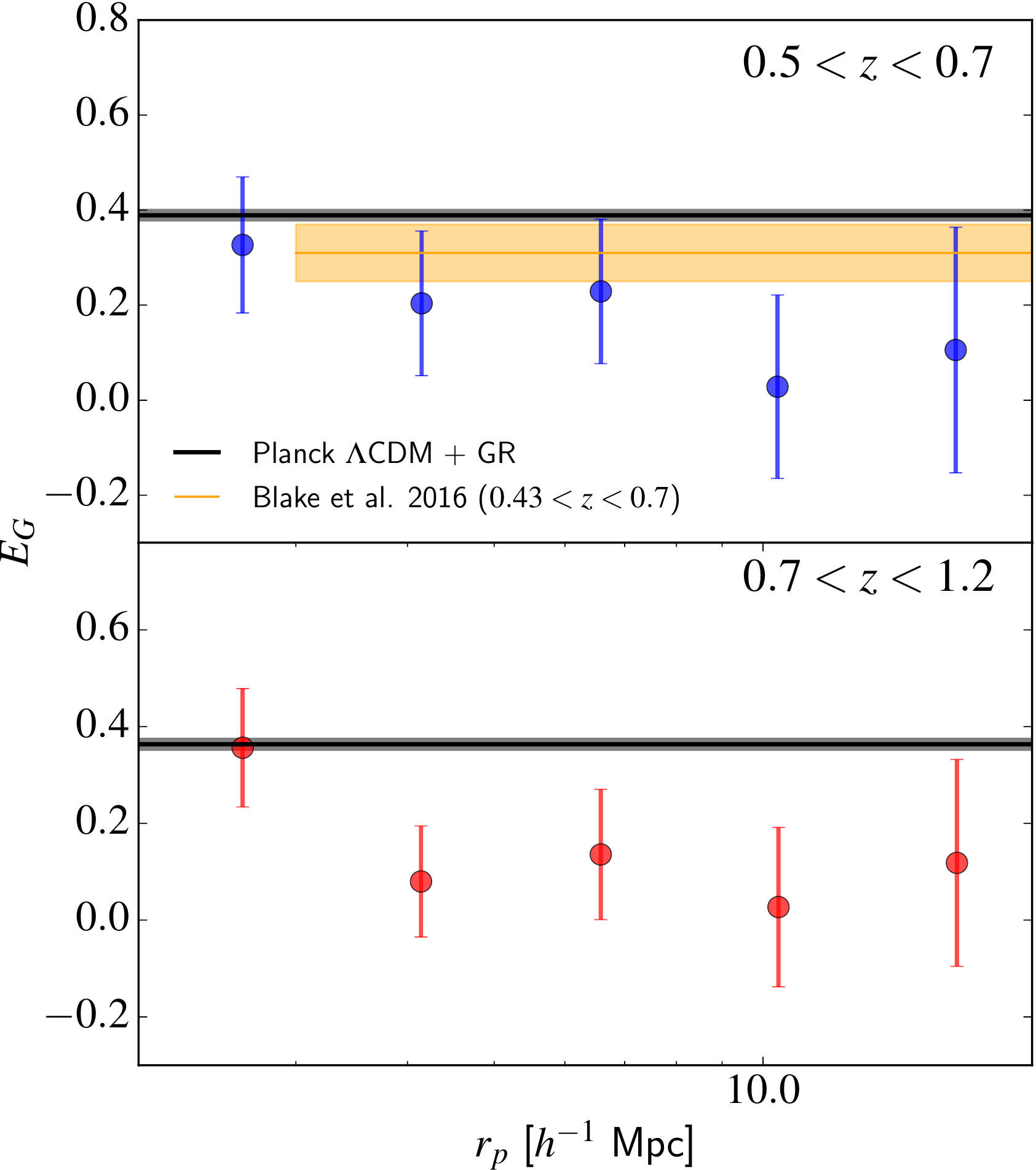}}
\caption{Gravitational slip parameter as a function of scale as
  measured at $0.5<z<0.7$ (top panel) and $0.7<z<1.2$ (bottom
  panel). In both panels, the solid curves and associated shaded areas
  correspond to the expectations and $68\%$ uncertainties for General
  Relativity in a $\Lambda \rm{CDM}$ background model set to
  TT+lowP+lensing Planck 2015 predictions \citep{planck15}. In the top
  panel, the horizontal stripe shows the averaged
  $\smash{\overline{E}_G}$ over the range $3\mhmpc$<$r_p$<$50\mhmpc$
  obtained by \citet{blake16} at $0.43<z<0.7$. $E_G$ asymptotes to
  $\Omega_m/f$ in the standard model, and the simplest way of erasing
  the modest discrepancy with the model prediction would be to lower
  the density parameter.}
\label{fig:Eg}
\end{figure}

The $E_G(r_p)$ measurements are presented in Fig. \ref{fig:Eg} for the
two redshift intervals under consideration, and compared with the
linear predictions for ${\rm \Lambda CDM+GR}$ (horizontal line and
associated $68\%$ contour). We find that our measurements at $z=0.6$
are compatible within $1\sigma$ with the standard model, although the
central values tend to be slightly lower. We also report in this
figure the averaged gravitational slip parameter over the range
$3\mhmpc$<$r_p$<$50\mhmpc$, $\smash{\overline{E}_G}$, obtained by
\citet{blake16} in the similar redshift range $0.43<z<0.7$. It is
represented with a stripe in the figure, with horizontal extent
corresponding to the range of $r_p$ used to measure
$\smash{\overline{E}_G}$ and vertical extent showing the $\pm1\sigma$
error on the measurement. By averaging our $E_G$ over
$3\mhmpc$<$r_p$<$20\mhmpc$ we obtain
$\smash{\overline{E}_G}(z=0.6)=0.16\pm0.09$ and
$\smash{\overline{E}_G}(z=0.86)=0.09\pm0.07$. Our results are in good
agreement with this measurement and also with that by \citet{pullen16}
at much higher scales, which also exhibits a slightly lower value
compared with ${\rm \Lambda CDM+GR}$ prediction. The $E_G$
measurements are lower than ${\rm \Lambda CDM+GR}$ at $r_p>3\mhmpc$
but remain within $1-2\sigma$, depending on the scale. At $z=0.86$,
the agreement with ${\rm \Lambda CDM+GR}$ is poorer than at lower
redshift.
 
The origin of the tendency of our $E_G$ measurements to be smaller
than expected at $r_p>3\mhmpc$ remains unclear. Particularly given our
statistical errors, we have to be cautious in interpreting this
trend. In any case, such a result could arise as a result of residual
observational systematics or a misinterpretation of the observables,
rather than any break-down of standard gravitational physics.  From
the construction of $E_G$, these low values of $E_G$ seem to be most
probably caused by the rather low measured amplitude of
$\Upsilon_{gm}$ at $r_p>3\mhmpc$, and seen in Fig. \ref{fig:ups}. If
this discrepancy is upheld by further data, one possible
interpretation is that weak lensing prefers a lower value of
$\Omega_m$ than that determined by CMB data. It is worth noticing that
a similar tension has already been identified in the CFHTLenS cosmic
shear analysis of \citet{heymans13}, as well as in the more recent
analysis performed in the KiDS lensing survey
\citep{hildebrandt16}. It is clear that this point needs to be
investigated in detail in the future, in particular in the preparation
of next-generation very large surveys combining galaxy clustering and
weak lensing observables.

\section{Conclusion}

This paper has presented a combined analysis of redshift-space
distortions and galaxy-galaxy lensing in the final VIPERS dataset,
making use of complementary data from the CFHTLenS lensing survey over
the same area. We have built a consistent theoretical model of the two
observables, which includes prescriptions for non-linear, non-local
galaxy bias, as well as quasi-linear redshift-space distortions. This
model has been shown to enable robust measurements of the growth rate
of structure. The model robustness and adopted methodology have been
tested by using a series of realistic mock surveys constructed for
this purpose.

The main goal of VIPERS has been to provide an accurate measurement of
the growth rate of structure using redshift-space distortions in a
redshift regime where the growth is not well determined. With the
first data release we were able to provide an initial measurement of
$f\sigma_8$ at $z=0.8$ \citep{delatorre13a}. The final dataset
increases the survey volume by a factor of $1.6$, and by further
adding galaxy-galaxy lensing information, we have been able to provide
new accurate measurements of $f\sigma_8$ at both $z=0.6$ and
$z=0.86$. We have found values of $f\sigma_8(z=0.6)=0.48\pm0.12$ and
$f\sigma_8(z=0.86)=0.48\pm0.10$, which are consistent with previous
measurements at lower or similar redshifts.

The additional galaxy-galaxy lensing constraint and the specific
treatment of $\sigma_8$ to describe the non-linearity level of the
real-space power spectra entering the model alleviate the degeneracy
between the galaxy bias parameter, $\sigma_8$, and $f$, and has
allowed direct measurements of these two parameters. We have obtained
values of $\left(f,\sigma_8\right)=(0.93\pm0.22,0.52\pm0.06)$ and
$\left(f,\sigma_8\right)=(0.99\pm0.19,0.48\pm0.04)$ at $z=0.6$ and
$z=0.86$, respectively. These measurements put new constraints on
gravity at the epoch when the Universe was almost half its present
age. Our measurements are statistically consistent with a Universe
where the gravitational interactions between structures on
cosmological scales can be described by General Relativity, although
they are not yet accurate enough to rule out some commonly considered
alternatives to General Relativity.

In addition to measuring the growth rate of structure, we have been
able to measure the gravitational slip parameter, $E_G$, for the first
time at $z>0.6$. This quantity, which can be directly constructed from
galaxy clustering and galaxy-galaxy lensing observables, is sensitive
to the growth rate of structure and mean matter density in the
Universe. We have obtained averaged values of the gravitational slip
parameter of $\smash{\overline{E}_G}(z=0.6)=0.16\pm0.09$ and
$\smash{\overline{E}_G}(z=0.86)=0.09\pm0.07$. Our $E_G$ measurements
are consistent within $1-2\sigma$, although they exhibit slightly
lower values than expected in the standard model for gravity in a
$\Lambda \rm{CDM}$ background.

Overall, this analysis has demonstrated the importance of the
combination of galaxy clustering in redshift space and galaxy-galaxy
lensing in order to probe the origin of cosmic acceleration. This
combination can alleviate the inherent uncertainty related to galaxy
bias in RSD analyses and provide new insights into the gravitational
physics at work on cosmological scales. This analysis and adopted
methodology can be seen as a proof-of-concept in the context of the
preparation of next-generation cosmological surveys such as Euclid
\citep{laureijs11}, which will allow galaxy clustering and
galaxy-galaxy lensing to be combined with exquisite precision.

\begin{acknowledgements}
  
We acknowledge the crucial contribution of the ESO staff for the
management of service observations. In particular, we are deeply
grateful to M. Hilker for his constant help and support of this
program. Italian participation to VIPERS has been funded by INAF
through PRIN 2008, 2010, 2014 and 2015 programs. LG and BRG
acknowledge support from the European Research Council through grant
n.~291521. OLF acknowledges support from the European Research Council
through grant n.~268107. JAP acknowledges support of the European
Research Council through the COSFORM ERC Advanced Research Grant (\#
670193). GDL acknowledges financial support from the European Research
Council through grant n.~202781. RT acknowledges financial support
from the European Research Council through grant n.~202686. AP, KM,
and JK have been supported by the National Science Centre (grants
UMO-2012/07/B/ST9/04425 and UMO-2013/09/D/ST9/04030). EB, FM and LM
acknowledge the support from grants ASI-INAF I/023/12/0 and PRIN MIUR
2010-2011. LM also acknowledges financial support from PRIN INAF
2012. SDLT, EJ, and MP acknowledge the support of the OCEVU Labex
(ANR-11-LABX-0060) and the A*MIDEX project (ANR-11-IDEX-0001-02)
funded by the "Investissements d'Avenir" French government program
managed by the ANR. Research conducted within the scope of the HECOLS
International Associated Laboratory, supported in part by the Polish
NCN grant DEC-2013/08/M/ST9/00664. TM and SA acknowledge financial
support from the ANR Spin(e) through the French grant
ANR-13-BS05-0005.

The authors gratefully acknowledge the Gauss Centre for Supercomputing
e.V. (www.gauss-centre.eu) and the Partnership for Advanced
Supercomputing in Europe (PRACE, www.prace-ri.eu) for funding the
MultiDark simulation project by providing computing time on the GCS
Supercomputer SuperMUC at Leibniz Supercomputing Centre (LRZ,
www.lrz.de).

\end{acknowledgements}

\bibliographystyle{aa}
\bibliography{biblio,biblio_VIPERS}

\begin{thebibliography}{97}
\expandafter\ifx\csname natexlab\endcsname\relax\def\natexlab#1{#1}\fi

\bibitem[{{Alam} {et~al.}(2017){Alam}, {Ata}, {Bailey}, {Beutler}, {Bizyaev},
  {Blazek}, {Bolton}, {Brownstein}, {Burden}, {Chuang}, {Comparat}, {Cuesta},
  {Dawson}, {Eisenstein}, {Escoffier}, {Gil-Mar{\'{\i}}n}, {Grieb}, {Hand},
  {Ho}, {Kinemuchi}, {Kirkby}, {Kitaura}, {Malanushenko}, {Malanushenko},
  {Maraston}, {McBride}, {Nichol}, {Olmstead}, {Oravetz}, {Padmanabhan},
  {Palanque-Delabrouille}, {Pan}, {Pellejero-Ibanez}, {Percival}, {Petitjean},
  {Prada}, {Price-Whelan}, {Reid}, {Rodr{\'{\i}}guez-Torres}, {Roe}, {Ross},
  {Ross}, {Rossi}, {Rubi{\~n}o-Mart{\'{\i}}n}, {Saito}, {Salazar-Albornoz},
  {Samushia}, {S{\'a}nchez}, {Satpathy}, {Schlegel}, {Schneider},
  {Sc{\'o}ccola}, {Seo}, {Sheldon}, {Simmons}, {Slosar}, {Strauss}, {Swanson},
  {Thomas}, {Tinker}, {Tojeiro}, {Maga{\~n}a}, {Vazquez}, {Verde}, {Wake},
  {Wang}, {Weinberg}, {White}, {Wood-Vasey}, {Y{\`e}che}, {Zehavi}, {Zhai}, \&
  {Zhao}}]{alam16}
{Alam}, S., {Ata}, M., {Bailey}, S., {et~al.} 2017, \mnras, 470, 2617

\bibitem[{{Alcock} \& {Paczynski}(1979)}]{alcock79}
{Alcock}, C. \& {Paczynski}, B. 1979, \nat, 281, 358

\bibitem[{{Anderson} {et~al.}(2014){Anderson}, {Aubourg}, {Bailey}, {Beutler},
  {Bhardwaj}, {Blanton}, {Bolton}, {Brinkmann}, {Brownstein}, {Burden},
  {Chuang}, {Cuesta}, {Dawson}, {Eisenstein}, {Escoffier}, {Gunn}, {Guo}, {Ho},
  {Honscheid}, {Howlett}, {Kirkby}, {Lupton}, {Manera}, {Maraston}, {McBride},
  {Mena}, {Montesano}, {Nichol}, {Nuza}, {Olmstead}, {Padmanabhan},
  {Palanque-Delabrouille}, {Parejko}, {Percival}, {Petitjean}, {Prada},
  {Price-Whelan}, {Reid}, {Roe}, {Ross}, {Ross}, {Sabiu}, {Saito}, {Samushia},
  {S{\'a}nchez}, {Schlegel}, {Schneider}, {Scoccola}, {Seo}, {Skibba},
  {Strauss}, {Swanson}, {Thomas}, {Tinker}, {Tojeiro}, {Maga{\~n}a}, {Verde},
  {Wake}, {Weaver}, {Weinberg}, {White}, {Xu}, {Y{\`e}che}, {Zehavi}, \&
  {Zhao}}]{anderson14}
{Anderson}, L., {Aubourg}, {\'E}., {Bailey}, S., {et~al.} 2014, \mnras, 441, 24

\bibitem[{{Baldauf} {et~al.}(2010){Baldauf}, {Smith}, {Seljak}, \&
  {Mandelbaum}}]{baldauf10}
{Baldauf}, T., {Smith}, R.~E., {Seljak}, U., \& {Mandelbaum}, R. 2010, \prd,
  81, 063531

\bibitem[{{Ballinger} {et~al.}(1996){Ballinger}, {Peacock}, \&
  {Heavens}}]{ballinger96}
{Ballinger}, W.~E., {Peacock}, J.~A., \& {Heavens}, A.~F. 1996, \mnras, 282,
  877

\bibitem[{{Bel} {et~al.}(2017)}]{bel17}
{Bel}, J. {et~al.} 2017, \rm{in preparation}

\bibitem[{{Benjamin} {et~al.}(2013){Benjamin}, {Van Waerbeke}, {Heymans},
  {Kilbinger}, {Erben}, {Hildebrandt}, {Hoekstra}, {Kitching}, {Mellier},
  {Miller}, {Rowe}, {Schrabback}, {Simpson}, {Coupon}, {Fu},
  {Harnois-D{\'e}raps}, {Hudson}, {Kuijken}, {Semboloni}, {Vafaei}, \&
  {Velander}}]{benjamin13}
{Benjamin}, J., {Van Waerbeke}, L., {Heymans}, C., {et~al.} 2013, \mnras, 431,
  1547

\bibitem[{{Bernardeau} {et~al.}(2002){Bernardeau}, {Colombi}, {Gazta{\~n}aga},
  \& {Scoccimarro}}]{bernardeau02}
{Bernardeau}, F., {Colombi}, S., {Gazta{\~n}aga}, E., \& {Scoccimarro}, R.
  2002, \physrep, 367, 1

\bibitem[{{Beutler} {et~al.}(2011){Beutler}, {Blake}, {Colless}, {Jones},
  {Staveley-Smith}, {Campbell}, {Parker}, {Saunders}, \& {Watson}}]{beutler11}
{Beutler}, F., {Blake}, C., {Colless}, M., {et~al.} 2011, \mnras, 416, 3017

\bibitem[{{Beutler} {et~al.}(2012){Beutler}, {Blake}, {Colless}, {Jones},
  {Staveley-Smith}, {Poole}, {Campbell}, {Parker}, {Saunders}, \&
  {Watson}}]{beutler12}
{Beutler}, F., {Blake}, C., {Colless}, M., {et~al.} 2012, \mnras, 423, 3430

\bibitem[{{Beutler} {et~al.}(2014){Beutler}, {Saito}, {Seo}, {Brinkmann},
  {Dawson}, {Eisenstein}, {Font-Ribera}, {Ho}, {McBride}, {Montesano},
  {Percival}, {Ross}, {Ross}, {Samushia}, {Schlegel}, {S{\'a}nchez}, {Tinker},
  \& {Weaver}}]{beutler14}
{Beutler}, F., {Saito}, S., {Seo}, H.-J., {et~al.} 2014, \mnras, 443, 1065

\bibitem[{{Blake} {et~al.}(2012){Blake}, {Brough}, {Colless}, {Contreras},
  {Couch}, {Croom}, {Croton}, {Davis}, {Drinkwater}, {Forster}, {Gilbank},
  {Gladders}, {Glazebrook}, {Jelliffe}, {Jurek}, {Li}, {Madore}, {Martin},
  {Pimbblet}, {Poole}, {Pracy}, {Sharp}, {Wisnioski}, {Woods}, {Wyder}, \&
  {Yee}}]{blake12}
{Blake}, C., {Brough}, S., {Colless}, M., {et~al.} 2012, \mnras, 425, 405

\bibitem[{{Blake} {et~al.}(2016){Blake}, {Joudaki}, {Heymans}, {Choi}, {Erben},
  {Harnois-Deraps}, {Hildebrandt}, {Joachimi}, {Nakajima}, {van Waerbeke}, \&
  {Viola}}]{blake16}
{Blake}, C., {Joudaki}, S., {Heymans}, C., {et~al.} 2016, \mnras, 456, 2806

\bibitem[{{Cabr{\'e}} \& {Gazta{\~n}aga}(2009)}]{cabre09}
{Cabr{\'e}}, A. \& {Gazta{\~n}aga}, E. 2009, \mnras, 393, 1183

\bibitem[{{Cacciato} {et~al.}(2013){Cacciato}, {van den Bosch}, {More}, {Mo},
  \& {Yang}}]{cacciato13}
{Cacciato}, M., {van den Bosch}, F.~C., {More}, S., {Mo}, H., \& {Yang}, X.
  2013, \mnras, 430, 767

\bibitem[{{Cappi} {et~al.}(2015){Cappi}, {Marulli}, {Bel}, {Cucciati},
  {Branchini}, {de la Torre}, {Moscardini}, {Bolzonella}, {Guzzo}, {Abbas},
  {Adami}, {Arnouts}, {Bottini}, {Coupon}, {Davidzon}, {De Lucia}, {Fritz},
  {Franzetti}, {Fumana}, {Garilli}, {Granett}, {Ilbert}, {Iovino}, {Krywult},
  {Le Brun}, {Le F{\`e}vre}, {Maccagni}, {Ma{\l}ek}, {McCracken}, {Paioro},
  {Polletta}, {Pollo}, {Scodeggio}, {Tasca}, {Tojeiro}, {Vergani},
  {Zanichelli}, {Burden}, {Di Porto}, {Marchetti}, {Marinoni}, {Mellier},
  {Nichol}, {Peacock}, {Percival}, {Phleps}, {Schimd}, {Schlagenhaufer},
  {Wolk}, \& {Zamorani}}]{cappi15}
{Cappi}, A., {Marulli}, F., {Bel}, J., {et~al.} 2015, \aap, 579, A70

\bibitem[{{Chan} {et~al.}(2012){Chan}, {Scoccimarro}, \& {Sheth}}]{chan12}
{Chan}, K.~C., {Scoccimarro}, R., \& {Sheth}, R.~K. 2012, \prd, 85, 083509

\bibitem[{{Chuang} {et~al.}(2016){Chuang}, {Prada}, {Pellejero-Ibanez},
  {Beutler}, {Cuesta}, {Eisenstein}, {Escoffier}, {Ho}, {Kitaura}, {Kneib},
  {Manera}, {Nuza}, {Rodr{\'{\i}}guez-Torres}, {Ross},
  {Rubi{\~n}o-Mart{\'{\i}}n}, {Samushia}, {Schlegel}, {Schneider}, {Wang},
  {Weaver}, {Zhao}, {Brownstein}, {Dawson}, {Maraston}, {Olmstead}, \&
  {Thomas}}]{chuang16}
{Chuang}, C.-H., {Prada}, F., {Pellejero-Ibanez}, M., {et~al.} 2016, \mnras,
  461, 3781

\bibitem[{{Cole}(2011)}]{cole11}
{Cole}, S. 2011, \mnras, 416, 739

\bibitem[{{Cole} {et~al.}(2005){Cole}, {Percival}, {Peacock}, {Norberg},
  {Baugh}, {Frenk}, {Baldry}, {Bland-Hawthorn}, {Bridges}, {Cannon}, {Colless},
  {Collins}, {Couch}, {Cross}, {Dalton}, {Eke}, {De Propris}, {Driver},
  {Efstathiou}, {Ellis}, {Glazebrook}, {Jackson}, {Jenkins}, {Lahav}, {Lewis},
  {Lumsden}, {Maddox}, {Madgwick}, {Peterson}, {Sutherland}, \&
  {Taylor}}]{cole05}
{Cole}, S., {Percival}, W.~J., {Peacock}, J.~A., {et~al.} 2005, \mnras, 362,
  505

\bibitem[{{Colless} {et~al.}(2001){Colless}, {Dalton}, {Maddox}, {Sutherland},
  {Norberg}, {Cole}, {Bland-Hawthorn}, {Bridges}, {Cannon}, {Collins}, {Couch},
  {Cross}, {Deeley}, {De Propris}, {Driver}, {Efstathiou}, {Ellis}, {Frenk},
  {Glazebrook}, {Jackson}, {Lahav}, {Lewis}, {Lumsden}, {Madgwick}, {Peacock},
  {Peterson}, {Price}, {Seaborne}, \& {Taylor}}]{colless01}
{Colless}, M., {Dalton}, G., {Maddox}, S., {et~al.} 2001, \mnras, 328, 1039

\bibitem[{{Colless} {et~al.}(2003){Colless}, {Peterson}, {Jackson}, {Peacock},
  {Cole}, {Norberg}, {Baldry}, {Baugh}, {Bland-Hawthorn}, {Bridges}, {Cannon},
  {Collins}, {Couch}, {Cross}, {Dalton}, {De Propris}, {Driver}, {Efstathiou},
  {Ellis}, {Frenk}, {Glazebrook}, {Lahav}, {Lewis}, {Lumsden}, {Maddox},
  {Madgwick}, {Sutherland}, \& {Taylor}}]{colless03}
{Colless}, M., {Peterson}, B.~A., {Jackson}, C., {et~al.} 2003, ArXiv e-print:
  astro-ph/0306581

\bibitem[{{Coupon} {et~al.}(2015){Coupon}, {Arnouts}, {van Waerbeke},
  {Moutard}, {Ilbert}, {van Uitert}, {Erben}, {Garilli}, {Guzzo}, {Heymans},
  {Hildebrandt}, {Hoekstra}, {Kilbinger}, {Kitching}, {Mellier}, {Miller},
  {Scodeggio}, {Bonnett}, {Branchini}, {Davidzon}, {De Lucia}, {Fritz}, {Fu},
  {Hudelot}, {Hudson}, {Kuijken}, {Leauthaud}, {Le F{\`e}vre}, {McCracken},
  {Moscardini}, {Rowe}, {Schrabback}, {Semboloni}, \& {Velander}}]{coupon15}
{Coupon}, J., {Arnouts}, S., {van Waerbeke}, L., {et~al.} 2015, \mnras, 449,
  1352

\bibitem[{{de la Torre} \& {Guzzo}(2012)}]{delatorre12}
{de la Torre}, S. \& {Guzzo}, L. 2012, \mnras, 427, 327

\bibitem[{{de la Torre} {et~al.}(2013){de la Torre}, {Guzzo}, {Peacock},
  {Branchini}, {Iovino}, {Granett}, {Abbas}, {Adami}, {Arnouts}, {Bel},
  {Bolzonella}, {Bottini}, {Cappi}, {Coupon}, {Cucciati}, {Davidzon}, {De
  Lucia}, {Fritz}, {Franzetti}, {Fumana}, {Garilli}, {Ilbert}, {Krywult}, {Le
  Brun}, {Le F{\`e}vre}, {Maccagni}, {Ma{\l}ek}, {Marulli}, {McCracken},
  {Moscardini}, {Paioro}, {Percival}, {Polletta}, {Pollo}, {Schlagenhaufer},
  {Scodeggio}, {Tasca}, {Tojeiro}, {Vergani}, {Zanichelli}, {Burden}, {Di
  Porto}, {Marchetti}, {Marinoni}, {Mellier}, {Monaco}, {Nichol}, {Phleps},
  {Wolk}, \& {Zamorani}}]{delatorre13a}
{de la Torre}, S., {Guzzo}, L., {Peacock}, J.~A., {et~al.} 2013, \aap, 557, A54

\bibitem[{{de la Torre} \& {Peacock}(2013)}]{delatorre13b}
{de la Torre}, S. \& {Peacock}, J.~A. 2013, \mnras, 435, 743

\bibitem[{{Di Porto} {et~al.}(2016){Di Porto}, {Branchini}, {Bel}, {Marulli},
  {Bolzonella}, {Cucciati}, {de la Torre}, {Granett}, {Guzzo}, {Marinoni},
  {Moscardini}, {Abbas}, {Adami}, {Arnouts}, {Bottini}, {Cappi}, {Coupon},
  {Davidzon}, {De Lucia}, {Fritz}, {Franzetti}, {Fumana}, {Garilli}, {Ilbert},
  {Iovino}, {Krywult}, {Le Brun}, {Le F{\`e}vre}, {Maccagni}, {Ma{\l}ek},
  {McCracken}, {Paioro}, {Polletta}, {Pollo}, {Scodeggio}, {Tasca}, {Tojeiro},
  {Vergani}, {Zanichelli}, {Burden}, {Marchetti}, {Martizzi}, {Mellier},
  {Nichol}, {Peacock}, {Percival}, {Viel}, {Wolk}, \& {Zamorani}}]{diporto14}
{Di Porto}, C., {Branchini}, E., {Bel}, J., {et~al.} 2016, \aap, 594, A62

\bibitem[{{Dodelson} \& {Schneider}(2013)}]{dodelson13}
{Dodelson}, S. \& {Schneider}, M.~D. 2013, \prd, 88, 063537

\bibitem[{{Eisenstein} {et~al.}(2005){Eisenstein}, {Zehavi}, {Hogg},
  {Scoccimarro}, {Blanton}, {Nichol}, {Scranton}, {Seo}, {Tegmark}, {Zheng},
  {Anderson}, {Annis}, {Bahcall}, {Brinkmann}, {Burles}, {Castander},
  {Connolly}, {Csabai}, {Doi}, {Fukugita}, {Frieman}, {Glazebrook}, {Gunn},
  {Hendry}, {Hennessy}, {Ivezi{\'c}}, {Kent}, {Knapp}, {Lin}, {Loh}, {Lupton},
  {Margon}, {McKay}, {Meiksin}, {Munn}, {Pope}, {Richmond}, {Schlegel},
  {Schneider}, {Shimasaku}, {Stoughton}, {Strauss}, {SubbaRao}, {Szalay},
  {Szapudi}, {Tucker}, {Yanny}, \& {York}}]{eisenstein05}
{Eisenstein}, D.~J., {Zehavi}, I., {Hogg}, D.~W., {et~al.} 2005, \apj, 633, 560

\bibitem[{{Erben} {et~al.}(2013){Erben}, {Hildebrandt}, {Miller}, {van
  Waerbeke}, {Heymans}, {Hoekstra}, {Kitching}, {Mellier}, {Benjamin}, {Blake},
  {Bonnett}, {Cordes}, {Coupon}, {Fu}, {Gavazzi}, {Gillis}, {Grocutt}, {Gwyn},
  {Holhjem}, {Hudson}, {Kilbinger}, {Kuijken}, {Milkeraitis}, {Rowe},
  {Schrabback}, {Semboloni}, {Simon}, {Smit}, {Toader}, {Vafaei}, {van Uitert},
  \& {Velander}}]{erben13}
{Erben}, T., {Hildebrandt}, H., {Miller}, L., {et~al.} 2013, \mnras, 433, 2545

\bibitem[{{Fisher} {et~al.}(1994){Fisher}, {Davis}, {Strauss}, {Yahil}, \&
  {Huchra}}]{fisher94}
{Fisher}, K.~B., {Davis}, M., {Strauss}, M.~A., {Yahil}, A., \& {Huchra}, J.~P.
  1994, \mnras, 267, 927

\bibitem[{{Garilli} {et~al.}(2014){Garilli}, {Guzzo}, {Scodeggio},
  {Bolzonella}, {Abbas}, {Adami}, {Arnouts}, {Bel}, {Bottini}, {Branchini},
  {Cappi}, {Coupon}, {Cucciati}, {Davidzon}, {De Lucia}, {de la Torre},
  {Franzetti}, {Fritz}, {Fumana}, {Granett}, {Ilbert}, {Iovino}, {Krywult}, {Le
  Brun}, {Le F{\`e}vre}, {Maccagni}, {Ma{\l}ek}, {Marulli}, {McCracken},
  {Paioro}, {Polletta}, {Pollo}, {Schlagenhaufer}, {Tasca}, {Tojeiro},
  {Vergani}, {Zamorani}, {Zanichelli}, {Burden}, {Di Porto}, {Marchetti},
  {Marinoni}, {Mellier}, {Moscardini}, {Nichol}, {Peacock}, {Percival},
  {Phleps}, \& {Wolk}}]{garilli14}
{Garilli}, B., {Guzzo}, L., {Scodeggio}, M., {et~al.} 2014, \aap, 562, A23

\bibitem[{{Gil-Mar{\'{\i}}n} {et~al.}(2016){Gil-Mar{\'{\i}}n}, {Percival},
  {Brownstein}, {Chuang}, {Grieb}, {Ho}, {Kitaura}, {Maraston}, {Prada},
  {Rodr{\'{\i}}guez-Torres}, {Ross}, {Samushia}, {Schlegel}, {Thomas},
  {Tinker}, \& {Zhao}}]{gilmarin16}
{Gil-Mar{\'{\i}}n}, H., {Percival}, W.~J., {Brownstein}, J.~R., {et~al.} 2016,
  \mnras, 460, 4188

\bibitem[{{Gil-Mar{\'{\i}}n} {et~al.}(2017){Gil-Mar{\'{\i}}n}, {Percival},
  {Verde}, {Brownstein}, {Chuang}, {Kitaura}, {Rodr{\'{\i}}guez-Torres}, \&
  {Olmstead}}]{gilmarin17}
{Gil-Mar{\'{\i}}n}, H., {Percival}, W.~J., {Verde}, L., {et~al.} 2017, \mnras,
  465, 1757

\bibitem[{{Gil-Mar{\'{\i}}n} {et~al.}(2014){Gil-Mar{\'{\i}}n}, {Wagner},
  {Nore{\~n}a}, {Verde}, \& {Percival}}]{gilmarin14}
{Gil-Mar{\'{\i}}n}, H., {Wagner}, C., {Nore{\~n}a}, J., {Verde}, L., \&
  {Percival}, W. 2014, \jcap, 12, 029

\bibitem[{{Giocoli} {et~al.}(2016){Giocoli}, {Jullo}, {Metcalf}, {de la Torre},
  {Yepes}, {Prada}, {Comparat}, {G{\"o}ttlober}, {Kyplin}, {Kneib}, {Petkova},
  {Shan}, \& {Tessore}}]{giocoli16}
{Giocoli}, C., {Jullo}, E., {Metcalf}, R.~B., {et~al.} 2016, \mnras

\bibitem[{{Goranova} {et~al.}(2009){Goranova}, {Hudelot}, {Magnard},
  {McCracken}, {Mellier}, {Monnerville}, {Schulthe is}, {S\'emah},
  {Cuillandre}, \& H.}]{goranova09}
{Goranova}, Y., {Hudelot}, P., {Magnard}, F., {et~al.} 2009, {The CFHTLS T0006
  Release}, {http://terapix.iap.fr/cplt/table\_syn\_T0006.html}

\bibitem[{{Goroff} {et~al.}(1986){Goroff}, {Grinstein}, {Rey}, \&
  {Wise}}]{goroff86}
{Goroff}, M.~H., {Grinstein}, B., {Rey}, S.-J., \& {Wise}, M.~B. 1986, \apj,
  311, 6

\bibitem[{{Granett} {et~al.}(2015){Granett}, {Branchini}, {Guzzo}, {Abbas},
  {Adami}, {Arnouts}, {Bel}, {Bolzonella}, {Bottini}, {Cappi}, {Coupon},
  {Cucciati}, {Davidzon}, {De Lucia}, {de la Torre}, {Fritz}, {Franzetti},
  {Fumana}, {Garilli}, {Ilbert}, {Iovino}, {Krywult}, {Le Brun}, {Le
  F{\`e}vre}, {Maccagni}, {Ma{\l}ek}, {Marulli}, {McCracken}, {Polletta},
  {Pollo}, {Scodeggio}, {Tasca}, {Tojeiro}, {Vergani}, {Zanichelli}, {Burden},
  {Di Porto}, {Marchetti}, {Marinoni}, {Mellier}, {Moutard}, {Moscardini},
  {Nichol}, {Peacock}, {Percival}, \& {Zamorani}}]{granett15}
{Granett}, B.~R., {Branchini}, E., {Guzzo}, L., {et~al.} 2015, \aap, 583, A61

\bibitem[{{Guzzo} {et~al.}(2008){Guzzo}, {Pierleoni}, {Meneux}, {Branchini},
  {Le F{\`e}vre}, {Marinoni}, {Garilli}, {Blaizot}, {De Lucia}, {Pollo},
  {McCracken}, {Bottini}, {Le Brun}, {Maccagni}, {Picat}, {Scaramella},
  {Scodeggio}, {Tresse}, {Vettolani}, {Zanichelli}, {Adami}, {Arnouts},
  {Bardelli}, {Bolzonella}, {Bongiorno}, {Cappi}, {Charlot}, {Ciliegi},
  {Contini}, {Cucciati}, {de la Torre}, {Dolag}, {Foucaud}, {Franzetti},
  {Gavignaud}, {Ilbert}, {Iovino}, {Lamareille}, {Marano}, {Mazure}, {Memeo},
  {Merighi}, {Moscardini}, {Paltani}, {Pell{\`o}}, {Perez-Montero}, {Pozzetti},
  {Radovich}, {Vergani}, {Zamorani}, \& {Zucca}}]{guzzo08}
{Guzzo}, L., {Pierleoni}, M., {Meneux}, B., {et~al.} 2008, \nat, 451, 541

\bibitem[{{Guzzo} {et~al.}(2014){Guzzo}, {Scodeggio}, {Garilli}, {Granett},
  {Fritz}, {Abbas}, {Adami}, {Arnouts}, {Bel}, {Bolzonella}, {Bottini},
  {Branchini}, {Cappi}, {Coupon}, {Cucciati}, {Davidzon}, {De Lucia}, {de la
  Torre}, {Franzetti}, {Fumana}, {Hudelot}, {Ilbert}, {Iovino}, {Krywult}, {Le
  Brun}, {Le F{\`e}vre}, {Maccagni}, {Ma{\l}ek}, {Marulli}, {McCracken},
  {Paioro}, {Peacock}, {Polletta}, {Pollo}, {Schlagenhaufer}, {Tasca},
  {Tojeiro}, {Vergani}, {Zamorani}, {Zanichelli}, {Burden}, {Di Porto},
  {Marchetti}, {Marinoni}, {Mellier}, {Moscardini}, {Nichol}, {Percival},
  {Phleps}, \& {Wolk}}]{guzzo14}
{Guzzo}, L., {Scodeggio}, M., {Garilli}, B., {et~al.} 2014, \aap, 566, A108

\bibitem[{{Hamilton}(1993)}]{hamilton93}
{Hamilton}, A.~J.~S. 1993, \apj, 417, 19

\bibitem[{{Hartlap} {et~al.}(2007){Hartlap}, {Simon}, \&
  {Schneider}}]{hartlap07}
{Hartlap}, J., {Simon}, P., \& {Schneider}, P. 2007, \aap, 464, 399

\bibitem[{{Hawken} {et~al.}(2017){Hawken}, {Granett}, {Iovino}, {Guzzo},
  {Peacock}, {de la Torre}, {Garilli}, {Bolzonella}, {Scodeggio}, {Abbas},
  {Adami}, {Bottini}, {Cappi}, {Cucciati}, {Davidzon}, {Fritz}, {Franzetti},
  {Krywult}, {Le Brun}, {Le Fevre}, {Maccagni}, {Ma{\l}ek}, {Marulli},
  {Polletta}, {Pollo}, {Tasca}, {Tojeiro}, {Vergani}, {Zanichelli}, {Arnouts},
  {Bel}, {Branchini}, {De Lucia}, {Ilbert}, {Moscardini}, \&
  {Percival}}]{hawken16}
{Hawken}, A.~J., {Granett}, B.~R., {Iovino}, A., {et~al.} 2017, \aap, in press,
  ArXiv e-print: 1611.07046

\bibitem[{{Heymans} {et~al.}(2013){Heymans}, {Grocutt}, {Heavens}, {Kilbinger},
  {Kitching}, {Simpson}, {Benjamin}, {Erben}, {Hildebrandt}, {Hoekstra},
  {Mellier}, {Miller}, {Van Waerbeke}, {Brown}, {Coupon}, {Fu},
  {Harnois-D{\'e}raps}, {Hudson}, {Kuijken}, {Rowe}, {Schrabback}, {Semboloni},
  {Vafaei}, \& {Velander}}]{heymans13}
{Heymans}, C., {Grocutt}, E., {Heavens}, A., {et~al.} 2013, \mnras, 432, 2433

\bibitem[{{Heymans} {et~al.}(2012){Heymans}, {Van Waerbeke}, {Miller}, {Erben},
  {Hildebrandt}, {Hoekstra}, {Kitching}, {Mellier}, {Simon}, {Bonnett},
  {Coupon}, {Fu}, {Harnois D{\'e}raps}, {Hudson}, {Kilbinger}, {Kuijken},
  {Rowe}, {Schrabback}, {Semboloni}, {van Uitert}, {Vafaei}, \&
  {Velander}}]{heymans12}
{Heymans}, C., {Van Waerbeke}, L., {Miller}, L., {et~al.} 2012, \mnras, 427,
  146

\bibitem[{{Hildebrandt} {et~al.}(2012){Hildebrandt}, {Erben}, {Kuijken}, {van
  Waerbeke}, {Heymans}, {Coupon}, {Benjamin}, {Bonnett}, {Fu}, {Hoekstra},
  {Kitching}, {Mellier}, {Miller}, {Velander}, {Hudson}, {Rowe}, {Schrabback},
  {Semboloni}, \& {Ben{\'{\i}}tez}}]{hildebrandt12}
{Hildebrandt}, H., {Erben}, T., {Kuijken}, K., {et~al.} 2012, \mnras, 421, 2355

\bibitem[{{Hildebrandt} {et~al.}(2017){Hildebrandt}, {Viola}, {Heymans},
  {Joudaki}, {Kuijken}, {Blake}, {Erben}, {Joachimi}, {Klaes}, {Miller},
  {Morrison}, {Nakajima}, {Verdoes Kleijn}, {Amon}, {Choi}, {Covone}, {de
  Jong}, {Dvornik}, {Fenech Conti}, {Grado}, {Harnois-D{\'e}raps}, {Herbonnet},
  {Hoekstra}, {K{\"o}hlinger}, {McFarland}, {Mead}, {Merten}, {Napolitano},
  {Peacock}, {Radovich}, {Schneider}, {Simon}, {Valentijn}, {van den Busch},
  {van Uitert}, \& {Van Waerbeke}}]{hildebrandt16}
{Hildebrandt}, H., {Viola}, M., {Heymans}, C., {et~al.} 2017, \mnras, 465, 1454

\bibitem[{{Howlett} {et~al.}(2015){Howlett}, {Ross}, {Samushia}, {Percival}, \&
  {Manera}}]{howlett15}
{Howlett}, C., {Ross}, A.~J., {Samushia}, L., {Percival}, W.~J., \& {Manera},
  M. 2015, \mnras, 449, 848

\bibitem[{{Kaiser}(1987)}]{kaiser87}
{Kaiser}, N. 1987, \mnras, 227, 1

\bibitem[{Kaufman {et~al.}(2008)Kaufman, Schervish, \& Nychka}]{kaufman08}
Kaufman, C.~G., Schervish, M.~J., \& Nychka, D.~W. 2008, Journal of the
  American Statistical Association, 103, 1545

\bibitem[{{Klypin} {et~al.}(2016){Klypin}, {Yepes}, {Gottl{\"o}ber}, {Prada},
  \& {He{\ss}}}]{klypin16}
{Klypin}, A., {Yepes}, G., {Gottl{\"o}ber}, S., {Prada}, F., \& {He{\ss}}, S.
  2016, \mnras, 457, 4340

\bibitem[{{Landy} \& {Szalay}(1993)}]{landy93}
{Landy}, S.~D. \& {Szalay}, A.~S. 1993, \apj, 412, 64

\bibitem[{{Laureijs} {et~al.}(2011){Laureijs}, {Amiaux}, {Arduini},
  {Augu{\`e}res}, {Brinchmann}, {Cole}, {Cropper}, {Dabin}, {Duvet}, {Ealet},
  \& et~al.}]{laureijs11}
{Laureijs}, R., {Amiaux}, J., {Arduini}, S., {et~al.} 2011, ArXiv e-print:
  1110.3193

\bibitem[{{Le F{\`e}vre} {et~al.}(2003){Le F{\`e}vre}, {Saisse}, {Mancini},
  {Brau-Nogue}, {Caputi}, {Castinel}, {D'Odorico}, {Garilli}, {Kissler-Patig},
  {Lucuix}, {Mancini}, {Pauget}, {Sciarretta}, {Scodeggio}, {Tresse}, \&
  {Vettolani}}]{lefevre03}
{Le F{\`e}vre}, O., {Saisse}, M., {Mancini}, D., {et~al.} 2003, in \procspie,
  ed. M.~{Iye} \& A.~F.~M. {Moorwood}, Vol. 4841, 1670--1681

\bibitem[{{Lesgourgues}(2011)}]{lesgourgues11}
{Lesgourgues}, J. 2011, ArXiv e-print: 1104.2932

\bibitem[{{Mandelbaum} {et~al.}(2013){Mandelbaum}, {Slosar}, {Baldauf},
  {Seljak}, {Hirata}, {Nakajima}, {Reyes}, \& {Smith}}]{mandelbaum13}
{Mandelbaum}, R., {Slosar}, A., {Baldauf}, T., {et~al.} 2013, \mnras, 432, 1544

\bibitem[{{Marulli} {et~al.}(2013){Marulli}, {Bolzonella}, {Branchini},
  {Davidzon}, {de la Torre}, {Granett}, {Guzzo}, {Iovino}, {Moscardini},
  {Pollo}, {Abbas}, {Adami}, {Arnouts}, {Bel}, {Bottini}, {Cappi}, {Coupon},
  {Cucciati}, {De Lucia}, {Fritz}, {Franzetti}, {Fumana}, {Garilli}, {Ilbert},
  {Krywult}, {Le Brun}, {Le F{\`e}vre}, {Maccagni}, {Ma{\l}ek}, {McCracken},
  {Paioro}, {Polletta}, {Schlagenhaufer}, {Scodeggio}, {Tasca}, {Tojeiro},
  {Vergani}, {Zanichelli}, {Burden}, {Di Porto}, {Marchetti}, {Marinoni},
  {Mellier}, {Nichol}, {Peacock}, {Percival}, {Phleps}, {Wolk}, \&
  {Zamorani}}]{marulli13}
{Marulli}, F., {Bolzonella}, M., {Branchini}, E., {et~al.} 2013, \aap, 557, A17

\bibitem[{{Matsubara} \& {Suto}(1996)}]{matsubara96}
{Matsubara}, T. \& {Suto}, Y. 1996, \apjl, 470, L1

\bibitem[{{McDonald} \& {Roy}(2009)}]{mcdonald09}
{McDonald}, P. \& {Roy}, A. 2009, \jcap, 8, 020

\bibitem[{{Metropolis} {et~al.}(1953){Metropolis}, {Rosenbluth}, {Rosenbluth},
  {Teller}, \& {Teller}}]{metropolis53}
{Metropolis}, N., {Rosenbluth}, A.~W., {Rosenbluth}, M.~N., {Teller}, A.~H., \&
  {Teller}, E. 1953, \jcp, 21, 1087

\bibitem[{{Micheletti} {et~al.}(2014){Micheletti}, {Iovino}, {Hawken},
  {Granett}, {Bolzonella}, {Cappi}, {Guzzo}, {Abbas}, {Adami}, {Arnouts},
  {Bel}, {Bottini}, {Branchini}, {Coupon}, {Cucciati}, {Davidzon}, {De Lucia},
  {de la Torre}, {Fritz}, {Franzetti}, {Fumana}, {Garilli}, {Ilbert},
  {Krywult}, {Le Brun}, {Le F{\`e}vre}, {Maccagni}, {Ma{\l}ek}, {Marulli},
  {McCracken}, {Polletta}, {Pollo}, {Schimd}, {Scodeggio}, {Tasca}, {Tojeiro},
  {Vergani}, {Zanichelli}, {Burden}, {Di Porto}, {Marchetti}, {Marinoni},
  {Mellier}, {Moutard}, {Moscardini}, {Nichol}, {Peacock}, {Percival}, \&
  {Zamorani}}]{micheletti14}
{Micheletti}, D., {Iovino}, A., {Hawken}, A.~J., {et~al.} 2014, \aap, 570, A106

\bibitem[{{Miller} {et~al.}(2013){Miller}, {Heymans}, {Kitching}, {van
  Waerbeke}, {Erben}, {Hildebrandt}, {Hoekstra}, {Mellier}, {Rowe}, {Coupon},
  {Dietrich}, {Fu}, {Harnois-D{\'e}raps}, {Hudson}, {Kilbinger}, {Kuijken},
  {Schrabback}, {Semboloni}, {Vafaei}, \& {Velander}}]{miller13}
{Miller}, L., {Heymans}, C., {Kitching}, T.~D., {et~al.} 2013, \mnras, 429,
  2858

\bibitem[{{Miyatake} {et~al.}(2015){Miyatake}, {More}, {Mandelbaum}, {Takada},
  {Spergel}, {Kneib}, {Schneider}, {Brinkmann}, \& {Brownstein}}]{miyatake15}
{Miyatake}, H., {More}, S., {Mandelbaum}, R., {et~al.} 2015, \apj, 806, 1

\bibitem[{{More} {et~al.}(2015){More}, {Miyatake}, {Mandelbaum}, {Takada},
  {Spergel}, {Brownstein}, \& {Schneider}}]{more15}
{More}, S., {Miyatake}, H., {Mandelbaum}, R., {et~al.} 2015, \apj, 806, 2

\bibitem[{{Nakajima} {et~al.}(2012){Nakajima}, {Mandelbaum}, {Seljak}, {Cohn},
  {Reyes}, \& {Cool}}]{nakajima12}
{Nakajima}, R., {Mandelbaum}, R., {Seljak}, U., {et~al.} 2012, \mnras, 420,
  3240

\bibitem[{{Navarro} {et~al.}(1996){Navarro}, {Frenk}, \& {White}}]{navarro96}
{Navarro}, J.~F., {Frenk}, C.~S., \& {White}, S.~D.~M. 1996, \apj, 462, 563

\bibitem[{{Navarro} {et~al.}(1997){Navarro}, {Frenk}, \& {White}}]{navarro97}
{Navarro}, J.~F., {Frenk}, C.~S., \& {White}, S.~D.~M. 1997, \apj, 490, 493

\bibitem[{{Okumura} {et~al.}(2016){Okumura}, {Hikage}, {Totani}, {Tonegawa},
  {Okada}, {Glazebrook}, {Blake}, {Ferreira}, {More}, {Taruya}, {Tsujikawa},
  {Akiyama}, {Dalton}, {Goto}, {Ishikawa}, {Iwamuro}, {Matsubara},
  {Nishimichi}, {Ohta}, {Shimizu}, {Takahashi}, {Takato}, {Tamura}, {Yabe}, \&
  {Yoshida}}]{okumura16}
{Okumura}, T., {Hikage}, C., {Totani}, T., {et~al.} 2016, \pasj, 68, 38

\bibitem[{{Paz} \& {S{\'a}nchez}(2015)}]{paz15}
{Paz}, D.~J. \& {S{\'a}nchez}, A.~G. 2015, \mnras, 454, 4326

\bibitem[{{Peacock} {et~al.}(2001){Peacock}, {Cole}, {Norberg}, {Baugh},
  {Bland-Hawthorn}, {Bridges}, {Cannon}, {Colless}, {Collins}, {Couch},
  {Dalton}, {Deeley}, {De Propris}, {Driver}, {Efstathiou}, {Ellis}, {Frenk},
  {Glazebrook}, {Jackson}, {Lahav}, {Lewis}, {Lumsden}, {Maddox}, {Percival},
  {Peterson}, {Price}, {Sutherland}, \& {Taylor}}]{peacock01}
{Peacock}, J.~A., {Cole}, S., {Norberg}, P., {et~al.} 2001, \nat, 410, 169

\bibitem[{{Percival} {et~al.}(2010){Percival}, {Reid}, {Eisenstein}, {Bahcall},
  {Budavari}, {Frieman}, {Fukugita}, {Gunn}, {Ivezi{\'c}}, {Knapp}, {Kron},
  {Loveday}, {Lupton}, {McKay}, {Meiksin}, {Nichol}, {Pope}, {Schlegel},
  {Schneider}, {Spergel}, {Stoughton}, {Strauss}, {Szalay}, {Tegmark},
  {Vogeley}, {Weinberg}, {York}, \& {Zehavi}}]{percival10}
{Percival}, W.~J., {Reid}, B.~A., {Eisenstein}, D.~J., {et~al.} 2010, \mnras,
  401, 2148

\bibitem[{{Percival} {et~al.}(2014){Percival}, {Ross}, {S{\'a}nchez},
  {Samushia}, {Burden}, {Crittenden}, {Cuesta}, {Magana}, {Manera}, {Beutler},
  {Chuang}, {Eisenstein}, {Ho}, {McBride}, {Montesano}, {Padmanabhan}, {Reid},
  {Saito}, {Schneider}, {Seo}, {Tojeiro}, \& {Weaver}}]{percival14}
{Percival}, W.~J., {Ross}, A.~J., {S{\'a}nchez}, A.~G., {et~al.} 2014, \mnras,
  439, 2531

\bibitem[{{Pezzotta} {et~al.}(2017){Pezzotta}, {de la Torre}, {Bel}, {Granett},
  {Guzzo}, {Peacock}, {Garilli}, {Scodeggio}, {Bolzonella}, {Abbas}, {Adami},
  {Bottini}, {Cappi}, {Cucciati}, {Davidzon}, {Franzetti}, {Fritz}, {Iovino},
  {Krywult}, {Le Brun}, {Le F{\`e}vre}, {Maccagni}, {Ma{\l}ek}, {Marulli},
  {Polletta}, {Pollo}, {Tasca}, {Tojeiro}, {Vergani}, {Zanichelli}, {Arnouts},
  {Branchini}, {Coupon}, {De Lucia}, {Koda}, {Ilbert}, {Mohammad}, {Moutard},
  \& {Moscardini}}]{pezzotta16}
{Pezzotta}, A., {de la Torre}, S., {Bel}, J., {et~al.} 2017, \aap, 604, A33

\bibitem[{{Planck Collaboration} {et~al.}(2016){Planck Collaboration}, {Ade},
  {Aghanim}, {Arnaud}, {Ashdown}, {Aumont}, {Baccigalupi}, {Banday},
  {Barreiro}, {Bartlett}, \& et~al.}]{planck15}
{Planck Collaboration}, {Ade}, P.~A.~R., {Aghanim}, N., {et~al.} 2016, \aap,
  594, A13

\bibitem[{{Pope} \& {Szapudi}(2008)}]{pope08}
{Pope}, A.~C. \& {Szapudi}, I. 2008, \mnras, 389, 766

\bibitem[{{Pullen} {et~al.}(2016){Pullen}, {Alam}, {He}, \& {Ho}}]{pullen16}
{Pullen}, A.~R., {Alam}, S., {He}, S., \& {Ho}, S. 2016, \mnras, 460, 4098

\bibitem[{{Reyes} {et~al.}(2010){Reyes}, {Mandelbaum}, {Seljak}, {Baldauf},
  {Gunn}, {Lombriser}, \& {Smith}}]{reyes10}
{Reyes}, R., {Mandelbaum}, R., {Seljak}, U., {et~al.} 2010, \nat, 464, 256

\bibitem[{{Rota} {et~al.}(2017){Rota}, {Granett}, {Bel}, {Guzzo}, {Peacock},
  {Wilson}, {Pezzotta}, {de la Torre}, {Garilli}, {Bolzonella}, {Scodeggio},
  {Abbas}, {Adami}, {Bottini}, {Cappi}, {Cucciati}, {Davidzon}, {Franzetti},
  {Fritz}, {Iovino}, {Krywult}, {Le Brun}, {Le F{\`e}vre}, {Maccagni},
  {Ma{\l}ek}, {Marulli}, {Percival}, {Polletta}, {Pollo}, {Tasca}, {Tojeiro},
  {Vergani}, {Zanichelli}, {Arnouts}, {Branchini}, {Coupon}, {De Lucia},
  {Ilbert}, {Moscardini}, \& {Moutard}}]{rota16}
{Rota}, S., {Granett}, B.~R., {Bel}, J., {et~al.} 2017, \aap, 601, A144

\bibitem[{{Saito} {et~al.}(2014){Saito}, {Baldauf}, {Vlah}, {Seljak},
  {Okumura}, \& {McDonald}}]{saito14}
{Saito}, S., {Baldauf}, T., {Vlah}, Z., {et~al.} 2014, \prd, 90, 123522

\bibitem[{{Samushia} {et~al.}(2012){Samushia}, {Percival}, \&
  {Raccanelli}}]{samushia12}
{Samushia}, L., {Percival}, W.~J., \& {Raccanelli}, A. 2012, \mnras, 420, 2102

\bibitem[{{Samushia} {et~al.}(2014){Samushia}, {Reid}, {White}, {Percival},
  {Cuesta}, {Zhao}, {Ross}, {Manera}, {Aubourg}, {Beutler}, {Brinkmann},
  {Brownstein}, {Dawson}, {Eisenstein}, {Ho}, {Honscheid}, {Maraston},
  {Montesano}, {Nichol}, {Roe}, {Ross}, {S{\'a}nchez}, {Schlegel}, {Schneider},
  {Streblyanska}, {Thomas}, {Tinker}, {Wake}, {Weaver}, \&
  {Zehavi}}]{samushia14}
{Samushia}, L., {Reid}, B.~A., {White}, M., {et~al.} 2014, \mnras, 439, 3504

\bibitem[{{Scoccimarro} {et~al.}(1999){Scoccimarro}, {Couchman}, \&
  {Frieman}}]{scoccimarro99}
{Scoccimarro}, R., {Couchman}, H.~M.~P., \& {Frieman}, J.~A. 1999, \apj, 517,
  531

\bibitem[{{Scodeggio} {et~al.}(2009){Scodeggio}, {Franzetti}, {Garilli}, {Le
  F{\`e}vre}, \& {Guzzo}}]{scodeggio09}
{Scodeggio}, M., {Franzetti}, P., {Garilli}, B., {Le F{\`e}vre}, O., \&
  {Guzzo}, L. 2009, The Messenger, 135, 13

\bibitem[{{Scodeggio} {et~al.}(2017){Scodeggio}, {Guzzo}, {Garilli}, {Granett},
  {Bolzonella}, {de la Torre}, {Abbas}, {Adami}, {Arnouts}, {Bottini}, {Cappi},
  {Coupon}, {Cucciati}, {Davidzon}, {Franzetti}, {Fritz}, {Iovino}, {Krywult},
  {Le Brun}, {Le F{\'e}vre}, {Maccagni}, {Malek}, {Marchetti}, {Marulli},
  {Polletta}, {Pollo}, {Tasca}, {Tojeiro}, {Vergani}, {Zanichelli}, {Bel},
  {Branchini}, {De Lucia}, {Ilbert}, {McCracken}, {Moutard}, {Peacock},
  {Zamorani}, {Burden}, {Fumana}, {Jullo}, {Marinoni}, {Mellier}, {Moscardini},
  \& {Percival}}]{scodeggio16}
{Scodeggio}, M., {Guzzo}, L., {Garilli}, B., {et~al.} 2017, \aap, in press,
  ArXiv e-print: 1611.07048

\bibitem[{{Simpson} {et~al.}(2013){Simpson}, {Heymans}, {Parkinson}, {Blake},
  {Kilbinger}, {Benjamin}, {Erben}, {Hildebrandt}, {Hoekstra}, {Kitching},
  {Mellier}, {Miller}, {Van Waerbeke}, {Coupon}, {Fu}, {Harnois-D{\'e}raps},
  {Hudson}, {Kuijken}, {Rowe}, {Schrabback}, {Semboloni}, {Vafaei}, \&
  {Velander}}]{simpson13}
{Simpson}, F., {Heymans}, C., {Parkinson}, D., {et~al.} 2013, \mnras, 429, 2249

\bibitem[{{Smith} {et~al.}(2003){Smith}, {Peacock}, {Jenkins}, {White},
  {Frenk}, {Pearce}, {Thomas}, {Efstathiou}, \& {Couchman}}]{smith03}
{Smith}, R.~E., {Peacock}, J.~A., {Jenkins}, A., {et~al.} 2003, \mnras, 341,
  1311

\bibitem[{{Takahashi} {et~al.}(2012){Takahashi}, {Sato}, {Nishimichi},
  {Taruya}, \& {Oguri}}]{takahashi12}
{Takahashi}, R., {Sato}, M., {Nishimichi}, T., {Taruya}, A., \& {Oguri}, M.
  2012, \apj, 761, 152

\bibitem[{{Taruya} {et~al.}(2010){Taruya}, {Nishimichi}, \& {Saito}}]{taruya10}
{Taruya}, A., {Nishimichi}, T., \& {Saito}, S. 2010, \prd, 82, 063522

\bibitem[{{Taylor} \& {Joachimi}(2014)}]{taylor14}
{Taylor}, A. \& {Joachimi}, B. 2014, \mnras, 442, 2728

\bibitem[{{Tegmark} {et~al.}(2004){Tegmark}, {Blanton}, {Strauss}, {Hoyle},
  {Schlegel}, {Scoccimarro}, {Vogeley}, {Weinberg}, {Zehavi}, {Berlind},
  {Budavari}, {Connolly}, {Eisenstein}, {Finkbeiner}, {Frieman}, {Gunn},
  {Hamilton}, {Hui}, {Jain}, {Johnston}, {Kent}, {Lin}, {Nakajima}, {Nichol},
  {Ostriker}, {Pope}, {Scranton}, {Seljak}, {Sheth}, {Stebbins}, {Szalay},
  {Szapudi}, {Verde}, {Xu}, {Annis}, {Bahcall}, {Brinkmann}, {Burles},
  {Castander}, {Csabai}, {Loveday}, {Doi}, {Fukugita}, {Gott}, {Hennessy},
  {Hogg}, {Ivezi{\'c}}, {Knapp}, {Lamb}, {Lee}, {Lupton}, {McKay}, {Kunszt},
  {Munn}, {O'Connell}, {Peoples}, {Pier}, {Richmond}, {Rockosi}, {Schneider},
  {Stoughton}, {Tucker}, {Vanden Berk}, {Yanny}, {York}, \& {SDSS
  Collaboration}}]{tegmark04}
{Tegmark}, M., {Blanton}, M.~R., {Strauss}, M.~A., {et~al.} 2004, \apj, 606,
  702

\bibitem[{{Tinker} {et~al.}(2008){Tinker}, {Kravtsov}, {Klypin}, {Abazajian},
  {Warren}, {Yepes}, {Gottl{\"o}ber}, \& {Holz}}]{tinker08}
{Tinker}, J., {Kravtsov}, A.~V., {Klypin}, A., {et~al.} 2008, \apj, 688, 709

\bibitem[{{Tinker} {et~al.}(2010){Tinker}, {Robertson}, {Kravtsov}, {Klypin},
  {Warren}, {Yepes}, \& {Gottl{\"o}ber}}]{tinker10}
{Tinker}, J.~L., {Robertson}, B.~E., {Kravtsov}, A.~V., {et~al.} 2010, \apj,
  724, 878

\bibitem[{{Velander} {et~al.}(2014){Velander}, {van Uitert}, {Hoekstra},
  {Coupon}, {Erben}, {Heymans}, {Hildebrandt}, {Kitching}, {Mellier}, {Miller},
  {Van Waerbeke}, {Bonnett}, {Fu}, {Giodini}, {Hudson}, {Kuijken}, {Rowe},
  {Schrabback}, \& {Semboloni}}]{velander14}
{Velander}, M., {van Uitert}, E., {Hoekstra}, H., {et~al.} 2014, \mnras, 437,
  2111

\bibitem[{{Wilson} {et~al.}(2017)}]{wilson16}
{Wilson}, M. {et~al.} 2017, \rm{in preparation}

\bibitem[{{Xu} {et~al.}(2013){Xu}, {Cuesta}, {Padmanabhan}, {Eisenstein}, \&
  {McBride}}]{xu13}
{Xu}, X., {Cuesta}, A.~J., {Padmanabhan}, N., {Eisenstein}, D.~J., \&
  {McBride}, C.~K. 2013, \mnras, 431, 2834

\bibitem[{{Zhang} {et~al.}(2007){Zhang}, {Liguori}, {Bean}, \&
  {Dodelson}}]{Zhang07}
{Zhang}, P., {Liguori}, M., {Bean}, R., \& {Dodelson}, S. 2007, Physical Review
  Letters, 99, 141302

\end{thebibliography}

\appendix

\section{Theoretical power spectra for biased tracers}\label{appendix}

This appendix presents the models describing the real-space
galaxy-galaxy, galaxy-velocity divergence, and galaxy-matter power
spectra, which enter the modelling of RSD and galaxy-galaxy
lensing. We adopt the non-linear non-local bias model of
\citet{mcdonald09} that relates the galaxy overdensity $\delta_{g}$
and matter overdensity $\delta$ as:
\begin{eqnarray} \label{eq:bias2}
\delta_{g}({\bf x}) &=& b_1 \delta({\bf x}) + \frac{1}{2}b_2
      [\delta^2({\bf x})-\sigma^2]+\frac{1}{2}b_{s^2}[s^2({\bf
          x})-\langle s^2 \rangle] \nonumber \\
      && + O(s^3({\bf x})).
\end{eqnarray}
In this equation, $b_1$ and $b_2$ are the linear and
second-order non-linear bias terms, $b_{s^2}$ the non-local bias term,
$s$ is the tidal tensor term from which non-locality originates. The
$\sigma^2$ and $\langle s^2\rangle$ terms ensure the condition
$\langle \delta_g\rangle=0$. From the bias model of Eq. \ref{eq:bias2}
one can derive the following power spectra for galaxy-galaxy,
galaxy-velocity divergence ($\theta$), and galaxy-matter correlations:
\begin{eqnarray}
  P_{gg}(k)&=&b_1^2 P_{\delta\delta}(k)+2b_2b_1P_{b2,\delta}(k)+2b_{s^2}b_1P_{bs2,\delta}(k) \nonumber \\
  && +b_2^2P_{b22}(k) +2b_2b_{s^2}P_{b2s2}(k)+b_{s^2}^2P_{bs22}(k) \nonumber \\
  && +2b_1b_{3\rm nl}\sigma_3^2(k)P_{\rm lin}(k) + N, \\
  P_{g\theta}(k)&=&b_1P_{\delta\theta}(k)+b_2P_{b2,\theta}(k)+b_{s^2}P_{bs2,\theta}(k) \nonumber \\
  && +b_{3\rm nl}\sigma_3^2(k)P_{\rm lin}(k), \\
  P_{gm}(k) &=& b_1 P_{\delta\delta}(k) + b_2 P_{b2,\delta}(k) + b_{s^2} P_{bs2,\delta}(k) \nonumber \\
  && + b_{3nl}\sigma^2_3(k)P_{\rm lin}(k),
\end{eqnarray}
where \citep[e.g.][]{beutler14,gilmarin14}
\begin{eqnarray}
  P_{b2,\delta}(k)&=&\int \frac{d^3q}{(2\pi)^3}\, P_{\rm lin}(q)P_{\rm lin}(|{\bf k}-{\bf q}|) F_2({\bf q},{\bf k-q}),\\
  P_{bs2,\delta}(k)&=& \int \frac{d^3q}{(2\pi)^3}\, P_{\rm lin}(q)P_{\rm lin}(|{\bf k}-{\bf q}|) F_2({\bf q},{\bf k-q}) \nonumber \\
  && S_2({\bf q},{\bf k-q}),  \\
  P_{b2,\theta}(k)&=&\int\frac{d^3q}{(2\pi)^3}\,P_{\rm lin}(q)P_{\rm lin}(|{\bf k}-{\bf q}|)G_2({\bf q},{\bf k}-{\bf q}),\\
  P_{bs2,\theta}(k)&=&\int\frac{d^3q}{(2\pi)^3}\,P_{\rm lin}(q)P_{\rm lin}(|{\bf k}-{\bf q}|)G_2({\bf q},{\bf k}-{\bf q}) \nonumber \\
  && S_2({\bf q},{\bf k}-{\bf q}), \\
  P_{b2s2}(k)&=& -\frac{1}{2}\int \frac{d^3q}{(2\pi)^3}\, P_{\rm lin}(q)\left[ \frac{2}{3}P_{\rm lin}(q) - P_{\rm lin}(|{\bf k}-{\bf q}|) \right. \nonumber \\
    && \left. S_2({\bf q},{\bf k}-{\bf q}) \right],  \\
  P_{bs22}(k)&=& -\frac{1}{2} \int \frac{d^3q}{(2\pi)^3}\, P_{\rm lin}(q)\left[  \frac{4}{9}P_{\rm lin}(q) - P_{\rm lin}(|{\bf k}-{\bf q}|) \right. \nonumber \\
    && \left. S_2({\bf q},{\bf k}-{\bf q})^2\right], \\
  P_{b22}(k)&=& -\frac{1}{2} \int \frac{d^3q}{(2\pi)^3}\, P_{\rm lin}(q)\left[ P_{\rm lin}(q)-P_{\rm lin}({\bf k}-{\bf q}|)\right],\\
  \sigma^2_3(k)&=&\int\frac{d^3{\bf q}}{(2\pi)^3}\,P_{\rm lin}(q)\left[ \frac{5}{6}+\frac{15}{8}S_2({\bf q},{\bf k}-{\bf q}) \right. \nonumber \\
    && \left. S_2(-{\bf q},{\bf k}) -\frac{5}{4}S_2({\bf q},{\bf k}-{\bf q})  \right].
\end{eqnarray}
In the above equations, $S_2$, $F_2$, $G_2$ perturbation theory
kernels are defined by \citep[e.g.][]{goroff86,bernardeau02}
\begin{eqnarray}
  S_2({\bf k}_i,{\bf k}_j)&=&\frac{({\bf k}_i\cdot{\bf k}_j)^2}{(k_ik_j)^2}-\frac{1}{3}, \\
  F_2({\bf k}_i,{\bf k}_j)&=& \frac{5}{7}+\frac{1}{2}\frac{{\bf k}_i\cdot{\bf k}_j}{k_ik_j}\left(\frac{k_i}{k_j}+\frac{k_j}{k_i}\right)+\frac{2}{7}\left[ \frac{{\bf k}_i\cdot{\bf k}_j}{k_ik_j} \right]^2, \\
  G_2({\bf k}_i,{\bf k}_j)&=& \frac{3}{7}+\frac{1}{2}\frac{{\bf k}_i\cdot{\bf k}_j}{k_ik_j}\left(\frac{k_i}{k_j}+\frac{k_j}{k_i}\right)+\frac{4}{7}\left[ \frac{{\bf k}_i\cdot{\bf k}_j}{k_ik_j} \right]^2.
\end{eqnarray}

\section{Posterior likelihood contours}\label{appendix2}

\begin{figure*}[ht]
\centering
\includegraphics[width=9cm]{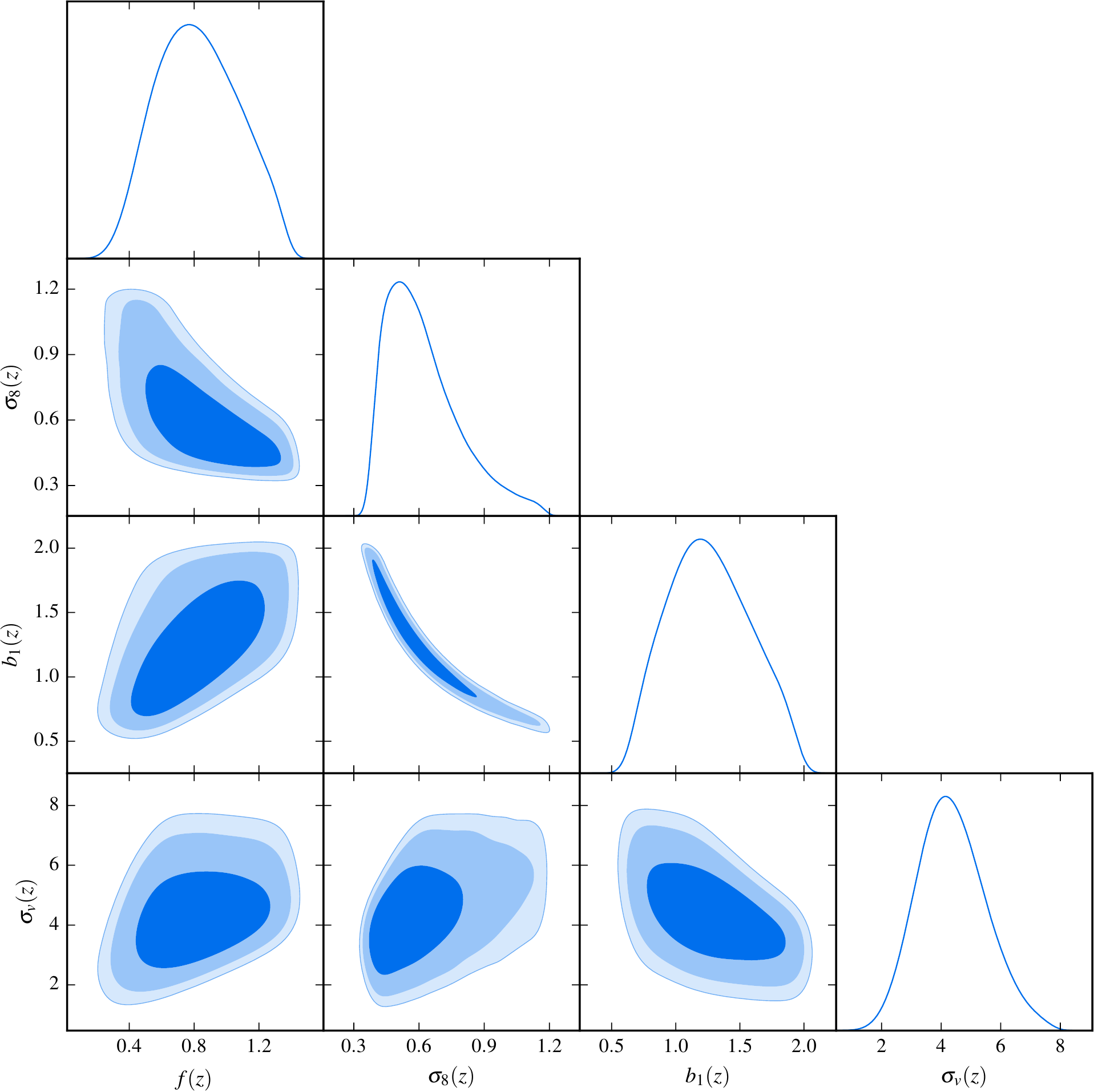}
\includegraphics[width=9cm]{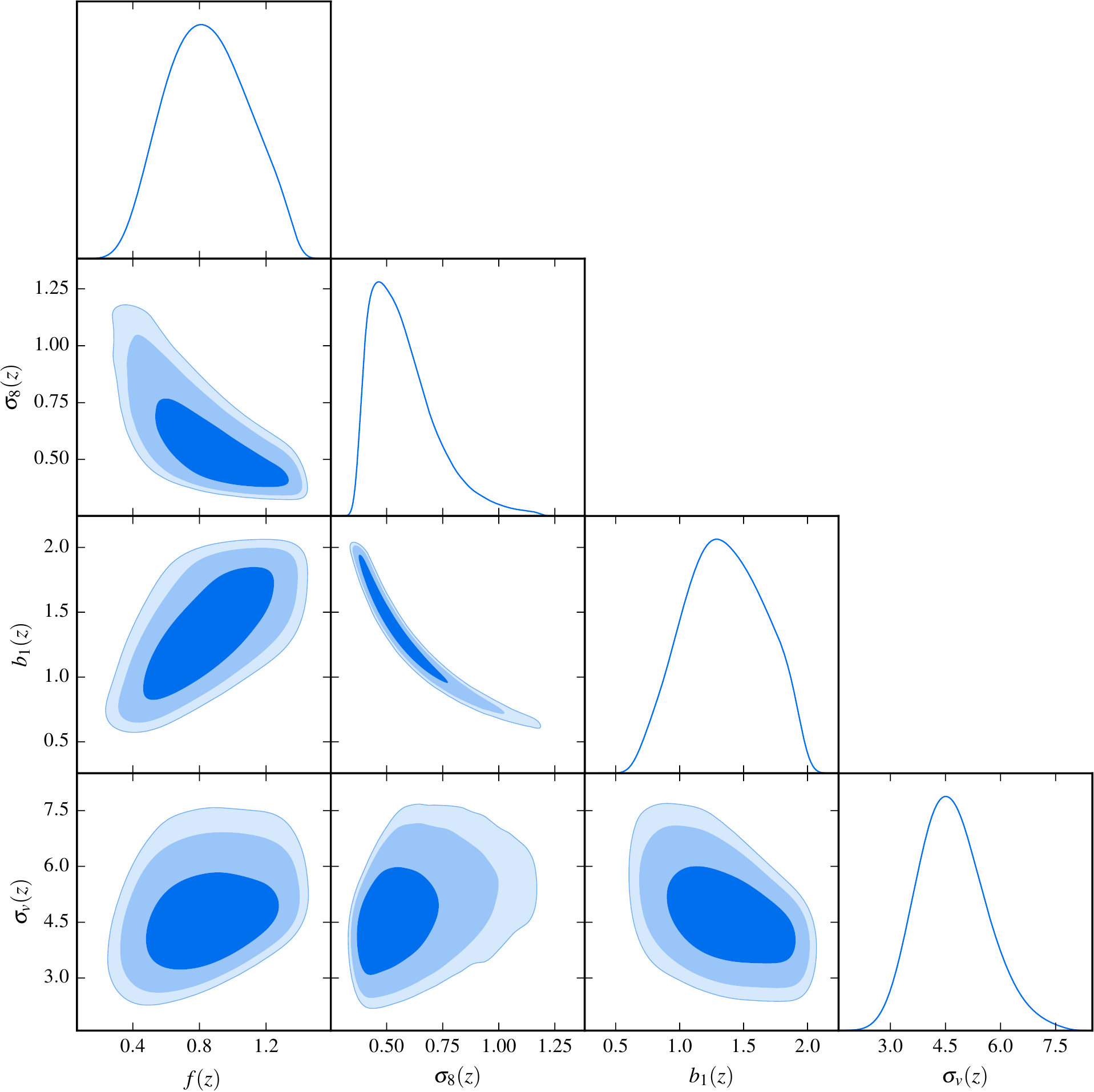}
\caption{Posterior likelihood contours for $f$, $\sigma_8$, $b_1$, and
  $\sigma_v$ parameters at $z=0.6$ (left panel) and $z=0.86$ (right
  panel) in the case where RSD are considered alone (see
  Sect. \ref{sec:fsig8meas}).}
\label{fig:cont1}
\end{figure*}

\begin{figure*}[ht]
\centering
\includegraphics[width=9cm]{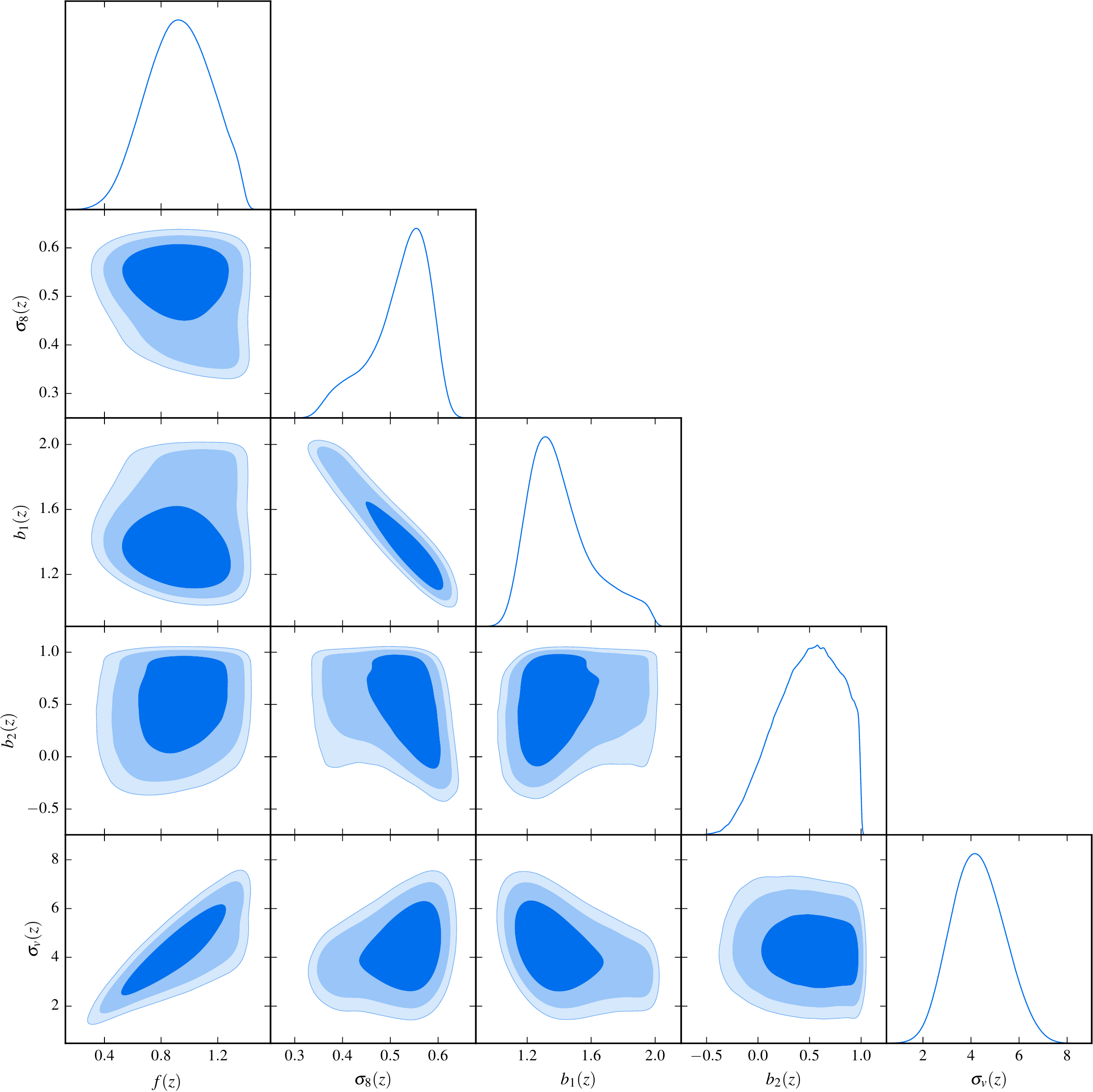}
\includegraphics[width=9cm]{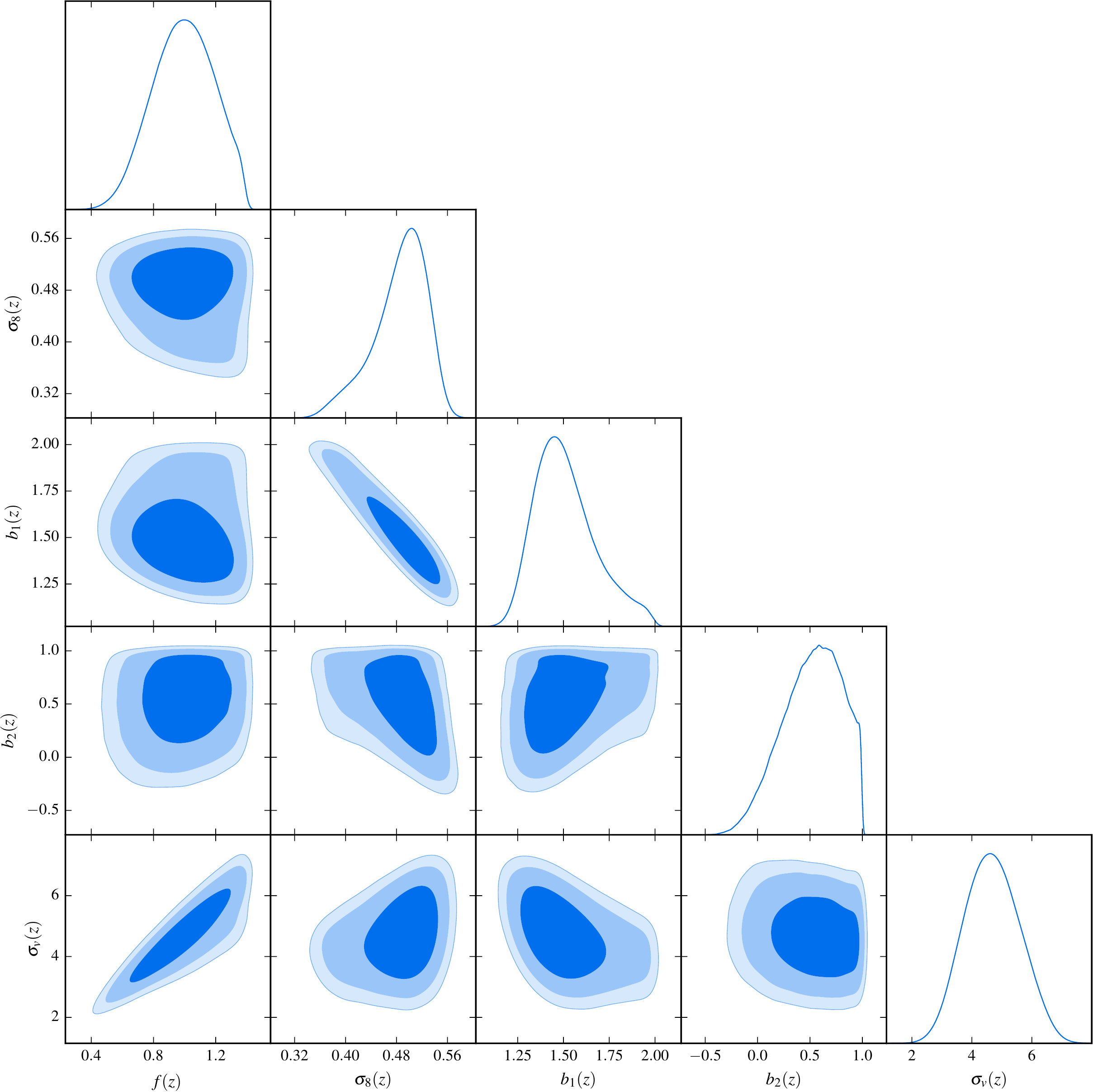}
\caption{Posterior likelihood contours for $f$, $\sigma_8$, $b_1$,
  $b_2$, and $\sigma_v$ parameters at $z=0.6$ (left panel) and
  $z=0.86$ (right panel) in the case where RSD and galaxy-galaxy
  lensing are combined (see Sect. \ref{sec:fsig8meas}).}
\label{fig:cont2}
\end{figure*}

In this appendix are provided the posterior likelihood contours of all
pairs of parameters appearing in the likelihood analyses presented in
Sect. \ref{sec:fsig8meas}. Fig. \ref{fig:cont1} shows the posterior
likelihood contours in the case of the RSD-only analysis, while
Fig. \ref{fig:cont2} in the case where RSD and galaxy-galaxy lensing
are combined. In both figures, the three types of shaded regions in
each subpanel correspond to the posterior likelihood contours at
$68\%$, $95\%$, and $99\%$.

\end{document}